\documentclass[aps,pra,reprint,amsmath,amssymb,superscriptaddress,onecolumn,longbibliography, notitlepage, 11pt]{revtex4-2}
\usepackage{mathtools}
\usepackage{siunitx}
\usepackage[hidelinks]{hyperref}
\usepackage{svg}
\usepackage{caption}
\usepackage{float}
\usepackage{subcaption}
\usepackage[braket, qm]{qcircuit}
\usepackage{xurl}
\usepackage[mathlines]{lineno}

\usepackage{bbm}

\usepackage[labelfont=bf, justification=RaggedRight, figurename={Figure }, tablename={Table}, labelsep=period]{caption}

\usepackage{xcolor}

\begin{document}
\title{Image sensing with multilayer, nonlinear optical neural networks }


\author{Tianyu~Wang}
\email{These authors contributed equally.}
\affiliation{School of Applied and Engineering Physics, Cornell University, Ithaca, NY 14853, USA}

\author{Mandar~M.~Sohoni}
\email{These authors contributed equally.}
\affiliation{School of Applied and Engineering Physics, Cornell University, Ithaca, NY 14853, USA}

\author{Logan~G.~Wright} 
\affiliation{School of Applied and Engineering Physics, Cornell University, Ithaca, NY 14853, USA}
\affiliation{NTT Physics and Informatics Laboratories, NTT Research, Inc., Sunnyvale, CA 94085, USA}

\author{Martin~M.~Stein}
\affiliation{School of Applied and Engineering Physics, Cornell University, Ithaca, NY 14853, USA}

\author{Shi-Yuan~Ma}
\affiliation{School of Applied and Engineering Physics, Cornell University, Ithaca, NY 14853, USA}

\author{Tatsuhiro~Onodera}
\affiliation{School of Applied and Engineering Physics, Cornell University, Ithaca, NY 14853, USA}
\affiliation{NTT Physics and Informatics Laboratories, NTT Research, Inc., Sunnyvale, CA 94085, USA}

\author{Maxwell~Anderson}
\affiliation{School of Applied and Engineering Physics, Cornell University, Ithaca, NY 14853, USA}

\author{Peter~L.~McMahon}
\email{To whom correspondence should be addressed:\newline tw329@cornell.edu, mms477@cornell.edu, lgw32@cornell.edu, pmcmahon@cornell.edu}
\affiliation{School of Applied and Engineering Physics, Cornell University, Ithaca, NY 14853, USA}
\affiliation{Kavli Institute at Cornell for Nanoscale Science, Cornell University, Ithaca, NY 14853, USA}

\begin{abstract}
Optical imaging is commonly used for both scientific and technological applications across industry and academia. In image sensing, a measurement, such as of an object's position or contour, is performed by computational analysis of a digitized image. An emerging image-sensing paradigm \cite{duarte2008single, chang2018hybrid, martel2020neural, wetzstein2020inference, brady2020smart, Musk2021aiDay} breaks this delineation between data collection and analysis by designing optical components to perform not imaging, but \textit{encoding}. By optically encoding images into a compressed, low-dimensional latent space suitable for efficient computational post-analysis, these image sensors can operate with fewer pixels and fewer photons, allowing higher-throughput, lower-latency operation. Optical neural networks (ONNs), primarily developed as accelerators for deep-neural-network inference, offer a platform for processing data in the analog, optical domain \cite{farhat1985optical, denz1998onns, Shen2017, Lin2018, hamerly2019large, wetzstein2020inference, nahmias2020photonic, xu202111, feldmann2021parallel, shastri2021photonics, wang2021optical, bernstein2022single, tait2022quantifying, huang2022prospects, sludds2022delocalized}. ONN-based sensors have however been limited to linear processing, which is equivalent to a single-layer neural network (NN) \cite{matic1989comparison, kubala2003reducing, duarte2008single, stork2008theoretical, chang2018hybrid, colburn2019optical, martel2020neural, wetzstein2020inference, Mennel2020, pad2020efficient, zheng2022meta, shi2022loen, chen2022photonic, bezzam2022learning}.  Nonlinearity is a prerequisite for depth, and multilayer NNs significantly outperform shallow NNs on many tasks, including image processing and compression \cite{hinton2006reducing, krizhevsky2012imagenet, Goodfellow-et-al-2016}. Here, we realize a multilayer ONN pre-processor for image sensing, using a commercial image intensifier as a parallel optoelectronic, optical-to-optical nonlinear activation function. We demonstrate that the nonlinear ONN pre-processor can achieve compression ratios of up to 800:1 -- corresponding to compressing the input to just a two-dimensional output vector -- while still enabling high accuracy across several representative computer-vision tasks, including machine-vision benchmarks, flow-cytometry image classification, and measurement and identification of objects in real scenes. In all cases we find that the ONN's nonlinearity and depth allowed it to outperform a purely linear ONN encoder. Although our experimental demonstrations are specialized to ONN sensors for incoherent-light images, the growing multitude of ONN platforms should facilitate a range of ONN sensors. These ONN sensors may surpass conventional sensors by pre-processing optical information in spatial, temporal, and/or spectral dimensions, potentially with coherent and quantum qualities, all natively in the optical domain.
\end{abstract}

\maketitle

\section{Introduction}
\label{sec:intro}

Optical images are widely used to capture and convey information about the state or dynamics of physical systems, both in fundamental science and in technology. They are used to guide autonomous machines, to assess manufacturing processes, and to inform medical diagnoses and procedures. In all these applications, an optical system such as a microscope forms an image of a subject on a camera, which converts the photonic, analog image into an electronic, digital image. Digital images are typically many megabytes. For most applications however, nearly all this information is redundant or irrelevant. There are three main reasons: (i) natural images contain sparse information, and are therefore compressible \cite{duarte2008single, sterling_principles_2015, gibson2020single}; (ii) most applications involve images of subjects with additional underlying commonalities beyond sparsity, and (iii) most information in an image is irrelevant to the image's end use. Here, we refer to machine-vision applications, for which factor (iii) is applicable, as \textit{image sensing} -- for these applications only a specific subset of information from each image is sought, as depicted in Figure \ref{Fig1}a.

The information inefficiency of conventional imaging has inspired machine-vision paradigms in which optics are designed not as a conventional imaging system, but instead as an optical encoder -- a computational pre-processor that extracts relevant information from an image \cite{duarte2008single, chang2018hybrid, martel2020neural, wetzstein2020inference,  Mennel2020, pad2020efficient, Musk2021aiDay, zheng2022meta, shi2022loen, chen2022photonic, bezzam2022learning}. Techniques include end-to-end optimization \cite{matic1989comparison, kubala2003reducing, stork2008theoretical, sitzmann2018end, colburn2019optical, martel2020neural, pad2020efficient, kim2020multi, Musk2021aiDay, markley2021physics, vargas2021time, zheng2022meta, chen2022photonic, bezzam2022learning}, compressed sensing and single-pixel imaging \cite{duarte2008single, liutkus2014imaging, gibson2020single, li2021spectrally}, coded apertures \cite{asif2016flatcam, vargas2021time, baek2021single, shi2022loen, bezzam2022learning}, and related approaches for computational lensless imaging \cite{sinha2017lensless, boominathan2022recent}. A common feature and limitation of these techniques is that the optics perform only linear operations. In end-to-end optimization, where the processing algorithm is usually a DNN, the optical system is effectively a linear encoder equivalent to a single-layer NN \cite{colburn2019optical, martel2020neural,  pad2020efficient, wetzstein2020inference, zheng2022meta, shi2022loen}. Since these approaches may emulate the hierarchical information distillation of biological vision, some are considered neuromorphic \cite{serre2010neuromorphic, sterling_principles_2015, mead2020we}. Related trends include the broader fields of smart cameras \cite{brady2020smart}, in- and near-sensor computing \cite{burt1988smart, zhou2020near, Mennel2020}, variational quantum sensors \cite{marciniak2022optimal}, and machine-learning-enabled smart sensors \cite{ballard2021machine, ma2022intelligent}.

Optical pre-processing allows image-sensing systems to overcome a fundamental bottleneck in performance, enabling faster, smaller, and more energy-efficient image sensors. In a conventional image sensor, using a camera with $C$-fold fewer pixels typically leads to a $C$-fold improvement in achievable frame rate, number of photons per pixel, system power, size, weight, and cost, and at least a $C$-fold-lower decision latency. The reason for this is that these performance metrics are directly bottlenecked by the speed and energy cost of transducing images from the optical to digital electronic domain, of transporting them from the sensor to post-processor, and of post-processing high-dimensional digital image data. Nevertheless, conventional imaging pixel resolution cannot be reduced significantly without losing important content. This fundamental trade-off can be circumvented by using optical encoders to compress images into a low-dimensional latent feature space. In most applications, compression by $\gg$10 times is in principle feasible. However, while such high compression is routinely achieved with electronic DNNs, the computational capacity of simple optical encoders (such as single random or optimized masks) is rarely sufficient.  

Fortuitously, more computationally capable optical encoders are now within reach due to recent advances in optical neural networks (ONNs) \cite{wetzstein2020inference, shastri2021photonics, huang2022prospects}. ONNs are optoelectronic systems that optically perform mathematical operations involved in typical DNN inference calculations. By taking advantage of the large number of optical modes available in space, time, or frequency \cite{miller2019waves, wetzstein2020inference}, ONNs allow completely parallel, optical-domain computation of wide and densely connected layers in DNNs \cite{Lin2018, hamerly2019large, wetzstein2020inference, wang2021optical}, with the potential be orders of magnitude faster than electronic DNNs \cite{xu202111, shastri2021photonics, huang2022prospects}. While most experimentally demonstrated ONNs have involved only linear operations, such as matrix-vector multiplications and convolutions, being performed optically, it is widely appreciated that ONNs should also incorporate nonlinear activation functions \cite{wetzstein2020inference}. Compelling proposals for and early demonstrations of this have been reported \cite{wagner1987multilayer, Zuo:19, fard2020experimental, ryou2021free, li2022all}, but the problem of developing a suitable nonlinearity that can enable large-scale, deep ONNs is still considered a major outstanding challenge in the field \cite{wetzstein2020inference}.  

ONNs are thus ideal for enabling a new class of image sensing devices, \textit{ONN sensors} \cite{li1993optical, chang2018hybrid, martel2020neural, wetzstein2020inference, pad2020efficient, Mennel2020, huang2022prospects, zheng2022meta}, in which an ONN pre-processes data from and in the analog optical domain, prior to its conversion into digital-electronic signals. While linear ONNs still expand the capabilities of end-to-end-optimized image sensors compared to simpler optical systems, the pre-processing they provide is nonetheless still mathematically equivalent to at most a single NN layer. Depth -– and nonlinearity –- are essential for high-performance, efficient NN image processing: deep, nonlinear networks are exponentially (in the number of neurons) more efficient than single-layer NNs at approximating practically relevant functions \cite{lin2017does, poole2016exponential}. 

Here, we demonstrate an optical neural network image sensor that uses an optoelectronic optical-to-optical nonlinear activation (OONA) to perform multilayer ONN pre-processing for a variety of image sensing applications. Our multilayer, nonlinear ONN pre-processor conditionally compresses image data into a low-dimensional latent feature space in a single shot, achieving compression ratios up to 800:1. At high compression ratios, our device consistently outperforms conventional image sensing and linear optical pre-processing on experiments based on standard machine vision datasets, on flow cytometry image classification, and for real-scene object detection and measurement. The OONA used in our experiments is based on a commercial image intensifier typically used, e.g., in night-vision goggles or low-light scientific imaging. The linear layers (matrix-vector multiplications) in our ONN are implemented using a technique designed to facilitate incoherent images as direct inputs, with a microlens array for all-optical fan-out. Broadly, our findings support the use of multilayer ONNs with nonlinear activations as optical-domain pre-processors for sensors. Given the numerous ONN platforms and optical nonlinear activations now being developed, we expect that a multitude of deep ONN sensors is possible; these future sensors may detect information encoded in light’s spatial, spectral, and/or temporal degrees of freedom.  

\section{An ONN-based image sensor with optical-to-optical nonlinearity }

\begin{figure}
\includegraphics [width = 1.0\textwidth]{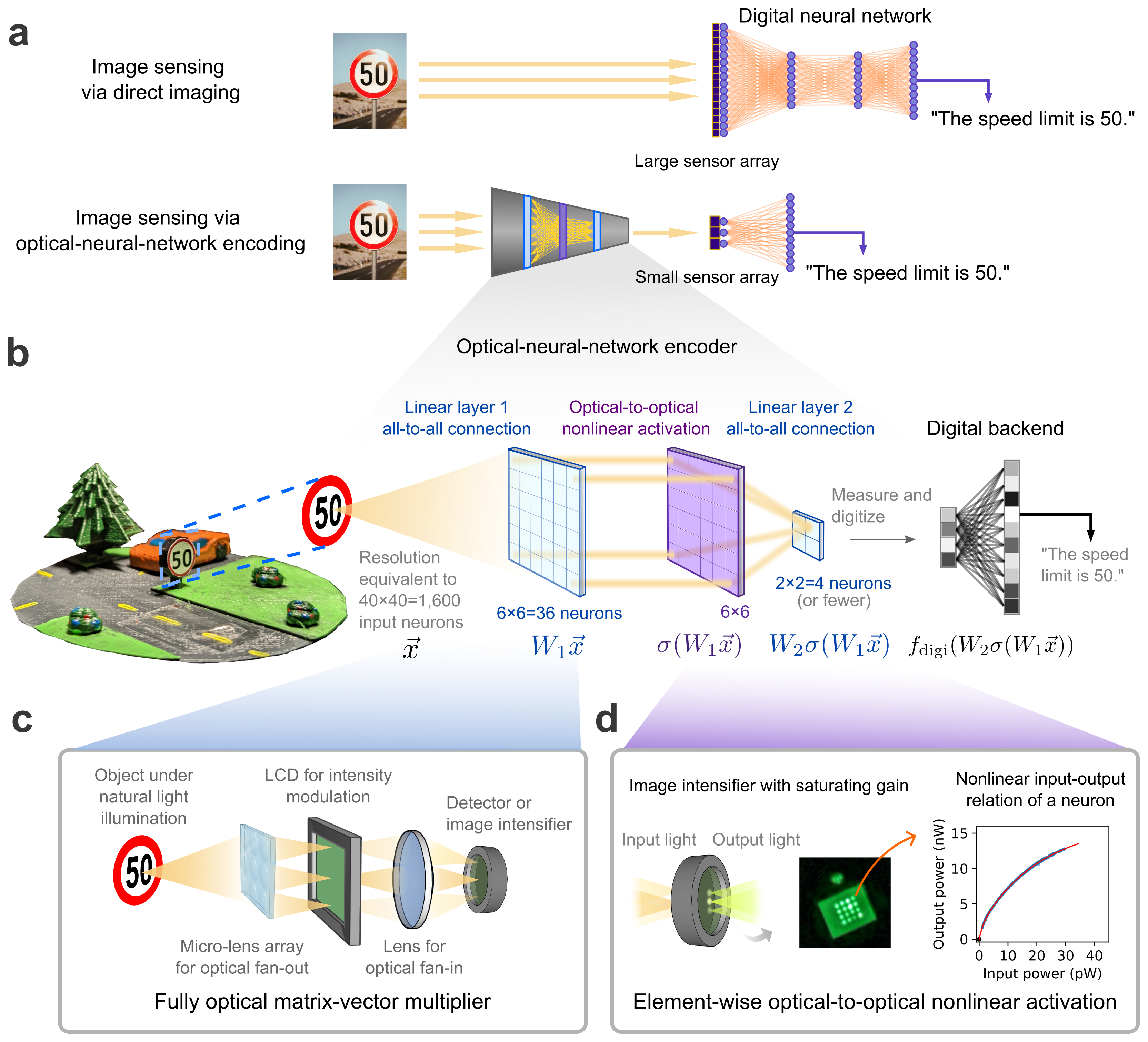}
\caption{\textbf{A multilayer optical-neural-network encoder as a frontend for image sensing.} \textbf{a}, Image sensing via direct imaging vs optical encoding. In conventional image sensing, an image is collected by a camera, and processed, often using a neural network (NN), to extract a small piece of relevant information, such as the speed limit or text of a sign. Rather than faithfully reproducing the full image of a scene onto the sensor array, an optical-neural-network (ONN) encoder instead pre-processes the image, compressing and extracting only the image information necessary for its end use, allowing a much smaller (fewer pixel) sensor array.  \textbf{b}, The neural-network diagram and corresponding mathematical operations of the ONN encoder used in this study. The ONN encoder consists of interleaved linear and nonlinear layers before the compressed signal is captured by a small photodetector array. $\vec{x}$: input image; $W_i$: weight matrix of fully connected layer $i$; $\sigma$: optical-to-optical nonlinear activation function; $f_\text{digi}$: digital backend function. \textbf{c}, The schematic of the fully optical matrix-vector multiplier used for constructing both linear layers in (b). \textbf{d}, The schematic of the optical-to-optical nonlinear activation layer realized with a saturating image intensifier. The inset plot shows that the output light intensity of a single spatial mode begins to saturate as the input light intensity increases, resembling the sigmoid activation function.}
\label{Fig1}
\end{figure}

Our experimental optical-neural-network (ONN)-based image sensor consists of a 2-layer, fully connected optical neural network in which the optical-to-optical, element-wise nonlinear activation function following the first layer is realized by a saturating microchannel plate inside a commercial image intensifier. The sensor read-out is implemented by a CMOS camera and subsequent digital post-processing. The optical pre-processor implements a nonlinear encoder – it maps input images carried by incoherent light to an abstract, lower-dimensional latent space. In our work, the encoder is not intended to perform lossless image compression. Instead, the optical encoder is trained to preserve only the information most relevant to a given image-sensing task (Fig.~\ref{Fig1}a). 

Imaging applications typically involve broadband, incoherent ambient light, so our device was designed to operate directly on broadband, incoherent visible light images. Starting with fanout -- which effectively creates copies of the input image -- implemented by an array of microlenses, optical fully connected matrix-vector multiplications are performed using a method similar to previous works \cite{wang2021optical, bernstein2022single, chen2022photonic} (see \hyperref[methods]{Methods} and Supplementary Note 2). Briefly, multiplication is achieved by attenuating the copies of the input image in proportion to the components of the weight matrix, and the summation of each output vector element is realized by focusing the attenuated light components using a lens (Fig.~\ref{Fig1}d). To realize the optical-to-optical nonlinear activation (OONA) operations applied to each element of this output vector, light is focused onto an image intensifier tube. Incident light generates free electrons from a photocathode, which are locally amplified by a microchannel plate (MCP) and then produce new, amplified bright spots as they strike a phosphor screen \cite{Zemel1991}. The local saturation of the MCP’s amplification leads to a saturating nonlinear response, qualitatively similar to the positive half of the sigmoid function (Fig.~\ref{Fig1}e and Supplementary Figure 8). Although the OONA is optoelectronic, rather than all-optical, its local, in-place realization preserves the spatial parallelism of the ONN, and avoids the time and energy costs required for read-out/in when the nonlinear activation is computed on a separate electronic processor, as in some previous work \cite{li1993optical, wang2021optical, bernstein2022single}.  To implement the second layer of the ONN, the light produced by the intensifier is processed by a second copy of the optical matrix-vector multiplier depicted in Figure \ref{Fig1}d. The output from this layer is detected by up to 4 binned superpixels of a camera (see \hyperref[methods]{Methods}), but in principle can be captured by an array of 4 photodetectors. 

To compare the multilayer, nonlinear ONN encoder to conventional (direct) image sensing and to a single-layer, linear ONN encoder, we included a beamsplitter prior to the intensifier OONA, which allowed us to reconfigure the image sensor for direct imaging (by setting the LCD from the first linear layer to be transparent) and for single-layer ONN pre-processing (by applying linear-layer weights to the first LCD).

\subsection{Nonlinear ONN encoders extract information more effectively than linear encoders}

\begin{figure}
\includegraphics[width = 1.05\textwidth]{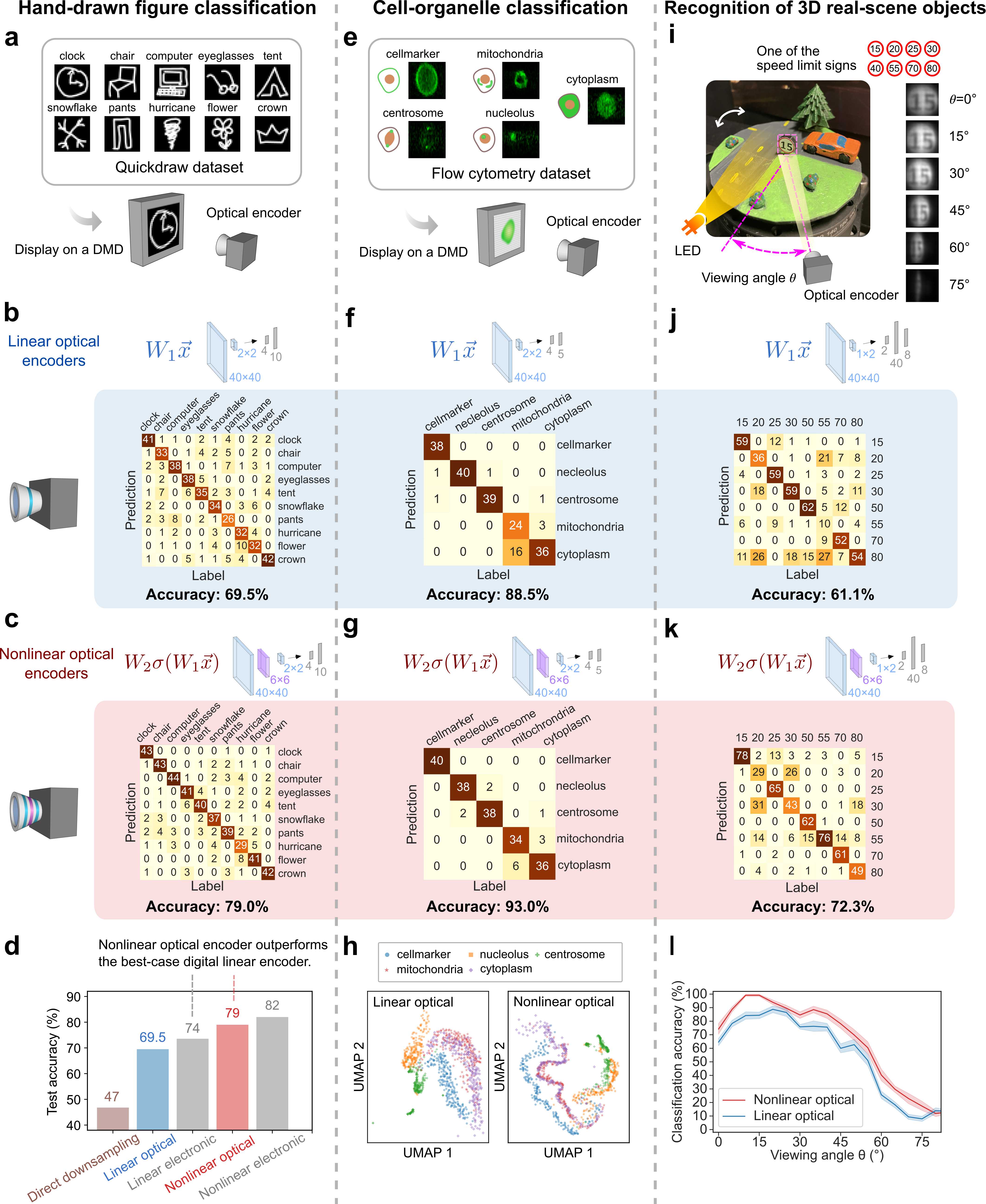}
\end{figure}

\begin{figure}
\caption{\textbf{Comparison between linear ONN and nonlinear ONN encoders on diverse image classification tasks.} \textbf{a}, Classification of hand-drawn figures from 10 different classes in the Quick, Draw! (QuickDraw) dataset. \textbf{b}, The results of QuickDraw \cite{JongejanJ.RowleyH.KawashimaT.KimJ.Fox-Gieg} classification with a linear or, \textbf{c}, nonlinear ONN encoder as the frontend. The neural-network architecture with corresponding mathematical operations is placed above the confusion matrix it produces (blue slabs: layers of linear optical neurons, purple slabs: layers of nonlinear activation function, grey bars: layers of digital neurons. $\vec{x}$: input image; $W_i$: weight matrix of fully connected layer $i$; $\sigma$: optical-to-optical nonlinear activation function). \textbf{d}, Comparison of the accuracy derived from classifiers equipped with different frontends, including direct imaging (with downsampling), a linear ONN encoder, a linear digital electronic encoder, a nonlinear ONN encoder, and a nonlinear digital electronic encoder (from left to right). In all cases, the encoder's output dimension (number of pixels) is 4. \textbf{e}, Classification of HeLa cells labeled for different organelles from a dataset acquired from flow cytometry experiments \cite{schraivogel2022high}. \textbf{f}, The results of cell-organelle classification with a linear, or a nonlinear \textbf{g} ONN encoder as the frontend. The neural-network architecture is placed above the confusion matrix it resulted in. \textbf{h}, Visualization of the compressed cell-organelle data with density uniform manifold approximation and projection (DensMAP) \cite{narayan2020density}. The data points are labeled as the ground truth. \textbf{i}, Recognizing 3D objects: each of 8 3D-printed speed-limit signs are viewed from different perspectives by an optical encoder, to classify the speed-limit number on the sign. \textbf{j}, The results of classifying speed limits with a linear or, \textbf{k}, a nonlinear ONN encoder as the frontend. The neural-network architecture is shown above the confusion matrix it produces. \textbf{l}, Classification accuracy as a function of the viewing angle. The shaded area denotes one standard deviation from the mean for repeated classification tests.}
\label{Fig2}
\end{figure}

To evaluate the performance of the multilayer, nonlinear ONN encoder, we first performed several image classification tasks, summarized in Figure \ref{Fig2}. As a benchmark, we trained classifiers for 10 pre-selected classes of the Quick, Draw! (QuickDraw) image dataset \cite{JongejanJ.RowleyH.KawashimaT.KimJ.Fox-Gieg}. Input images (28×28 pixels) were binarized and displayed on a digital micromirror display (DMD), which was placed in front of each image sensor (nonlinear multilayer, linear single-layer, and direct imaging). For a direct comparison, the vector dimension at the optical-electronic bottleneck in each sensor is the same, a 2×2 array or 4-dimensional latent space, which represents a 196:1 image-compression ratio. The nonlinear, multilayer ONN sensor achieves better classification accuracy than the other sensors (Fig.~\ref{Fig2}b-c). Since our ONN components are not perfectly calibrated, they typically perform slightly worse than digital neural networks with similar architectures. To ensure that the accuracy advantage of the nonlinear, multilayer ONN encoder is consistently better than any possible linear encoder with the same bottleneck dimension, we also trained all-digital (with real-number weights and biases) single-layer linear encoders for the same task, without image downsampling (Fig.~\ref{Fig2}d). Despite the constraint of non-negative weights and the non-analytical form of our OONA, the experimental, multilayer nonlinear ONN encoder’s performance (79\% test accuracy) robustly surpasses that of linear encoders, beating both the optical (69.5\%) and optimized digital (74\%) single-layer encoders. Compared to an ideal digital multilayer encoder with real-valued weights plus biases and sigmoid nonlinear function, the experimental nonlinear ONN encoder has a reduced test accuracy but not by much (79\% vs 82\%).

To explore the potential of ONN image sensors for a more practically important application, we next tested our image sensors on the task of classifying fluorescent images of cell organelles acquired in a flow-cytometry device \cite{schraivogel2022high}. Image-based flow cytometry is an emerging technique in which cells are pulled through a microfluidic tube and imaged, ideally one-by-one, e.g., by fluorescence and/or phase imaging \cite{han2016imaging, li2019deep, lee2021toward, schraivogel2022high}. The cells can be autonomously sorted if each image can be analyzed quickly to determine the type of cell or the cell's characteristics. In order to process statistically useful collections of cells, so as to detect, e.g., extremely rare cancerous cells, it is essential to minimize the latency of each sorting decision to maintain a high throughput, such as 100,000 cells per second \cite{li2019deep, lee2021toward, schraivogel2022high}. In our experiments, we displayed binarized images from the dataset in Ref.~\cite{schraivogel2022high} on the DMD and performed classification with each sensor, as in the QuickDraw experiments (Fig.~\ref{Fig2}e). The multilayer, nonlinear ONN encoder compresses the input images by an effective ratio of 400:1, and results in a classification accuracy for the 5 considered classes that is better than the linear-ONN sensor (93\% vs 88.5\% test accuracy, Fig.~\ref{Fig2}f and g; higher local density within clusters, Fig.~\ref{Fig2}h). 

The two tasks considered so far are effectively experimental simulations of image-sensing tasks; real image-sensing tasks involve directly processing photons arriving from real 3D objects. To test this setting, we applied the image sensors to the task of classifying traffic signs in a real-model scene, the 3D-printed intersection shown in Fig.~\ref{Fig2}i. Due to the limited field-of-view of the particular microlens array used in this experiment, the input images to the image sensors (insets of Fig.~\ref{Fig2}g) contain primarily only the speed limit sign being classified. The nonlinear, multilayer ONN encoder results in better identification of the speed limit than the linear ONN encoder across a range of viewing angles from 0 to 80 degrees (Fig.~\ref{Fig2}j-l).

\subsection{The same optically compressed features enable a variety of image-sensing applications }

\begin{figure}
\includegraphics [width=1.0\textwidth]{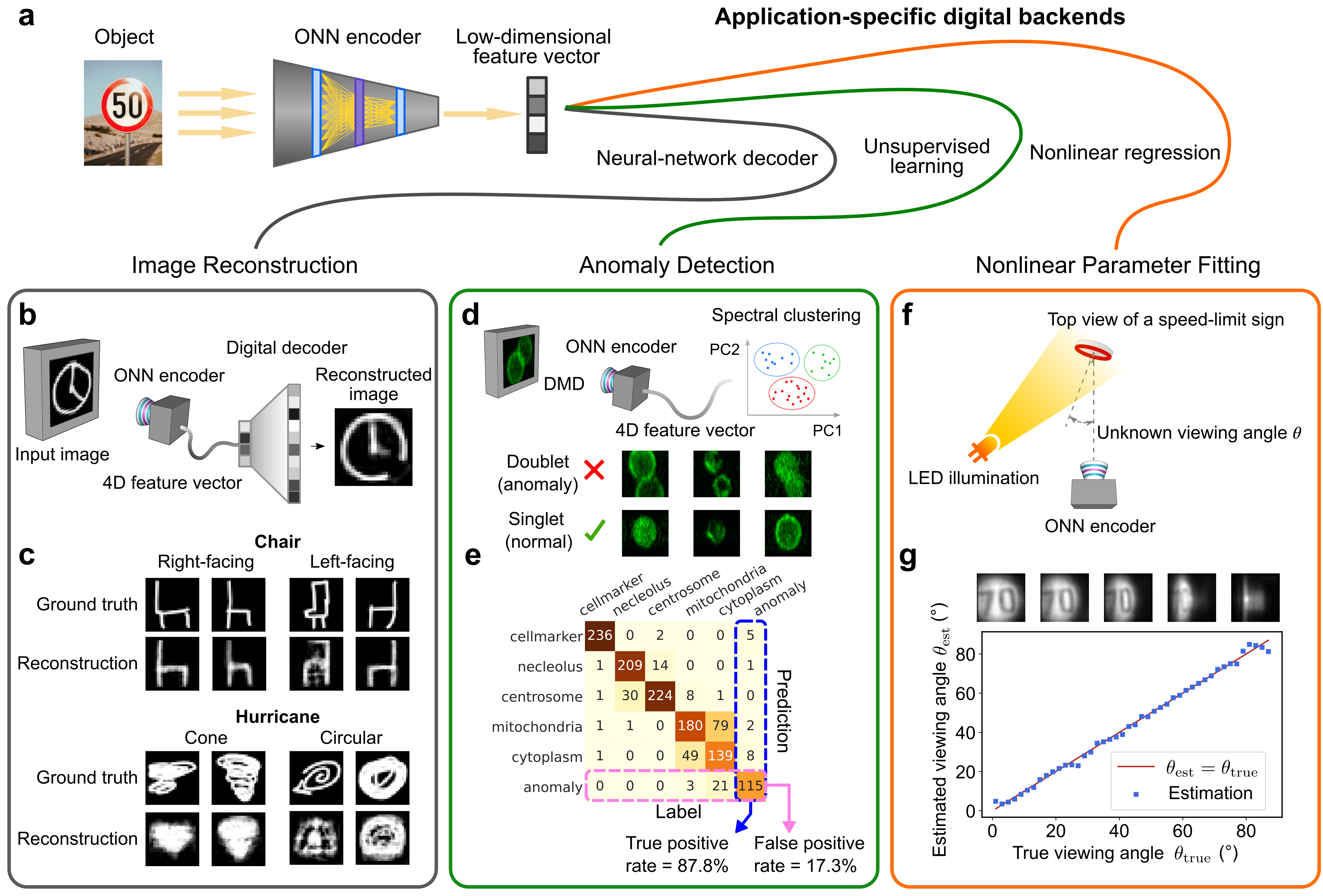}
\caption{\textbf{Nonlinear ONN encoders trained for classification can be re-used for diverse image sensing tasks by training only new digital backends.} \textbf{a}, By using the feature vectors produced by the nonlinear ONN encoders trained for classification as input to new digital backends, a diverse range of new image sensing tasks can be effectively performed. \textbf{b}, Images from the QuickDraw dataset were reconstructed by training a new digital decoder to reconstruct, rather than classify, images from the feature vectors. The encoder is the exact same ONN, including weights, as in Figure 2c-d. \textbf{c}, Although the encoder was only trained to preserve class information, randomly selected reconstructed images show that feature information such as the direction or shape of chairs and hurricanes is preserved. \textbf{d}, By performing unsupervised clustering (see \hyperref[methods]{Methods}) on the feature vector produced by the cell-organelle-classifier ONN frontend from Figure 2g, we can accurately detect anomalous doublet images which were not part of, (\textbf{e}), the encoder’s training set. \textbf{f}, We trained a new digital backend to, rather than classify the content of a speed-limit sign, use the speed-sign classifier’s feature vector to infer the viewing angle of the sign (\textbf{g}). The speed-limit images above the viewing angle inference plot (g) refer to the ground truth images at a few different viewing angles.}
\label{Fig3}
\end{figure}

By training new digital post-processers only, the same optical encoders trained for classification in the previous section can be re-used for a variety of other image-sensing tasks. If suitably trained (see \hyperref[methods]{Methods}), encoders can produce robust representations of high-dimensional images in the low-dimensional latent space, which preserve far more information than the bare minimum required for classification. For example, although the QuickDraw-classification encoder (Fig.~\ref{Fig2}a-c) was trained only to facilitate classification, the feature space evidently preserves more complex attributes of the original images beyond just the figure's class. When a digital decoder is trained to reconstruct QuickDraw images from the classification encoder’s features (Fig.~\ref{Fig3}b-c), it produces reconstructions that, while often lacking specific details (like the position of the clock’s hands), captures coarse intra-class details like the orientation or shape of chairs and hurricanes. 

Similarly, using the same multilayer ONN encoder previously trained for traffic-sign classification (Fig.~\ref{Fig2}i-l), we trained a new digital backend to predict the angle at which a traffic sign was viewed (Fig.~\ref{Fig3}f and \ref{Fig3}g). The resulting predictions are very accurate, although the performance is reduced if the network is required to predict viewing angle for all, rather than just one, speed-limit class at a time (Supplementary Figure 24). 

Finally, in many image-sensing applications, initial device training will not be able to account for edge cases that may be encountered in deployment. To test the capacity for anomalies not previously observed (and on which the optical encoder was not trained) to be detected, we introduced anomalous images of doublet cell clusters to the ONN image sensor (Fig.~\ref{Fig3}d). To detect these anomalies, we applied spectral clustering to the normalized 4-dimensional feature vectors produced by the ONN encoder previously trained for cell-organelle classification (see \hyperref[methods]{Methods}). By identifying the six most prominent clusters as the five trained classes, plus one class corresponding to anomalous images, we were able to adapt the digital decoder to reliably identify anomalous images in the test set (Fig.~\ref{Fig3}e). These results show that the nonlinear ONN encoder does not overfit to the initial training dataset, but instead preserves important data structure beyond the initially chosen classes, while still compressing the original images to a low-dimensional space.

\section{Multilayer ONN sensors scale favorably to complex image sensing tasks}
\begin{figure}[ht!]
\includegraphics[width=0.8\textwidth]{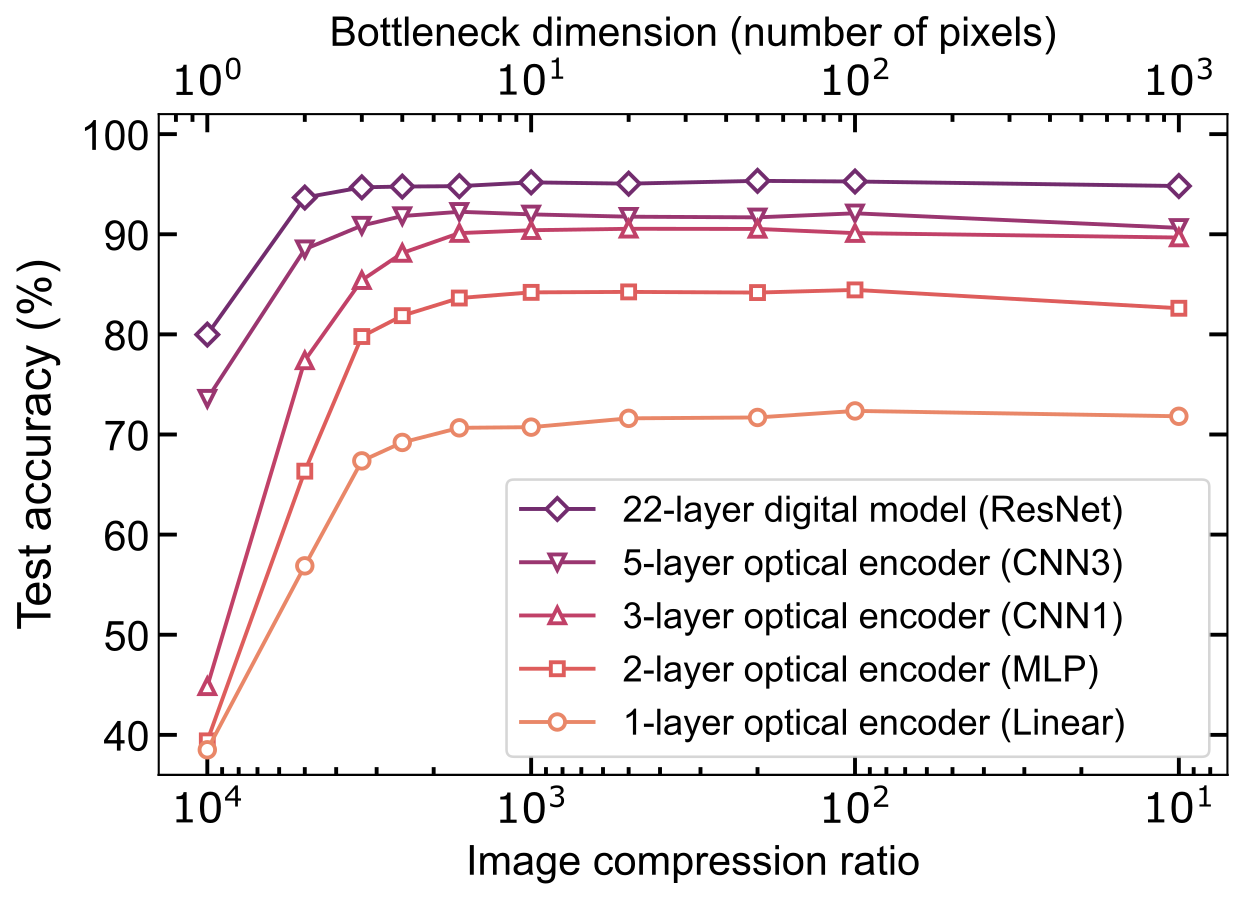}
\caption{\textbf{Simulations of performance scaling with deeper nonlinear optical neural network encoders for 10-class cell-organelle classification.} Classification accuracy as a function of image compression ratio (or bottleneck feature vector dimension) for all the models. Linear is a single-layer (linear) ONN encoder, MLP is a nonlinear encoder with two fully connected layers, CNN1 a 3-layer nonlinear ONN encoder with a convolutional layer followed by two fully connected layers, and CNN3 is a 5-layer nonlinear ONN encoder with 3 convolutional and two fully connected layers. The ResNet-based model is a state-of-the-art digital model shown here as an estimate of the upper bound on performance at each compression ratio. In general, deeper models produce higher accuracy, especially at higher compression ratios. }
\label{Fig4}
\end{figure}

The results presented in Figs \ref{Fig2} and \ref{Fig3} illustrate that a 2-layer nonlinear ONN pre-processor enables consistently better image-sensing performance than conventional imaging with direct downsampling or linear-ONN pre-processing, across a wide range of tasks. Nonetheless, ONN encoders with two fully connected layers is merely a first step. A key motivation for using an OONA is that it will facilitate even deeper ONN encoders and, in turn, facilitate more sophisticated optical pre-processing. To explore what may soon be possible with deeper, nonlinear ONN encoders, we performed realistic simulations of four different optical pre-processors, each with a different number of nonlinear and linear optical layers (see Figure \ref{Fig4}), performing an extended (10-class) version of the organelle-classification task considered in Figs \ref{Fig2}-\ref{Fig3} (Supplementary Figure 25). We chose this dataset, which is more challenging than the 5-class cell-organelle classification demonstrated in earlier experiments, so that we could study the performance for more complicated ONN encoders. Our simulations (see \hyperref[methods]{Methods} for details) consider physical noise, and involve strictly non-negative weights, which is a critical constraint for ONNs operating on incoherent light, such as fluorescence. 

Figure \ref{Fig4} shows how the classification accuracy of the different ONN pre-processors varies as the compression ratio is changed. The compression ratio is changed by modifying the number of output neurons in the final optical layer, which determines the number of pixels or photodetectors required on the photosensor. As a reference for achievable performance, we also performed the task with a fully digital classifier based on a ResNet model (18-layer pretrained ResNet plus 4 additional adapting layers) \cite{he2016deep}. All networks, including the all-digital reference, have the same single-layer digital decoder architecture. 

The key result in Fig.~\ref{Fig4} is that deeper ONNs, incorporating multiple nonlinear layers, lead to progressively better classification performance across a wide range of compression ratios. The benefit of pre-processor depth becomes especially evident at very high compression ratios: for a compression ratio of $10^4$ (bottleneck dimension 1) the 5-layer pre-processor (CNN3) achieves nearly double the accuracy of shallower networks.

\section{Discussion}
\label{sec:discussion}
Our results show that nonlinear, multilayer optical neural network (ONN) image pre-processors enable image sensors that outperform both image sensors based on conventional direct imaging, and those with only linear-optical pre-processing at high-compression ratios. In this high-compression regime, through a diverse range of image-sensing tasks, we show that deep, nonlinear optical image pre-processing enables better extraction of image information, resulting in better image-sensing performance. Furthermore, we see that the performance advantages of nonlinear ONN encoders scale favorably with additional layers of ONN pre-processing. Such nonlinear optical encoders extend the paradigm of end-to-end image system optimization \cite{matic1989comparison, kubala2003reducing, stork2008theoretical, sitzmann2018end, colburn2019optical, martel2020neural, kim2020multi, pad2020efficient, vargas2021time, markley2021physics, zheng2022meta, chen2022photonic, bezzam2022learning} to include more powerful nonlinear optical image pre-processing. As their capabilities are improved, we anticipate nonlinear ONN sensors will provide versatile, almost universal frontends for processing images in the analog, optical domain, facilitating both quantitative and qualitative advances throughout image sensing and machine vision.   

Although the performance advantages possible with ONN-based optical frontends might appear to come at the cost of increasing overall device complexity, the opposite may turn out to be true. In traditional optical sensors, the optical system must be optimized to agnostically preserve as much information in the incident optical signal as possible, since any of the image’s content could in principle be relevant to the end use. In an ONN-based sensor, the amount of information that must be preserved can be far less (only the relevant information), and distortions (including many manufacturing imperfections) may simply be adapted to by parameter adjustments, enabling, in principle, smaller, cheaper, and easier-to-manufacture optoelectronic systems for image sensing.  

Perhaps the most compelling evidence supporting the potential future impact of ONN-based sensors is the wide-ranging, rapid modern development of relevant optoelectronic technology. While ONNs, and even ONN-based sensors, were considered three decades ago (e.g., Refs~\cite{farhat1985optical, wagner1987multilayer, Psaltis1988, li1993optical}), bringing these inspiring concepts beyond the laboratory is only recently starting to become feasible \cite{Shen2017, wetzstein2020inference}. This is in part due to advances in relevant optoelectronics and nanophotonics needed to create low-cost, mass-manufacturable and small-footprint realizations of ONN devices, such as 2D-material optoelectronics \cite{Mennel2020, tan20202d, ma2022intelligent}, optical metasurfaces \cite{kildishev2013planar, quevedo2019roadmap, luo2021metasurface, zheng2022meta}, large-scale VCSEL arrays \cite{heuser2020developing, chen2022deep}, as well as silicon nanophotonics and micro-optoelectronics. Additionally, a key driver for this technology today is the anticipated demand for optical deep learning accelerator hardware \cite{Shen2017, wetzstein2020inference}. This application, motivated by the surging computation demands required for scaling deep-neural-network models, has inspired ONN hardware platforms that process information encoded not just in optical spatial modes, but also in time, frequency or mixed domains \cite{Lin2018, hamerly2019large, wetzstein2020inference, nahmias2020photonic, shastri2021photonics, xu202111, feldmann2021parallel}. 

The diversity of emerging ONN platforms suggests a multitude of ONN sensors will soon be possible to realize. ONN processors that process information encoded in different spatial degrees-of-freedom could be used to adapt the image-sensing paradigm to other optical degrees-of-freedom, extracting information not from incoherent-light images but rather from optical spectroscopic traces, or hyperspectral, LiDAR, or coherent image inputs.  For high-resolution image sensing, a free-space-facing ONN platform will still be necessary for early network layers, but integrated, compact ONN platforms could be used to partially or entirely replace the digital backend used in our work, taking as inputs directly the low-dimensional optical outputs of the free-space-facing layers. By entirely bypassing electronic bottlenecks on speed, sensitivity, and resolution, these all-optical intelligent sensors could one day operate with multi-THz bandwidth, gigapixel effective spatial resolution, and picosecond-scale latency.

\section*{Data and code availability}
Demonstration code for data gathering, and the code for training for all optical/digital neural networks is available at: \url{https://github.com/mcmahon-lab/Image-sensing-with-multilayer-nonlinear-optical-neural-networks}. All data generated and code used in this work is available at: \url{https://doi.org/10.5281/zenodo.6888985}.

\section*{Acknowledgements}
The authors wish to thank NTT Research for their financial and technical support. Portions of this work were supported by the National Science Foundation (award CCF-1918549) and a David and Lucile Packard Foundation Fellowship. P.L.M. acknowledges membership of the CIFAR Quantum Information Science Program as an Azrieli Global Scholar. We acknowledge helpful discussions with Alen Senanian, Benjamin Malia, Federico Presutti, Vladimir Kremenetski, Sridhar Prabhu, Anna Barth, and Rachel Oliver. We also acknowledge Sonalee Sohoni for help with figure design. 

\section*{Author Contributions}
T.W., L.G.W., Mandar M.S., and P.L.M. conceived the project and designed the experiments. Mandar M.S. and T.W. built and performed the experiments on the nonlinear and linear ONN encoders, and analyzed the data. T.W. performed the extended cell-organelle simulations. Martin M.S. performed the neural architecture search for QuickDraw reconstruction. S.M. and T.O. aided in simulations of deep optical encoders. M.G.A. assisted with 3D-scene modelling. L.G.W., T.W., Mandar M.S. and P.L.M. wrote the manuscript. P.L.M. and L.G.W. supervised the project. 

\section*{Competing Interests}
T.W., Mandar M.S., L.G.W. and P.L.M. are listed as inventors on a U.S. provisional patent application (Serial No. 63/392,042) on nonlinear optical neural network pre-processors for imaging and image sensing.

\bibliographystyle{npjqi.bst}
\bibliography{references_main}

\newpage
\section*{Methods}
\label{methods}
\subsection*{Multilayer optical-neural-network image pre-processor}
The optical neural network (ONN) pre-processor (Supplementary Figure \ref{Fig1}a) consists of an optical matrix-vector multiplier unit, an optical-to-optical nonlinear activation (OONA) unit, a second optical matrix-vector multiplier, and finally a camera. Light is detected in the compressed, low-dimensional latent space on the camera, and is subsequently digitally post-processed. For additional details, readers may consult Supplementary Note 1. 

The optical matrix-vector multiplier treats an image with $N$ pixels as an $N$-dimensional vector and multiplies it with a user-specified matrix. To implement an $N$  by $N^{\prime}$ matrix $W$ multiplying the $N$-dimensional input vector, the following steps occur, which are also illustrated graphically in Supplementary Figure \ref{Fig3}a. First, the input image (vector) is fanned-out to create $N^{\prime}$ identical copies. This is done by using a microlens array (MLA) to form $N^{\prime}$ identical images on regions of a spatial light modulator (SLM). Second, each optically fanned out copy of the image covers $N$ pixels on the SLM, and the intensity of each image was modulated in an element-wise fashion according to a different column of matrix $W$. Finally, after the intensity modulation by the SLM, which implements the weight multiplication, the intensity-modulated image copies are optically fanned-in by forming a demagnified image of the $N^{\prime}$  copies onto an image intensifier or a camera. Provided the size of the focused image of each attenuated copy is smaller than the resolution of the image intensifier (or the size of the camera superpixel), the photoelectrons generated by each optical copy are pooled to achieve the summation step of the matrix-vector multiplication for each row, producing the $N^{\prime}$-dimensional output vector. 

For the optical matrix-vector multiplier that implements the first fully connected layer, the square MLA array has a pitch of 1.1 ± 0.001 mm and a focal length of 128.8 mm (APO-Q-P1100-F105, OKO Optics). For the second fully connected layer, the MLA has a rectangular pitch of 4 mm × 3 mm and a focal length of 38.10 mm (\#63-230, Edmund Optics). The weights of each layer are stored as pixel values on a liquid crystal display (LCD, Sony LCX029, with LCX017 controllers by bild- und lichtsysteme GmbH). The LCDs were operated as transmissive intensity-modulation SLMs by placing two polarizers, oriented at +45 and -45 degrees relative to the pixel grid of the LCD, before and after the LCD panel. The transmission as a function of the LCD pixel value was calibrated. The calibration procedure for the LCD-based matrix-vector multipliers is described in Supplementary Note 2. Under white-light illumination, the extinction ratio of the LCD pixels was measured to be at least 400, and the LCD can provide 256 discrete modulation levels. 

The optical fan-in for the first layer was implemented by demagnifying optical fan-out copies after they were modulated by an LCD. The demagnification factor was 30x, implemented by a $4f$ imaging system composed of a singlet lens (LA1484-A-ML, Thorlabs Inc., f = 300 mm) and an objective lens (MY20X-804, 20x, Mitutoyo, f = 10 mm). The optical fan-in of the second layer was done using a zoom lens (Zoom 7000, Navitar Inc.) and imaged onto a camera (Prime 95B Scientific CMOS Camera, Teledyne Photometrics). The pixels values were summed digitally after read-out, but could equivalently be summed in an analog fashion by binning camera pixels, or by using larger pixels/photodetectors.

The optical-to-optical nonlinearity after the first matrix-vector multiplication was realized with a commercial image intensifier tube (MCP125/Q/S20/P46/GL, Photek Inc.). The image intensifier provides large input-output gains (around 800 in our work), a crucial feature for multilayer networks and low-light operation. A more subtle feature of the image intensifier optical-to-optical nonlinear activation (OONA) is that it resets the number of spatial optical modes: even though the number of modes incident to the photocathodes is equal to the number of weights in the weight matrix $NN^{\prime}$, the number of distinct output beams is only equal to the output vector size $N^{\prime}$.  

The device, and its local nonlinearity, operates as follows. In the image intensifier, light is collected on a photocathode, which produces photoelectrons in proportion to the local input light intensity. These photoelectrons are then locally amplified with a multi-channel plate (MCP). The amplified photoelectrons in each channel then excite photons on a phosphor screen, producing the light input to the next layer. The saturation of this input-output response results in the nonlinearity used in our ONN encoders. The image intensifier used in our experiments is from Photek, Inc., and includes a S20 photocathode, 1-stage MCP and P46 phosphor. We find that the nonlinearity of intensifier varies slightly from channel to channel, so we calibrated the input-output response for all 36 illuminated regions separately (Supplementary Figure 8), fitting them each to a curve of the form $y=a(1-e^{-bx} )+c(1-e^{-dx})$, where $a,b,c,d$ are fit parameters for each region. The intensifier’s response time was measured to be approximately 20 µs (Supplementary Figure 7). 

For most experiments in this work, the ONN device and architecture are similar: the input is a 1,600 (40×40 pixels) image, and the first fully connected layer consists of a 1,600 by 36 weight matrix, while the second fully connected layer, after the optical-to-optical nonlinearity, usually consisted of a 36 by 4 weight matrix, except for the traffic sign classification task, which used a 36 by 2 weight matrix. The convention for matrix size used throughout this paper is: the first dimension is the length of input vector or the number of neurons in the input layer, and the second dimension is the output vector dimension or the number of neurons in the output layer. The effective input image size equals the number of LCD pixels each optically fanned out copy of the input image covers on the first LCD, which is used as a transmissive SLM for element-wise multiplication.

In order to monitor the light at intermediate locations in the ONN pre-processor, and to enable us to perform experiments with direct-imaging and single-layer ONN pre-processing, we included a beamsplitter (BP245B1, Thorlabs Inc.) after the first LCD, and another (BS013, Thorlabs, Inc.) immediately after the image intensifier. Each beamsplitter directs part of the light to a monitoring camera, which enabled us to observe several intermediate steps of computation. The full experimental setup including these details is depicted in Supplementary Figure 2.

\subsection*{QuickDraw image classification}
We chose the Quick, Draw! (QuickDraw) dataset \cite{JongejanJ.RowleyH.KawashimaT.KimJ.Fox-Gieg, chang2018hybrid} to benchmark the performance of the  encoders because it (a) is significantly harder than the MNIST dataset, and (b) can be binarized and displayed on a digital micromirror display (DMD) without significant loss of image information. 10 classes (clock, chair, computer, eyeglasses, tent, snowflake, pants, hurricane, flower, crown) were chosen arbitrarily (by hand, but with no deliberate rationale other than to ensure the classes were not too similar) from the available 250+ classes. Inappropriate images or images that were obviously not of the intended class were removed by hand. The first 300 images remaining for each class were used for the training set (total size 3,000) with a random train-validation split of 250:50, while the next 50 were used for testing (total size 500). This dataset is included with all other data for this manuscript at \url{https://doi.org/10.5281/zenodo.6888985}.

For experiments, the QuickDraw images were resized to 100×100 pixels, binarized, and then displayed on a DMD (V650L Vialux GmbH). The DMD was illuminated by a white-light source (MNWHL4 - 4900 K, Thorlabs Inc.). 

To train the ONN’s weights, we needed to first measure the input images that were seen by the ONN device. This was necessary since the optically fanned out images formed on the LCD differed slightly from the digital image loaded onto the DMD, due to the imaging resolution limit and aberrations of the MLA. To measure these, we displayed each QuickDraw image on the DMD, leaving the LCD pixels all at their highest transmission, and then inserted a pellicle beamsplitter (BP245B1, Thorlabs) after the LCD to reflect part of light to a monitoring camera (see Supplementary Figure 2 for details). An image of the LCD panel was formed on the camera so that each fanned out copy of the input image could be captured by the monitoring camera as the effective ground truth of input images. These ground truth images were used for training the weights of ONN pre-processor on a computer (Supplementary Note 5) and for checking the accuracy of optical matrix-vector multipliers (Supplementary Figure 5 and 6). Each ground truth image of the fanned-out copies was re-sized to 40×40=1,600 pixels, corresponding to the 40×40 LCD pixels used as the weights for each image. 

For the QuickDraw image classification task shown in Figure \ref{Fig2}a, the multilayer ONN encoder consisted of a matrix-vector multiplication with a weight matrix size 1,600 by 36, the 36 optical-to-optical nonlinear activations, and a final matrix-vector multiplication with a weight matrix size of 36 by 4 (Supplementary Figure 9). The digital decoder consisted of a single matrix-vector multiplication with a weight matrix size of 4 by 10. The linear ONN pre-processor involved just a single optical matrix-vector multiplication with a weight matrix size of 1,600 by 4, followed by a 4 by 10 digital decoder. For direct imaging, the 40×40 ground truth images were resized to 2×2 images by averaging pixel values and sent to a digital decoder consisting of a 4 by 10 weight matrix. The linear digital neural network shown in Figure \ref{Fig2}d consists of a linear layer with a 1,600 by 4 weight matrix followed by another linear layer with a 4 by 10 weight matrix. There is no nonlinear activation function between the two linear layers, and both have real-valued weights and bias terms. The nonlinear digital neural network shown in Figure \ref{Fig2}d has a linear layer with a 1,600 by 36 weight matrix, followed by element-wise nonlinear activations (sigmoid), followed by another linear layer with a 36 by 4 weight matrix, and finally a linear layer with a 4 by 10 weight matrix. There is no nonlinear activation between the 36 by 4 linear layer and 4 by 10 linear layer. All layers have real-valued weights and bias terms. 

\subsection*{Optical-neural-network training}
Training of the ONN layers was achieved primarily by creating an accurate model (digital twin) of the optical layers, and training the model’s parameters in silico, including the digital post-processing layer(s). The digital model treated each optical fully connected layer as matrix-vector multiplication, and included the 36 individually calibrated nonlinear curves for the image intensifier activation functions. Since our optical matrix-vector multiplier was engineered to exactly perform matrix-vector multiplication, our digital models are composed of mathematical operations like those in regular digital neural networks, but do not require simulation of any physical process such as optical diffraction. To improve the robustness of the model and allow it to be accurately implemented experimentally despite the imperfection of this calibration, we made use of three key techniques: an accurate calibrated digital model as described above, data augmentation for modeling physical noise and errors, and a layer-by-layer fine-tuning with experimentally collected data.

We performed data augmentation on training data with random image misalignments and convolutions, which were intended to mimic realistic optical aberrations and misalignments. This included translations (±5\% of the image size in each direction) and mismatched zoom factor (±4\% image scale). To manage the computational cost of this augmentation, we found that it was sufficient to only apply these augmentations to the input layer. We also added noise on the forward pass during training of about 2\% to each activation, after both the first and second layer (more details in Supplementary Note 5). 

We first trained models entirely digitally. We used a stochastic gradient optimizer (AdamW \cite{loshchilov2017decoupled}) for training. The training parameters, such as learning rate, vary from task to task and are only included in training code deposited in GitHub or Zenodo. Generally, each model was trained for multiple times with each training parameter randomly generated within a range. The parameters were fine-tuned from trial to trial by using the package Optuna \cite{optuna_2019}, until the best training result was achieved (e.g., the highest validation accuracy without obvious overfitting). 

After this digital training step, we fine-tuned the trained models using a layer-by-layer training scheme that incorporated data collected from the experimental device. We first uploaded the weights for the first optical layer obtained by training the digital model, and collected the nonlinear activations for each training image after the image intensifier using the monitoring imaging systems (see Supplementary Note 3). Using the images after the image intensifier as the input, we then retrained the second optical layer. We then uploaded the obtained weights for the second optical layer, and for each image in the training set collected the output from this second layer experimentally, which was used to finally retrain the last digital linear layer. Only after this layer-by-layer fine-tuning did we perform experimental testing with the test dataset.

\subsection*{Flow-cytometry image classification}
We performed an experimental benchmark of image-based cell-organelle classification using a procedure mostly similar to the QuickDraw benchmarks, including the experimental collection of input ground truth images, and the training procedures. Images from Ref. \cite{schraivogel2022high} (S-BSST644, available from \url{https://www.ebi.ac.uk/biostudies/}) were filtered into 5 classes based on the organelles (nucleolus, cytoplasm, centrosomes, cell mask, mitochondria) and the first 200 valid images per class were selected by hand for training (1,000 images in total) with a random train-validation split of 160:40, and the next 40 valid images per class were used for testing. Our selection criterion was to discard invalid images that involve multiple or no cells. Incidentally, images with multiple cells were added back later for the anomaly detection benchmark shown in Figure \ref{Fig3}. Like the QuickDraw images, these images were binarized and displayed on the DMD with a 100×100 resolution (in terms of DMD pixels), illuminated by the white light source.

\subsection*{Real-scene image classification}
For classification of objects in a real scene, we 3D-printed a small scene consisting of a road intersection centered around a traffic sign holder, in which different speed-limit signs could be placed. We used a zoom lens (Zoom 7000, Navitar Inc.) to image the speed-limit sign onto the input of the ONN image processor (Supplementary Figure 1). The demagnification of this lens was chosen so that the image of the sign relayed in front of the ONN encoder approximately spanned about 1 mm by 1 mm, which was the same physical size of the images displayed on the DMD. The scene was illuminated by two green LED lights (M530L4-C1, Thorlabs, Inc.) from different angles for more uniform illumination. 

To train the ONN weights for the real-scene tasks, we collected ground truth input images using a procedure similar to the classification tasks performed with the DMD input. These images were collected for each angle (0 to 88 degrees in 1-degree increments), for each of 8 classes (15, 20, 25, 30, 40, 55, 70 and 80 speed limits). Every 4th angle collected was used in the validation set, so the total dataset included 536 images for training and 176 images for validation. 

All other aspects of the training and network design are similar to the previous tasks, with only two exceptions: (a) The digital backend consisted of two layers instead of one layer. (b) The compressed dimension was 2, rather than 4. As with other tasks, this compression ratio was selected as the highest compression ratio for which the nonlinear, multilayer ONN was still able to perform the task with a reasonable accuracy. 

\subsection*{Additional image sensing tasks based on different digital backends}
\noindent\underline{Image reconstruction with autoencoders:} We reconstructed QuickDraw images (as shown in Fig.~\ref{Fig3}b-c) with a digital decoder in the following way. Starting with the 4-dimensional feature vectors produced by the ONN encoder previously trained for classification (Fig.~\ref{Fig2}), we trained a new digital decoder neural network that would produce an image whose structural similarity index (SSIM) was minimized relative to the ground truth training dataset images. The decoder neural network was chosen to be a multilayer perceptron, with batch normalization layers before each sigmoid activation function. The number and widths of the hidden layers were found by random neural architecture search, which produced a best-performing network with three hidden layers, where the final output dimension corresponds to the reconstructed 28 × 28 image. We found that larger (i.e., more powerful) decoders were unable to produce better reconstructions, suggesting that the 4-dimensional bottleneck is the limit on reconstruction accuracy here. The reconstructed images shown in Figure \ref{Fig3}c are randomly chosen test images in the hurricane and chair classes. All reconstructed images in the test set are shown in Supplementary Figure 13 to 22. For additional details, see Supplementary Note 9.

\noindent\underline{Anomaly detection with unsupervised learning:} We performed the anomaly detection shown in Figure \ref{Fig3}d-e as follows. First, we created a dataset consisting of 418 anomaly images by including images containing at least two cells from all the 5 original classes. These were previously excluded from training dataset for the cell-organelle classifier shown in Figure \ref{Fig2}e-h. Next, we displayed all images in this new, anomaly dataset on the DMD and, with the ONN encoder’s weights kept identical to those originally obtained for cell-organelle classification, collected the 4-dimensional feature vector for each image. Principal component analysis on these feature vectors (Supplementary Figure 23) shows that the anomalous images are distinct from the previously trained classes, occupying a part of the latent space that was previously not accessed by any of the trained classes. As a result, we were able to successfully perform spectral clustering on these feature vectors. This procedure involves computing the nearest-neighbor distances of the vectors to compute an affinity matrix, whose eigenvectors correspond to localized clusters. The largest five clusters (largest eigenvalues) of this matrix were found to correspond to each of the previously trained classes, while the sixth cluster was found to correspond to anomalous images. After assigning the most probable class label to each cluster for the maximum overall likelihood (Supplementary Figure 23), we computed the confusion matrix of classifying the original 5 classes plus the new anomaly class by comparing to the ground truth labels (Fig.~\ref{Fig3}e). The true positive rate was calculated as the percentage of anomalous images classified as anomalous images, and the false positive rate was calculated as the percentage of normal images in the total number of images classified as anomalies. 

\noindent\underline{Nonlinear parameter fitting:} We performed the estimation of speed-sign viewing angle as follows. Using the 2-dimensional feature vectors produced by the ONN encoder trained for speed limit classification in Figure \ref{Fig3}, we trained a new digital decoder neural network to predict sign viewing angle. The dataset split between training and validation here was that every even angle was used in the train set and every odd angle was used in the validation set, except we only considered one class (i.e., one speed limit) at a time (in other words, the sign angle estimation decoder only works for a given speed limit sign, rather than for an arbitrary sign). A multilayer perceptron with dimensions 2$\rightarrow$50$\rightarrow$100$\rightarrow$1 was found to perform well when trained with an L1 loss function, i.e., $|\theta_{\text{predicted}}-\theta_{\text{true}}|$. The angle prediction can be performed for all, rather than just one, speed-limit class at a time, albeit with reduced performance. The results are shown in Supplementary Figure 24.

\subsection*{Simulation of Deeper Optical Neural Networks for 10-class cell-organelle classification}
To explore the possible performance and applications of future, scaled-up nonlinear ONN encoders, we performed realistic physical simulations of optical neural networks based on our experiments, for a more challenging task: 10-class cell-organelle classification for image-based cytometry. 

A key motivation for using nonlinear, multilayer ONN encoders is that these encoders should be able to achieve much higher compression ratios than linear encoders, facilitating higher speed or potentially imaging with fewer detected photons. We therefore sought to compare the performance on the 10-class cell-organelle classification task for ONN encoders of varying depth and compression ratio. As Figure \ref{Fig4} shows, in general deeper networks produce higher quality compression (as evidenced by better classification accuracy), and this quality advantage is especially clear for very high compression ratios. 

The five networks considered are as follows. The first ONN pre-processor is a wide ($100\times100$=10,000-dimensional input vector), linear single-layer ONN (Linear). The second is a 2-layer multilayer perceptron (MLP) with 10,000-dimensional input, and a 200-dimensional hidden layer. Besides both the input and hidden dimension being much larger, this network is similar to the 2-layer fully connected ONN we realized experimentally. The third and fourth models extend this network deeper, adding one, for CNN1, or three, for CNN3, optical convolutional layers. Multi-channel optical convolutional layers of this kind have been realized before with $4f$ systems (e.g., \cite{chang2018hybrid, colburn2019optical}) , which are in many regards simpler and more amenable to compact implementation than fully connected optical layers. These convolutional neural networks (CNNs) also include a shifted ReLU activation (i.e., trained batch normalization layers followed by ReLU), which could be realized with a slight modification of the image intensifier electronics, or by the threshold-linear behavior of optically controlled VCSEL \cite{heuser2020developing} or LED arrays. We have primarily assumed pooling operations are AvgPool, which are straightforwardly implemented with optical summation. The MaxPool operation used once in CNN3 is more challenging but could plausibly be realized effectively by using a broad-area semiconductor laser or placing a master limit on the energy available to a VCSEL or LED array, such that the first unit to rise above threshold would suppress activity in others. These ONN designs are ultimately speculative; In general, we anticipate that practically realizing more powerful ONN encoders will require jointly designing compact, low-cost ONN hardware components and developing optics-friendly DNN architectures, rather than simply directly adapting existing digital DNN architectures. 

The decoders used for all networks are identical linear layers with dimensions N$\rightarrow$10, where N is the bottleneck dimension. Note that the compression ratio is taken to be 100×100=10,000, the original image resolution, divided by N. 

All optical networks were simulated with emulated physical noise on the forward pass (as described in detail in Supplementary Note 12), and weights were constrained to be non-negative (because we assumed incoherent light). We note that it is possible that intermediate layers could be realized with coherent light (and therefore with real-valued weights) even if the input light is strictly incoherent, i.e by using arrays of VCSELs \cite{heuser2020developing, chen2022deep}. We find that, while non-negative weights can generally be trained for various tasks, the performance of these networks is generally inferior to what is possible with real (i.e., both positive and negative) weights. Consequently, our results here are roughly a lower bound with respect to the performance of coherent-light-based ONN encoders. 

The dataset used for this task was adapted from Ref.~\cite{schraivogel2022high}’s data (S-BSST644, available from \url{https://www.ebi.ac.uk/biostudies/}) in the following way. First, we selected 10 of the 12 provided classes (the other two, Golgi and Control, had too few images and no fluorescent channel respectively). Unlike our dataset preparation for the 5-class version of this task performed experimentally, here we retained all images, including those with multiple or no cells. 

As a reference for achievable classification accuracy on this cell-organelle classification task (i.e., a practical upper-bound, in part due to the presence of anomalous multi- or no-cell images), we trained a purely digital classifier based on a ResNet-18 \cite{he2016deep} which was pretrained on ImageNet. This network includes four additional layers to adapt input images to the ResNet core, and to produce the final classification output. All weights of this network were fine-tuned by training with the training set. 

\end{document}


\setcounter{page}{1}
Supplementary Materials for 
\title{Image sensing with multilayer, nonlinear optical neural networks}

\author{Tianyu~Wang}
\email{Equal contribution}
\affiliation{School of Applied and Engineering Physics, Cornell University, Ithaca, NY 14853, USA}

\author{Mandar~M.~Sohoni}
\email{Equal contribution}
\affiliation{School of Applied and Engineering Physics, Cornell University, Ithaca, NY 14853, USA}

\author{Logan~G.~Wright} 
\affiliation{School of Applied and Engineering Physics, Cornell University, Ithaca, NY 14853, USA}
\affiliation{NTT Physics and Informatics Laboratories, NTT Research, Inc., Sunnyvale, CA 94085, USA}

\author{Martin~M.~Stein}
\affiliation{School of Applied and Engineering Physics, Cornell University, Ithaca, NY 14853, USA}

\author{Shi-Yuan~Ma}
\affiliation{School of Applied and Engineering Physics, Cornell University, Ithaca, NY 14853, USA}

\author{Tatsuhiro~Onodera}
\affiliation{School of Applied and Engineering Physics, Cornell University, Ithaca, NY 14853, USA}
\affiliation{NTT Physics and Informatics Laboratories, NTT Research, Inc., Sunnyvale, CA 94085, USA}

\author{Maxwell~Anderson}
\affiliation{School of Applied and Engineering Physics, Cornell University, Ithaca, NY 14853, USA}

\author{Peter~L.~McMahon}
\email{Contact: tw329@cornell.edu, mms477@cornell.edu, lgw32@cornell.edu, pmcmahon@cornell.edu}
\affiliation{School of Applied and Engineering Physics, Cornell University, Ithaca, NY 14853, USA}
\affiliation{Kavli Institute at Cornell for Nanoscale Science, Cornell University, Ithaca, NY 14853, USA}

\maketitle

\tableofcontents
\clearpage

\part{Experimental Setup}

\section{Overview of the experimental setup}
\label{overview_setup}

\begin{figure}[ht!]
\includegraphics [width=0.9\textwidth] {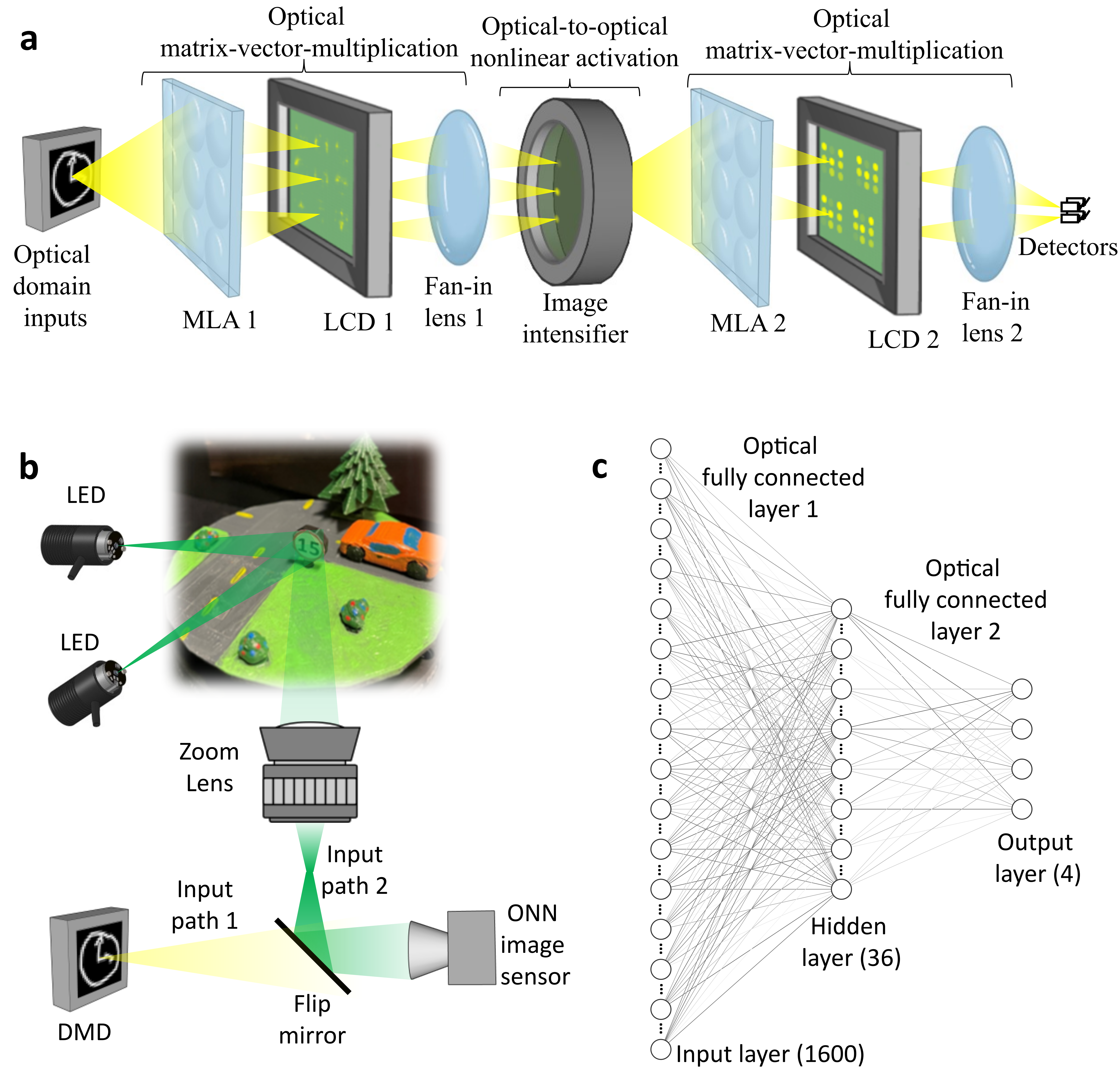}
\caption{\textbf{Schematic of the optical-neural-network encoder setup.}
\textbf{a}, The structure of the multi-layer optical neural network encoder. LCD: liquid crystal display. MLA: microlens array
\textbf{b}, The optical path for either routing light reflected off 3D objects in a real scene or a digital micromirror device (DMD) to the optical-neural-network (ONN) encoder. LED: light emitting diode.
\textbf{c}, The neural-network diagram implemented by (a).}
\label{figS_overview}
\end{figure}

\begin{figure}[ht!]
\includegraphics [width=0.77\textwidth] {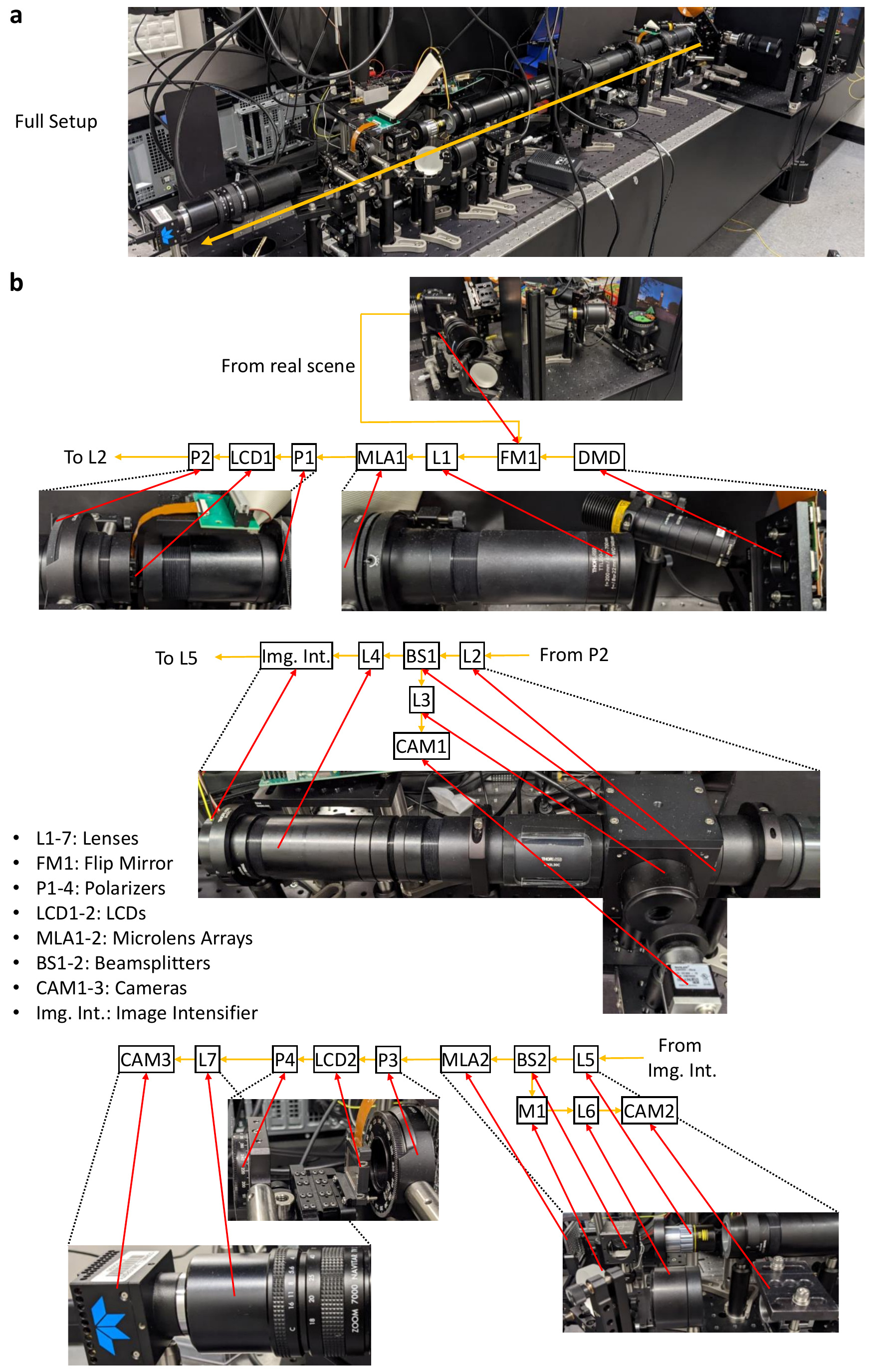}
\caption{\textbf{Experimental setup of the optical-neural-network encoder.} \textbf{a}, Image of the complete experimental setup. \textbf{b}, A detailed diagram of the optical path in the ONN encoder setup along with zoomed-in pictures of each part of the experimental setup. The details of all the experimental parts can be found in Supplementary Table \ref{parts}.}
\label{exp_setup}
\end{figure}

\noindent We constructed a 2-layer optical neural network for image encoding using two linear layers and an element-wise nonlinear layer in-between. The linear layers were each implemented with a fully optical matrix-vector multiplier (\ref{MVM}), and the nonlinear activations were implemented with an image intensifier (\ref{MCP}). Supplementary Figure \ref{figS_overview} shows a high-level schematic of the setup while Supplementary Figure \ref{exp_setup} shows a detailed diagram of each part and its location in the experimental setup. The parts seen in Supplementary Figure \ref{exp_setup} are listed in Supplementary Table \ref{parts}. The third column lists the distance of the part described in the row from another part. For example, in row 2 of Supplementary Table \ref{parts}, part L1 is at a distance of \SI{148}{\milli\meter} from part DMD.

The inputs to the optical-neural-network (ONN) encoder were light reflected either by a digital micromirror device (DMD) or a real physical object. For the task of recognizing hand-drawn figures (\ref{quickdraw}) and cell organelles (\ref{cell}), the input signal to the ONN encoder was generated by shining white-light (\SIrange[]{400}{700} {\nano\metre}, MNWHL4 - 4900 K, Thorlabs Inc.) onto the DMD (V-650L, Vialux GmBH) that displays the image. To reduce wavelength dispersion of the broadband illumination light caused by the fine structures on the DMD, we set up an illumination scheme that forms a real image of the LED on the DMD panel (Supplementary Figure \ref{figS_overview}b). The image of the DMD panel was further relayed onto the LCD panel of the first optical matrix-vector multiplier through a 4f imaging system, and then imaged onto the photocathode of the image intensifier through another 4f imaging system. 

\begin{table} 
\centering
\caption{Detailed list of parts from Supplementary Figure \ref{exp_setup}}
\label{parts}
\begin{tabular}{|l|r|r|}
\hline
\textbf{Part Number} & \textbf{Part Name} & \textbf{Distance from (Part Number, Distance)}\\
\hline
L1 & TTL200-A, Thorlabs Inc. & DMD, \SI{148}{\milli\meter}  \\
\hline
MLA1 & APO-Q-P1100-F105, OKO optics & L1, \SI{200}{\milli\meter}  \\
\hline
P1 & \#86-190, Edmund Optics & MLA1, $\sim$ \SI{5}{\milli\meter}  \\
\hline
LCD1 & LCX029, Sony & MLA1, \SI{128.8}{\milli\meter}  \\
\hline
P2 & \#86-190, Edmund Optics & LCD1, $\sim$ \SI{30}{\milli\meter}  \\
\hline
L2 & LA1484-A-ML, Thorlabs Inc. & LCD1, \SI{300}{\milli\meter}  \\
\hline
BS1 & BP245B1, Thorlabs Inc. & L2, $\sim$ \SI{30}{\milli\meter}  \\
\hline
L3 & TTL100-A, Thorlabs Inc. & BS1, $\sim$ \SI{30}{\milli\meter}  \\
\hline
CAM1 & acA4024 - \SI{29}{\micro\metre}, Basler Inc. & L3, \SI{60}{\milli\meter}  \\
\hline
L4 & MY20X-804, Mitutoyo & L2, $\sim$ \SI{300}{\milli\meter}  \\
\hline
Img.Int. & LA1484-A-ML, Thorlabs Inc. & L4, \SI{20}{\milli\meter}  \\
\hline
L5 & MY10X-803, Mitutoyo & Img. Int., \SI{34}{\milli\meter}  \\
\hline
BS2 & BS013, Thorlabs Inc. & L5, $\sim$ \SI{30}{\milli\meter}  \\
\hline
M1 & PF20-03-P01, Thorlabs Inc. & BS2, $\sim$ \SI{50}{\milli\meter}  \\
\hline
L6 & TTL100-A, Thorlabs Inc. & M1, $\sim$ \SI{30}{\milli\meter}  \\
\hline
CAM2 & acA3088 - \SI{57}{\micro\metre}, Basler Inc. & L6, $\sim$ \SI{60}{\milli\meter}  \\
\hline
MLA2 & \#63-230, Edmund Optics & BS2, $\sim$ \SI{50}{\milli\meter}  \\
\hline
P3 & \#86-190, Edmund Optics & MLA2, $\sim$ \SI{20}{\milli\meter}  \\
\hline
LCD2 & LCX029, Sony & MLA2, \SI{38.1}{\milli\meter}  \\
\hline
P4 & \#86-190, Edmund Optics & LCD2, $\sim$ \SI{50}{\milli\meter}  \\
\hline
L7 & Zoom 7000, Navitar Inc. & LCD2, $\sim$ \SI{100}{\milli\meter}  \\
\hline
CAM3 & Prime 95B Scientific CMOS Camera, Teledyne & L7, $\sim$ \SI{20}{\milli\meter}  \\
\hline
\end{tabular}
\end{table}

For real-scene image sensing, we first designed a CAD model of a scene with a speed-limit traffic sign placed at an intersection of roads, with a car and a tree in the background (Supplementary Figure \ref{figS_overview}b). The speed-limit sign can be swapped by plugging in and out of a slot in the base plate of the real scene model, and we 3D printed multiple speed-limit signs for swapping, each with a different speed limit number (15, 20, 25, 30, 40, 55, 70, 80). We 3D printed the model with polyactic acid plastic using a 3D printer (Original Prusa MINI, Prusa Research). The size of the whole printed model is about \SI{6}{\centi\metre} $\times$ \SI{6}{\centi\metre} $\times$ \SI{2}{\centi\metre}. 

The real scene model was placed on a rotation (PRMTZD, Thorlabs Inc.) and a translation mount (PT1-Z8, Thorlabs Inc.). The rotation and translation mounts were connected to servo-motor controllers (KDC101, Thorlabs Inc.) for automated motion control. The setup for imaging the real scene to the ONN encoder is shown in Supplementary Figure \ref{figS_overview}b. We used two green LEDs (M530L4-C1, Thorlabs Inc.) for illuminating the 3D-printed scene, each at a different incident angle to the traffic sign, so that the amount of light reflected off the traffic sign towards the ONN encoder could stay more uniform when the traffic sign was rotated to different angles. A de-magnified real image of the traffic sign was formed by a zoom lens (Zoom 7000, Navitar Inc.) in front of the ONN encoder. The image was of a similar size as the images displayed on the DMD, and the optical path length between the image and the ONN encoder is the same as the distance between the DMD panel and the ONN encoder (Supplementary Figure \ref{figS_overview}b). 

\section{A fully optical matrix-vector multiplier} \label{MVM}

\noindent We used the following scheme that decomposes arbitrary matrix-vector multiplication into 3 steps, namely optical fan-out, element-wise multiplication, and optical fan-in, that can be most conveniently implemented with optics, as shown in Supplementary Figure \ref{figs_optical_mvm}a and Ref. \cite{goodman1978fully, wang2021optical, bernstein2022single}. To multiply an $N$-dimensional row vector with an $N$ by $N'$ matrix $W$, optical fan-out first creates $N'$ identical optical copies of the same vector $\vec{x}$. In this study, we used a microlens array (MLA) to create $N'$ optical images of the same object (Supplementary Figure \ref{figs_optical_mvm}b), a technique that is commonly used in light-field imaging \cite{levoy2006lightfield, ng2006lightfield}. All the optical fan-out copies are imaged onto a liquid-crystal display (LCD) for intensity modulation (Supplementary Figure \ref{figs_optical_mvm}b and Supplementary Figure \ref{mnist_fanout}) \cite{tang2010partially}. Since our input light from the environment was incoherent, the modulation of the intensity of a spatial mode by an LCD pixel corresponds to product between two non-negative numbers. Each optical copy was imaged and aligned to a square patch of LCD pixels, each encoding a different column of the matrix (i.e., $W[:,j]$ or the $j$th column of matrix $W$). In our case, each patch consists of $40 \times 40$ LCD pixels (the pixel pitch is \SI{18}{\micro\metre} and the pixel size is \SI{12}{\micro\metre}.), and therefore $N=40\times40=1,600$, which is how we quantified the size of input vector $\vec{x}$ even when a real physical object without any pixelation was placed in front of the ONN encoder. After pixel-wise modulation by the LCD, the modulated optical copies (i.e., $W[:,j] \circ \vec{x}$) were demagnified and imaged onto either a camera sensor or the photocathode of the image intensifier (\ref{MCP}). When the size of each modulated optical copy is smaller than the size of the photodetector or the spatial resolution of the image intensifier, its optical energy is summed by pooling photoelectrons, which completes optical fan-in as the last step of optical matrix-vector multiplication \cite{wang2021optical, bernstein2022single, spall2020fully}. 

For fully connected layer 1, the MLA we used contains $26 \times 26 = 676$ square lenslets, each having a pitch size of $1.1 \pm 0.001$\SI{}{\milli\metre} and a focal length of $f=\text{\SI{128.8}{\milli\metre}}$ (APO-Q-P1100-F105, OKO optics). For fully connected layer 2, the MLA we used contains 63 lenslets, each has a  size of \SI{4}{\milli\metre} $\times$ \SI{3} {\milli\metre} with a focal length of \SI{38.10}{\milli\metre} (\#63-230, Edmund Optics). The LCD we used for both layers are LCX029 Sony controlled by an LCX017 controller provided by bbs bild- und lichtsysteme GmbH. To configure the LCD in intensity modulation mode,  two linear polarizers were placed, one on each side of the LCD screen: their polarization axes were rotated from the grid of the LCD by +45 and -45 degrees respectively, and thus are perpendicular to each other. Under white-light illumination, the extinction ratio of the LCD pixels was measured to be at least $400$, and the LCD can provide 256 discrete modulation levels. The optical fan-in of fully connected layer 1 was implemented by demagnifying the modulated optical fan-out copies on the LCD by a demagnification factor of 30$\times$, through a 4f imaging system composed of a singlet lens (LA1484-A-ML, Thorlabs Inc., $f=\text{\SI{300}{mm}}$) and an objective lens (MY20X-804, 20x, Mitutoyo, $f=\text{\SI{10}{mm}}$). The optical fan-in of fully connected layer 2 was performed by using a zoom lens (Zoom 7000, Navitar Inc.) to form a demagnified image of the second LCD onto a camera sensor (Prime 95B Scientific CMOS Camera,
Teledyne Photometrics -- CAM3 in Supplementary Table \ref{parts}). The pixels on the camera corresponding to each neuron at the output of fully connected layer 2 were summed digitally after read-out, but in principle, the summation could be performed by directly pooling photoelectrons in photodetectors given larger pixel sizes.

\section{Calibration of optical matrix-vector multipliers} \label{MVM_cali}

\begin{figure}[ht!]
\includegraphics [width=0.8\textwidth] {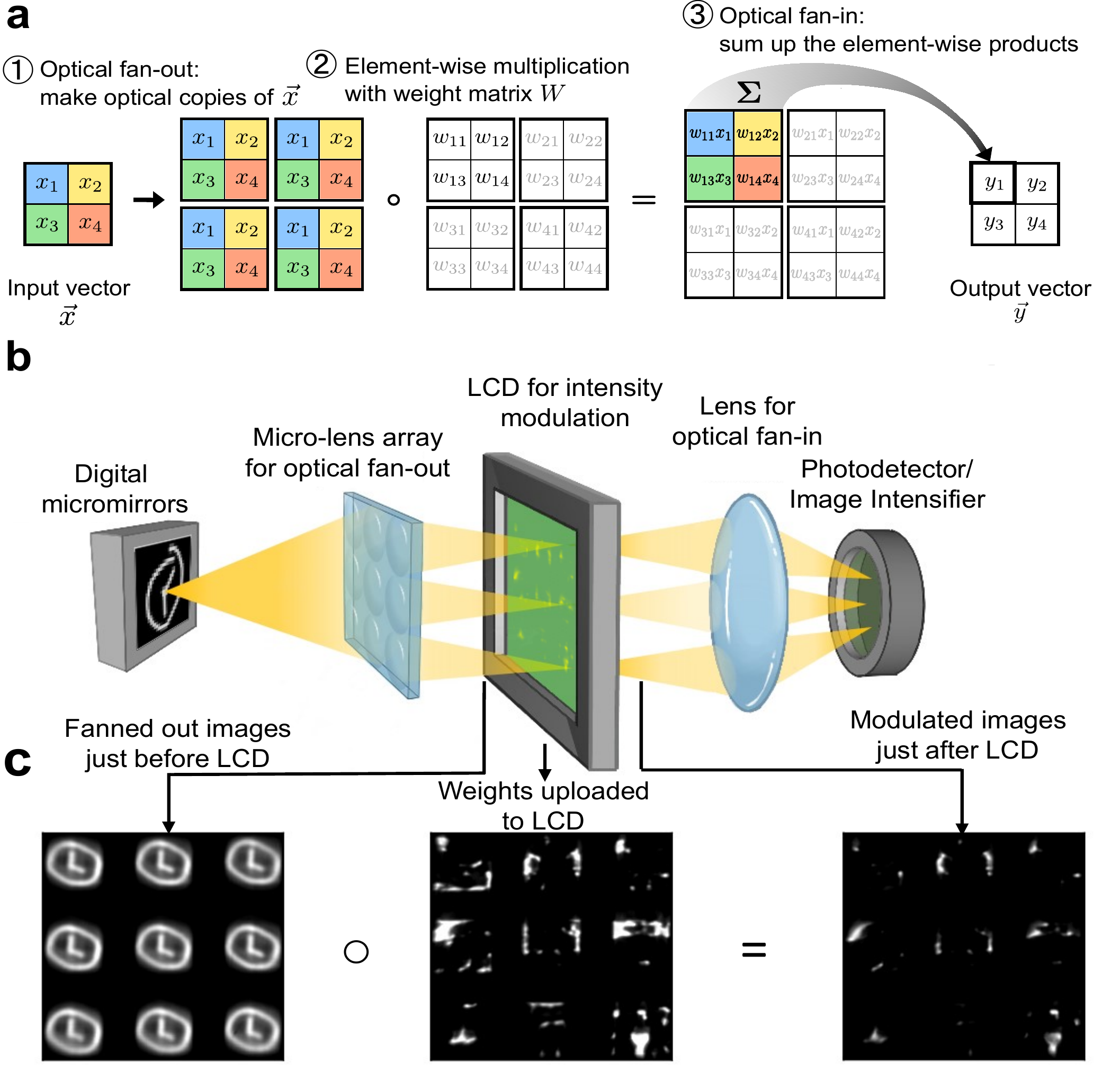}
\caption{ \textbf{The schematic of a fully optical matrix-vector multiplier.}
\textbf{a}, A step-by-step illustration of the three steps to compute arbitrary matrix-vector multiplication. \textbf{b}, The experimental schematic to implement optical matrix-vector multiplication as described in (a). The optical fan-out was implemented using a microlens array (MLA), the modulation of each copy was performed through pixel-wise intensity modulation by an LCD screen, and the optical fan-in was implemented using an objective lens. \textbf{c}, Examples of intermediate results of each step illustrated in (a). The images before and after modulation were captured on our experimental setup. Panel (a) is adapted from Ref. \cite{wang2021optical} with permission.}
\label{figs_optical_mvm}
\end{figure}

\begin{figure}[ht!]
\includegraphics [width = \textwidth] {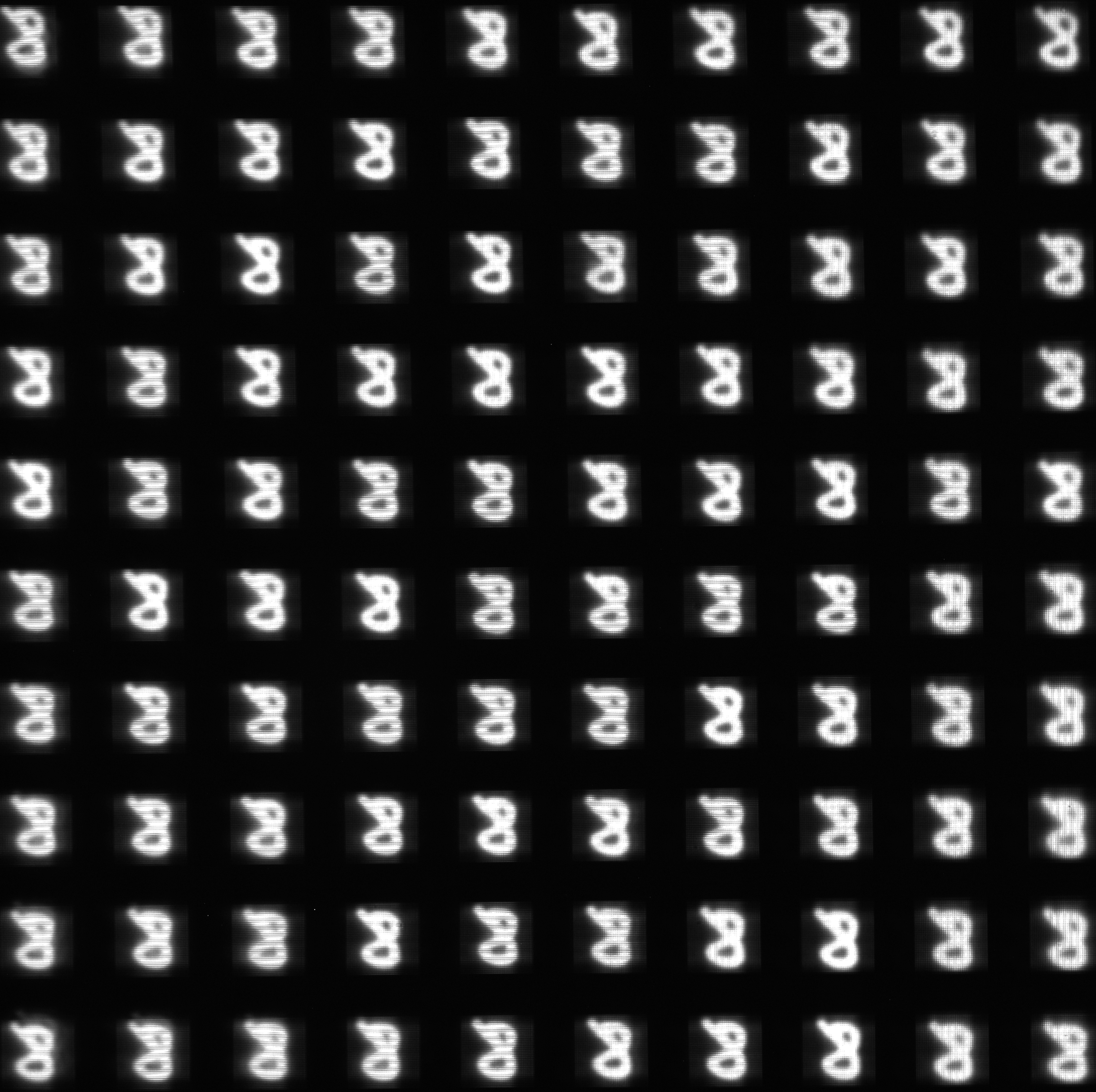}
\caption{\textbf{Optically fanned-out copies formed by MLA.} The image shows $10 \times 10 = 100$ optical copies formed on the plane of the LCD panel.}
\label{mnist_fanout}
\end{figure}

\begin{figure}[ht!]
\includegraphics [width=1.0\textwidth] {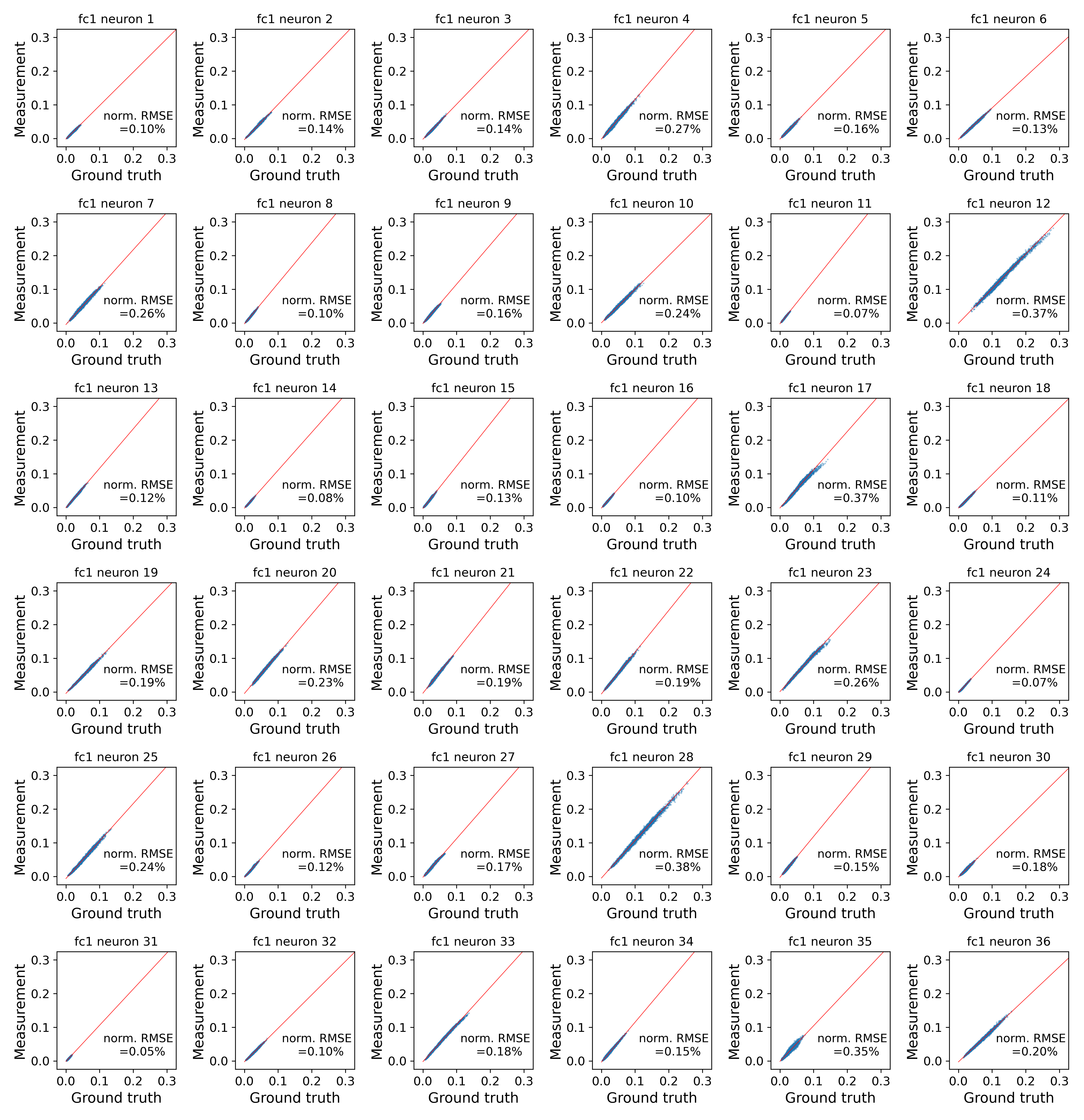}
\caption{\textbf{The calibration curve of each of the 36 neurons at the output of fully connected layer 1 (fc1).} The value of each neuron corresponds to the answers of vector-vector dot products (i.e., 'fc1 neuron $i$' reads out the dot product between $W_1[:,i]$ and $\vec{x}$), and the possible answer of each dot product is normalized to be in the range of $[0,1]$. The normalized root-mean-square error (norm. RMSE) is expressed in percentage, relative to the full possible answer range of 1. The data points shown here were measured from passing the training dataset of the 10-class QuickDraw dataset through the setup. Blue points: measured data points; Red lines: linear regression of the data points.}
\label{figs_mvm_cali_1}
\end{figure}

\begin{figure}[ht!]
\includegraphics [width=0.8\textwidth] {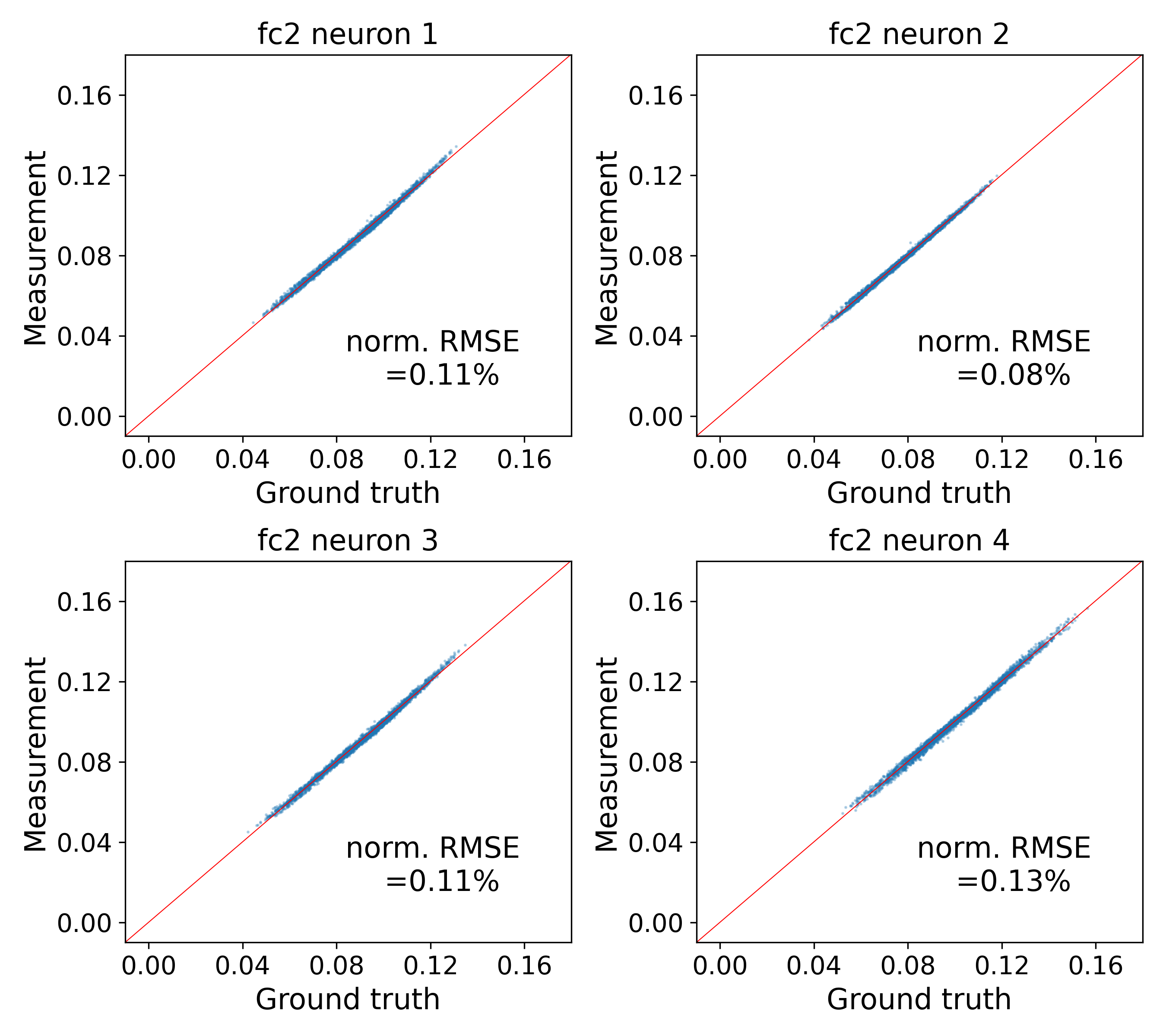}
\caption{\textbf{The calibration curve of each of the 4 neurons at the output of fully connected layer 2 (fc2).} The value of each neuron corresponds to the answers of vector-vector dot products (i.e., 'fc2 neuron $i$' reads out the dot product between $W_2[:,i]$ and $\vec{x}$), and the possible answer of each dot product is normalized to be in the range of $[0,1]$. The normalized root-mean-square error (norm. RMSE) is expressed in percentage, relative to the full possible answer range of 1. The data points shown here were measured from passing the training dataset of the 10-class QuickDraw dataset through the setup. Blue points: measured data points; Red lines: linear regression of the data points.
}
\label{figs_mvm_cali_2}
\end{figure}

\noindent The accuracy of optical matrix-vector multiplication was calibrated in the following way: We displayed a set of QuickDraw images on the DMD and captured optically fanned-out copies of the image in the plane of LCD 1 formed by MLA 1 (Supplementary Figure \ref{figS_overview}a). These images, instead of the original digital images, were treated as the ground truth of the input vectors to the optical matrix-vector multiplier, since the MLA could not resolve each DMD pixel, and therefore forms images blurrier than the digital images. These ground truth images were digitally multiplied by the weight matrices to obtain the ground truth of the output vectors of the optical matrix-vector multiplier, which was plotted as the horizontal axes in Supplementary Figure \ref{figs_mvm_cali_1}. To obtain the experimentally measured results of the optical matrix-vector multiplication, we placed a pellicle beam-splitter (BP245B1, Thorlabs Inc.) after LCD 1 and before the optical fan-in to the image intensifier. The reflected light was imaged onto the first monitoring camera (acA4024 - \SI{29}{\micro\metre}, Basler Inc. -- CAM1 in Supplementary Table \ref{parts}) and the image of each neuron was digitally summed as the measured results of the optical matrix-vector multiplication, which were plotted as the vertical axes in Supplementary Figure \ref{figs_mvm_cali_1}. Incidentally, the ground truth of the input vectors were obtained using the same setup except for setting LCD 1 to uniform transmission to capture optically fanned-out copies without any weight modulation. 

The calibration of optical matrix-vector multiplier in the fully connected layer 2 followed exactly the same procedure as the fully connected layer 1: To obtain the ground truth images for fully connected layer 2 we introduced another beam-splitter (BS013, Thorlabs Inc.) after the image intensifier and before the optical fan-out for optical fully connected layer 2. The reflected beam was imaged onto another monitoring camera (acA3088 - \SI{57}{\micro\metre}, Basler Inc. -- CAM2 in Supplementary Table \ref{parts}) that allowed us to measure the activation of the hidden-layer neurons after the optical-to-optical nonlinear activation function performed by the image intensifier. The activations of the hidden-layer neurons were multiplied with the weight matrix of the fully connected layer 2 to obtain the ground truth for the second linear optical layer (plotted as the horizontal axes in Supplementary Figure \ref{figs_mvm_cali_2}). To obtain the experimentally measured results of the second optical matrix-vector multiplication, the transmitted light was used for optical fan-out. The modulated images from the LCD 2 were imaged onto a final camera (Prime 95B Scientific CMOS Camera, Teledyne
Photometrics) using a zoom lens (Zoom 7000, Navitar Inc.), and the final vector was obtained by summing camera pixel values corresponding to each neuron (plotted as the vertical axes in Supplementary Figure \ref{figs_mvm_cali_2}).

The overall transmission of one optical matrix-vector multiplier was measured to be 2.9\% at best. This was measured in the following way: We measured the power of the light (with the white LED) reflected off of the DMD before the input lens (i.e., part L1 in Supplementary Figure \ref{exp_setup}) to the ONN encoder. This power was distributed over a large region of the MLA. The beam size that covered the MLA was about 2 inches in diameter. By scaling to the area of each lenslet, we computed the power per fan-out copy (neuron) before entering the ONN encoder. We then divided the output power per neuron (i.e., measured immediately in front of the photocathode of the image intensifier, see Supplementary Figure \ref{intensifier_impulse_response}) by this number to estimate the transmission through the matrix-vector-multiplier. We note that the optical loss of the current proof-of-concept experimental setup still cannot support low-light operation, however, the optical transmission of the setup can in principle be substantially improved by optimizing the optical imaging systems and the use of phase modulation as opposed to intensity modulation.

\section{Nonlinear activation functions via an image intensifier} \label{MCP}

\begin{figure}[ht!]
\includegraphics [width=1.0\textwidth] {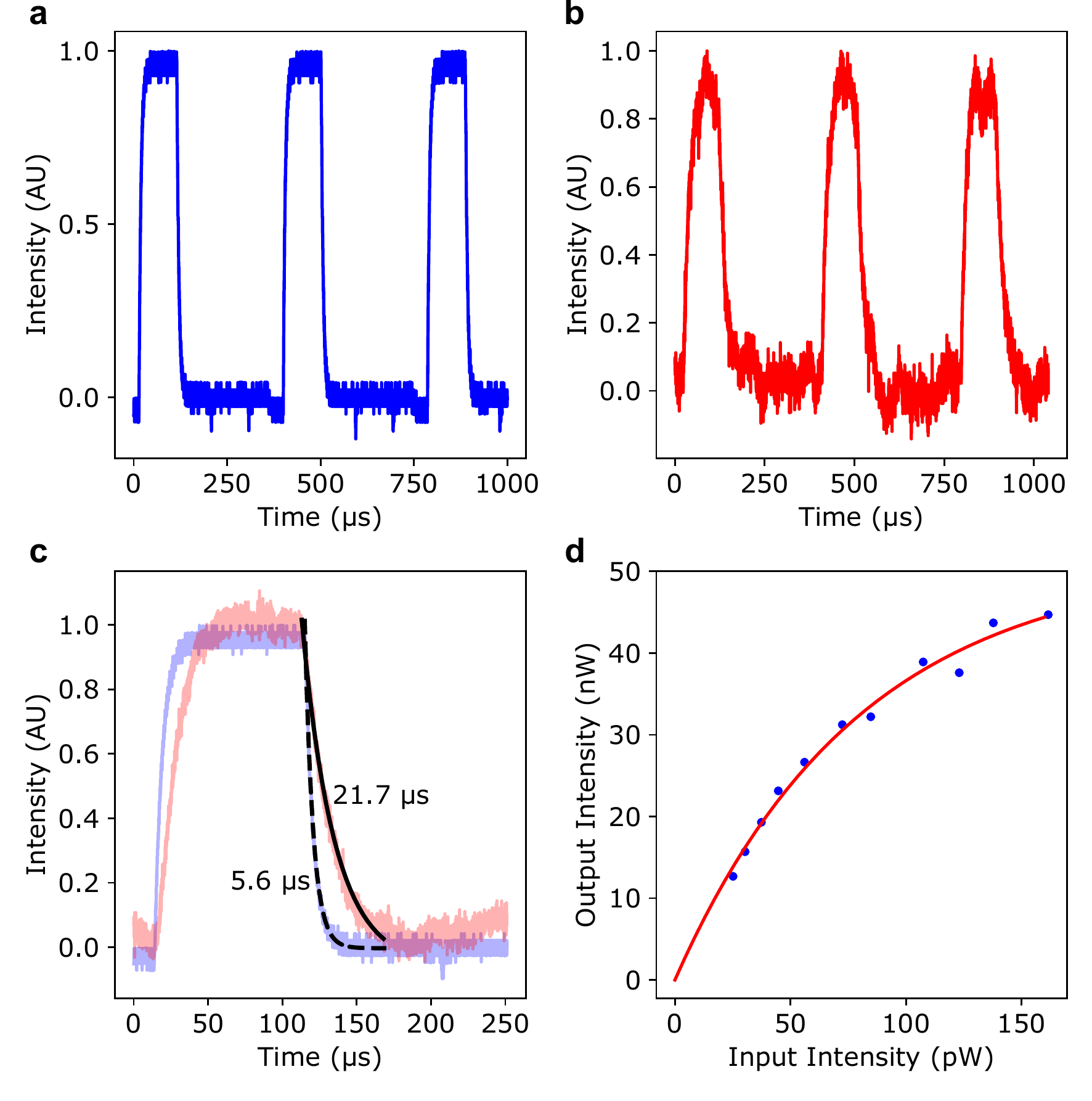}
\caption{\textbf{Measurement of the response time of the image intensifier}. \textbf{a}, Input pulses to the image intensifier using a DMD operating at \SI{10}{\kilo\hertz}. The on time was for $\sim$\SI{100}{\micro\second}, and the off time was for $\sim$ \SI{300}{\micro\second}. \textbf{b}, Output from the intensifier corresponding to the input in (a) at a gain control voltage $V_\text{gain}=\SI{3.35}{\volt}$. \textbf{c}, Fits to the exponential decay of the input and output light of the image intensifier. The dashed line shows the fit to the input light from the DMD, and the solid line shows the fit to the output light from the intensifier. The time constant for the image intensifier's decay was measured to be $\sim$ \SI{21.7}{\micro\second}. \textbf{d}, Optical-to-optical nonlinear activation at $>$\SI{10}{\kilo\hertz} bandwidth using white light illumination. The height of the output pulses of the image intensifier is plotted against its input pulse height. The red line depicts a fit to the nonlinear activation curve according to Equation \ref{eqn1}.}
\label{intensifier_impulse_response}
\end{figure}

\noindent An image intensifier is a device that receives an input optical image of low light level and outputs an amplified and brighter version of the same image by emitting light. Image intensifiers have been regularly applied to detecting weak optical signals involved in fast phenomena on nanosecond to microsecond time scale, since each constituent part of the image intensifier can be engineered to have $\sim$ \SI{}{\nano\second} response time and the gating time can be set to \SI{}{\nano\second} scale \cite{Zemel1991}. 

A typical image intensifier consists of a photocathode, one or two microchannel plate (MCP), and a phosphor screen. When an optical image is formed onto the photocathode, photoelectrons are generated according to the spatial pattern of the image, which are immediately amplified by the MCP in a spatially resolved fashion. As the amplified current reaches the phosphor screen, it excites photons on the phosphor screen following the same spatial pattern of the input image, and thus creating a brighter version of the input image. In this study, we used an image intensifier that can provide up to 3,500 times optical-to-optical amplification (S20 photocathode, 1-stage MCP, P46 phosphor, Photek Inc.). 

\begin{figure}[ht!]
\includegraphics [width=1.0\textwidth]{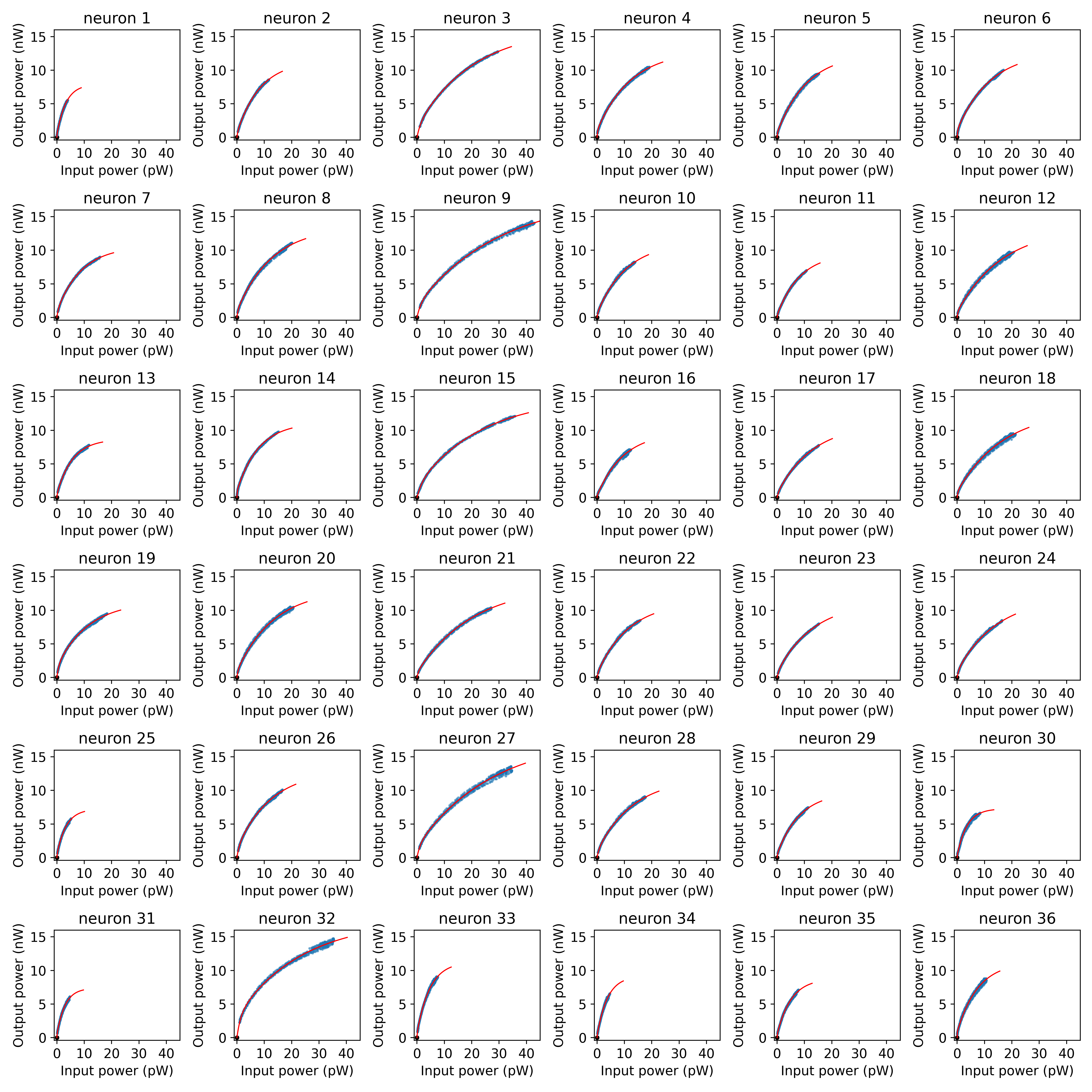}
\caption{ \textbf{Nonlinear activation functions induced by the image intensifier measured from all the 36 neurons at the output of fully connected layer 1 (fc1).} The data was measured from the training dataset of the 3D speed-limit sign recognition task. The gain control voltage $V_\text{gain}$ for the image intensifier was set to \SI{3.75}{\volt}. Blue points: measured data points; Red lines: fitting curves of the data points according to Equation \ref{eqn1}.
}
\label{figs_nonlinear_curves}
\end{figure}

We measured the response speed of the image intensifier and its nonlinear response to input light pulses. By flickering all the pixels of the DMD at its maximum update rate (Supplementary Figure \ref{figS_overview}b, we created a pulse train of light as the input to the image intensifier with $\sim \SI{100}{\micro\second}$ on time and $\sim \SI{300}{\micro\second}$ off time. Supplementary Figure \ref{intensifier_impulse_response}a shows the temporal profiles of the light pulse train measured before the intensifier by a photodiode (PDA100A2, Thorlabs), and Supplementary Figure \ref{intensifier_impulse_response}b shows the temporal profile of the light emission from the phosphor screen of the image intensifier measured by the same photodiode. The light emission pattern of the image intensifier completely followed the input light pattern created by the DMD. Each light pulse emitted by the image intensifier decays to the baseline level before the next pulse takes place, and its $1/e$ decay time was fitted to be \SI{21.7}{\micro\second} (Supplementary Figure \ref{intensifier_impulse_response}). Furthermore, we studied the relation between the input and output light intensity by changing the input light intensity (i.e., the pulse height in Supplementary Figure \ref{intensifier_impulse_response}a), and then measured the output light intensity (i.e., the pulse height in Supplementary Figure \ref{intensifier_impulse_response}b). Supplementary Figure \ref{intensifier_impulse_response}d shows that the output power of the image intensifier changes nonlinearly with its input power at a bandwidth of at least \SI{16} {\kilo\hertz} (3dB bandwidth for exponential decay can be calculated from the decay constant as 0.35/\SI{21.7} {\micro\second}=\SI{16} {\kilo\hertz}). Also note that the input power to the image intensifier to obtain nonlinear response was well below its damage threshold. The measured load current from the current monitor port of the image intensifier was about \SI{3}{\nano\A} which was well below the maximum rating of \SI{1}{\micro\A}.

Besides the nonlinear amplification of input light, it is also important to verify that the image intensifier can perform element-wise nonlinear activation by amplifying each spatial mode individually without significant cross-talk. To obtain spatially resolved nonlinear activation functions, we measured the brightness of the 36 spatial modes, corresponding to the 36 neurons of the hidden layer, before and after the image intensifier. The reason why only 36 neurons in a 6 by 6 array were chosen for the hidden layer was because we blocked every other optical copy formed by the lenslet on the MLA in order to prevent cross-talk between neighboring neurons on the image intensifier. The inputs to the intensifier were measured using the first monitoring camera CAM 1 (acA4024, Basler Inc., see \ref{MVM_cali}), i.e., the vector after the first optical fan-in. The output from the image intensifier was measured using the second monitoring camera CAM 2 (acA3088, Basler Inc., see \ref{MVM_cali}). A scatter plot was made using these two measurements and curves were fit to the data. The equation used for the fit was
\begin{equation} \label{eqn1}
    y = a(1 - e^{-bx}) + c(1 - e^{-dx})
\end{equation}
where $a, b, c, d$ are the fitting parameters. The fitted nonlinear curves can be seen in Supplementary Figure \ref{figs_nonlinear_curves}. We note that Equation \ref{eqn1} is purely empirical. In Supplementary Figure \ref{figs_nonlinear_curves}, the shape of the nonlinear activation curve is similar for each neuron, which indicates the nonlinear response is relatively uniform across the area of the image intensifier. 

For the QuickDraw and the cell-organelle classification experiments, the optical-to-optical gain was set to $\sim$ \SI{700}{\watt/\watt} at a gain control voltage $V_\text{gain}=\SI{3.3}{\volt}$; for the real-scene experiments, the gain was set to $\sim$  \SI{1000}{\watt/\watt} at a gain control voltage $V_\text{gain}=\SI{3.75}{\volt}$. No gating was applied during these experiments. Note that the optical-to-optical gains reported above are based on the product specification, the actual gain of the image intensifier also depends on multiple factors including the intensity and the wavelength of input light, especially when operating in the nonlinear regime. During the operation of all the experiments, the image intensifier along with the gain control circuits consumed $\sim \SI{500}{\milli\watt}$ average power.

\part{Datasets and Optical-neural-network Training}

\section{The general procedure for training optical-neural-network encoders} \label{onn_training}

\noindent For all the ONN encoders demonstrated in this study, we modeled the nonlinear ONN encoder as a 2-layer neural network with all-to-all connections (i.e., a multilayer perceptron, MLP), and trained its linear layers' weights digitally on a computer. Supplementary Figure \ref{quickdraw_nn_structure} shows the diagram of a typical neural network used in this study. The input images to the ONN were modeled as 8-bit grayscale images of size $40 \times 40 = 1,600$ total pixels, followed by one hidden layer with 36 neurons and one bottleneck layer with 4 neurons (or 2 neurons for the 3D speed-limit sign classification); each neuron of a layer is fully connected to each neuron of the previous layer. The nonlinear activation function for each of the 36 neurons in the hidden layer was modeled by a double-exponential function (Equation \ref{eqn1}) with parameters fitted to each of the individual neurons (Supplementary Figure \ref{figs_nonlinear_curves}). Depending on the task, different digital backends were used after the optical bottleneck layer. For example, a single digital fully connected layer was used for most of the classification tasks. The digital backend was always trained together with the ONN-encoder frontend as a whole during training.

Unlike training regular digital neural networks, training an ONN encoder operating with incoherent light requires maintaining all its weights non-negative throughout the training process. This was implemented by clamping all the weights of fully connected or convolutional layers in the range of $[0,1]$ during each forward propagation. To improve the robustness of model to physical noises and hardware errors, we employed the following techniques during training: 

\begin{enumerate}
    \item Adding noise to the inputs to each optical layer: To improve the noise resilience of the model, we added random noise in the range of $[0, 0.02]$ to each element of the input vector to every optical layer. Each element of the optical input vector is in the range of $[0,1]$, and therefore the relative error is around 2\%. 
    \item Data augmentation with random image transforms: To improve model tolerance to potential hardware imperfections, we augmented input images by random transformations composed of $\pm 5\%$ random translation and $\pm 4\%$ random scaling. These measures not only helped to improve model immunity to imaging errors, but also serves as a means for regularization to reduce overfitting.
    \item Tuning training parameters: We used AdamW optimizer \cite{loshchilov2017decoupled} for training. The training parameters (e.g., learning rate) were manually chosen, and fine-tuned by using a hyperparameter searching package Optuna \cite{optuna_2019}.
    \item Scheduled learning rate decay and stochastic weight averaging: For some model training, we used a decaying learning rate according to a cosine schedule for the AdamW optimizer and applied stochastic weight averaging \cite{izmailov2018averaging} after a certain number of training epochs when the learning rate diminishes. In some cases, these techniques helped the training to converge to flatter and more robust solutions.
\end{enumerate} 

To reduce the performance gap between digitally trained models and their execution on the physical ONN encoder, we used a layer-by-layer approach that is similar to Ref. \cite{zhou2021large} to train the weights for the ONN encoder, which we found greatly improved the accuracy as opposed to training both layers in simulation and then uploading the weights to the LCDs. The following is the protocol for training:
\begin{enumerate}
    \item Train both optical fully connected layers and the digital backend (Supplementary Figure \ref{quickdraw_nn_structure}) in simulation and upload the weights of optical fully connected layer 1 to the LCD of the first optical matrix-vector multiplier.
    \item Collect the data of the nonlinear activations after the image intensifier, i.e., the data from the second monitoring camera (CAM2 in Supplementary Table \ref{parts}, see \ref{MVM_cali}). 
    \item Retrain optical fully connected layer 2 and the digital backend in simulation using the nonlinear activations experimentally collected in (2) as the input data to optical fully connected layer 2. After maximizing validation accuracy, update the retrained weights of optical fully connected layer 2 to the LCD of the second optical matrix-vector multiplier.
    \item Collect the output of the entire ONN encoder from camera 3 (CAM3 in Supplementary Table \ref{parts}), which was placed at the output of the optical fully connected layer 2.
    \item Retrain the digital backend layer(s) to maximize validation accuracy.
\end{enumerate}

Once this entire training procedure was done, we fixed the weights of the entire ONN encoder together with its digital backend. For classification tasks, we then displayed test images, which had not been used for training, on the DMD. We collected the outputs of the ONN encoder and fed them through the digital backend to obtain test accuracy, which is the accuracy we used to compare between nonlinear and linear ONN classifiers.

All the neural-network models were implemented and trained in PyTorch (1.11.0) \cite{paszke2019pytorch}. The training hyperparameters, such as learning rate and batch size, can be found in the training code specific to each task (code available at: \url{https://github.com/mcmahon-lab/Image-sensing-with-multilayer-nonlinear-optical-neural-networks}).

\section{Recognition of hand-drawn figures} \label{quickdraw}

\begin{figure}[h!]
\includegraphics [width=0.92\textwidth]{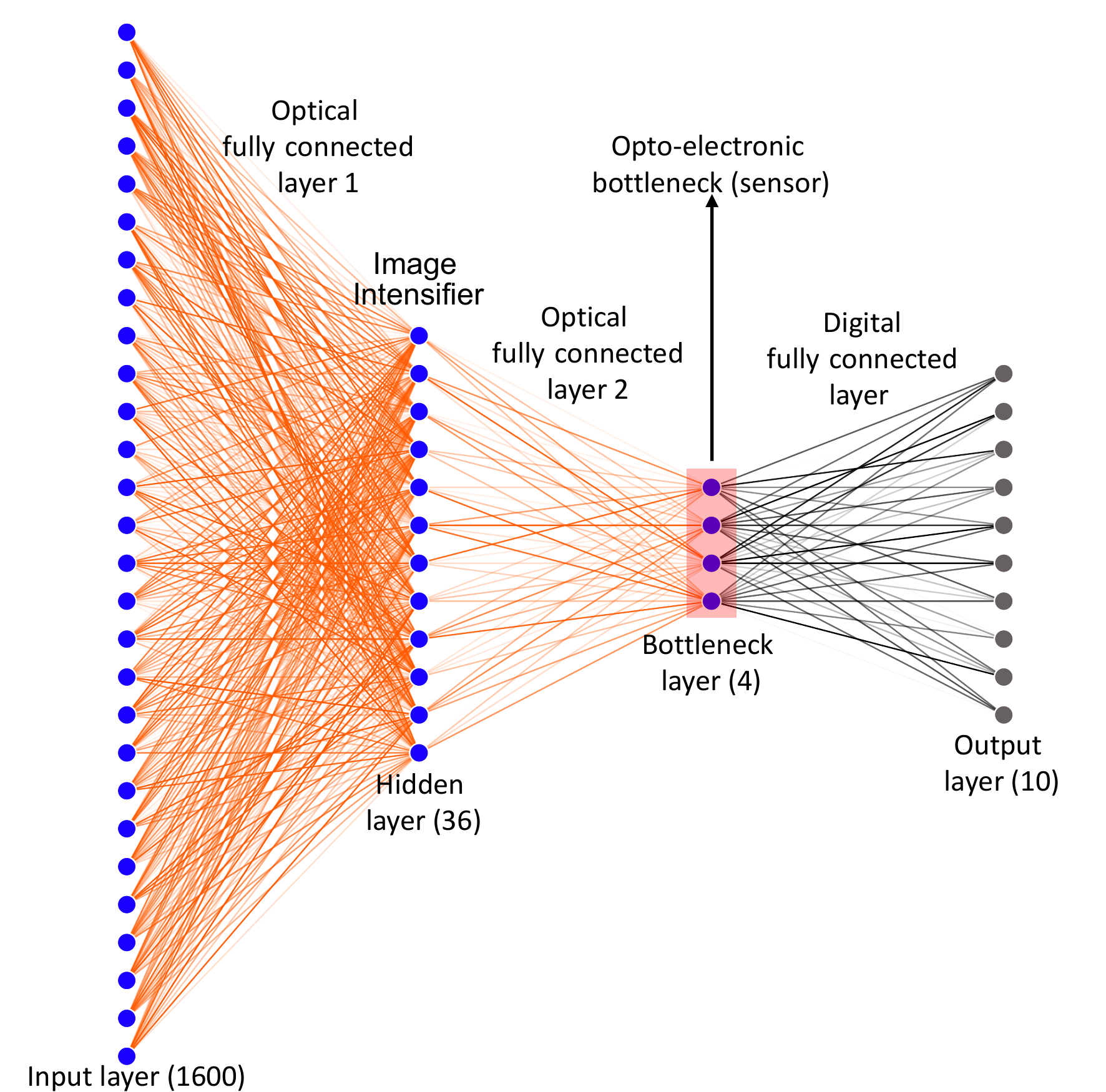}
\caption{\textbf{Neural-network diagram for the QuickDraw experiment} The input layer consists of $40 \times 40 = 1,600$ total pixels, followed by a hidden layer with 36 neurons and one bottleneck layer with 4 neurons; each neuron of a layer is fully connected to each neuron of the previous layer. An image intensifier was used as the nonlinear activation function for each of the 36 neurons in the hidden layer. A digital decoder serves as the backend after the bottleneck layer, which consists of a single fully connected linear layer. The digital output layer has an output dimension of 10, with each neuron corresponding to a class of QuickDraw objects.}
\label{quickdraw_nn_structure}
\end{figure}

\noindent We used the Google QuickDraw dataset \cite{quickdraw} to show that nonlinear ONN encoders outperform linear optical encoders for image classification. This dataset was chosen because it can be easily binarized and displayed on a DMD, but at the same time is much more difficult than MNIST since hand-drawn figures have much larger variation than digits. We chose 10 classes (clock, chair, computer, eyeglasses, tent, snowflake, pants, hurricane, flower, crown) from the 250+ available classes. These classes were chosen with no special rationale other than ensuring the classes were not too similar. For each class, images were taken according to their original sequence in the dataset, with the only filtering being done to remove improper or invalid images. For example, for the class 'chair', some images contain the text 'chair' instead of a drawing of a chair, and thus were removed. The train and validation dataset contains 300 images per class (3,000 images in total), and was randomly split to 250 and 50 images per class for training and validation respectively. The test set contains 50 images per class (500 images in total) that are completely different from those in the train and validation set. The dataset used in the experiment can be found on Zenodo (\verb|Figure_2bcd/Quickdraw_GT_images_April_7.npz| in \verb|Figure_2.zip| at \url{https://doi.org/10.5281/zenodo.6888985}).

We binarized these images and displayed them on the DMD (after resizing them to $100\times100$ from their original size of $28\times28$) and collected ground truth images. The ground truth images were taken according to the procedure described in \ref{MVM_cali}. These ground truth images needed to be collected because the MLA could not resolve each DMD pixel, and therefore we needed to use the blurrier images seen by the LCD instead of the original images displayed on the DMD to train the ONN encoder. The ground truth images spanned a size of $40\times40$ pixels on the LCD. Thus, 1,600 was used as the effective dimension of the input vector. These ground truth datasets were used for the training the weights to be uploaded to the LCDs. 

The digital model of ONN encoder consists of a 1,600-dimensional input layer, a hidden layer of 36 neurons and a bottleneck layer of 4 neurons. We appended a digital decoder after the bottleneck layer, which consists of a single fully connected linear layer. The digital output layer has an output dimension of 10, with each neuron corresponding to a class of QuickDraw objects. Supplementary Figure \ref{quickdraw_nn_structure} shows the entire neural network (optical + digital) diagram used for the QuickDraw experiments. The training procedure is described in \ref{onn_training}. The trained weights of optical fully connected layer 1 and 2 resulting from the training protocol described above are plotted in Supplemental Figure \ref{quickdraw_weights}.

\begin{figure}[h!]
\includegraphics [width=0.75\textwidth] {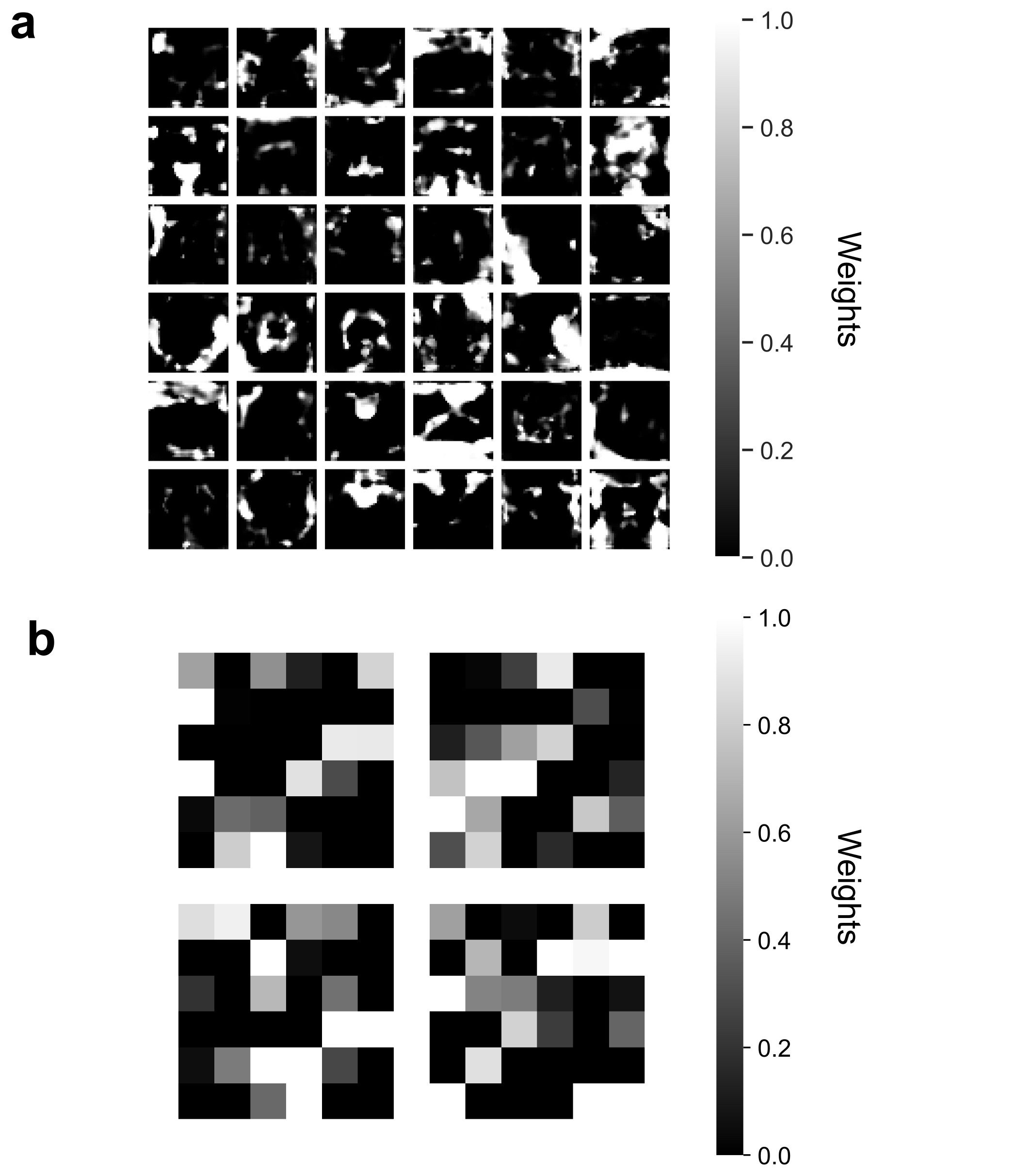}
\caption{\textbf{The trained weights of fully connected layers used in the nonlinear ONN encoder for the QuickDraw hand-drawn figure classification task.}  \textbf{a}, Non-negative weights of fully connected layer 1 (fc1), which maps the input layer consisting of $40 \times 40=1,600$ neurons to the hidden layer consisting of $6 \times 6=36$ neurons. \textbf{b}, Non-negative weights of fully connected layer 2 (fc2), which maps the hidden layer consisting of $6 \times 6=36$ neurons to the bottleneck layer consisting of $2 \times 2=4$ neurons.}
\label{quickdraw_weights}
\end{figure}

\section{Image-based cell-organelle classification} \label{cell}

\begin{figure}[h!]
\includegraphics [width=0.6\textwidth] {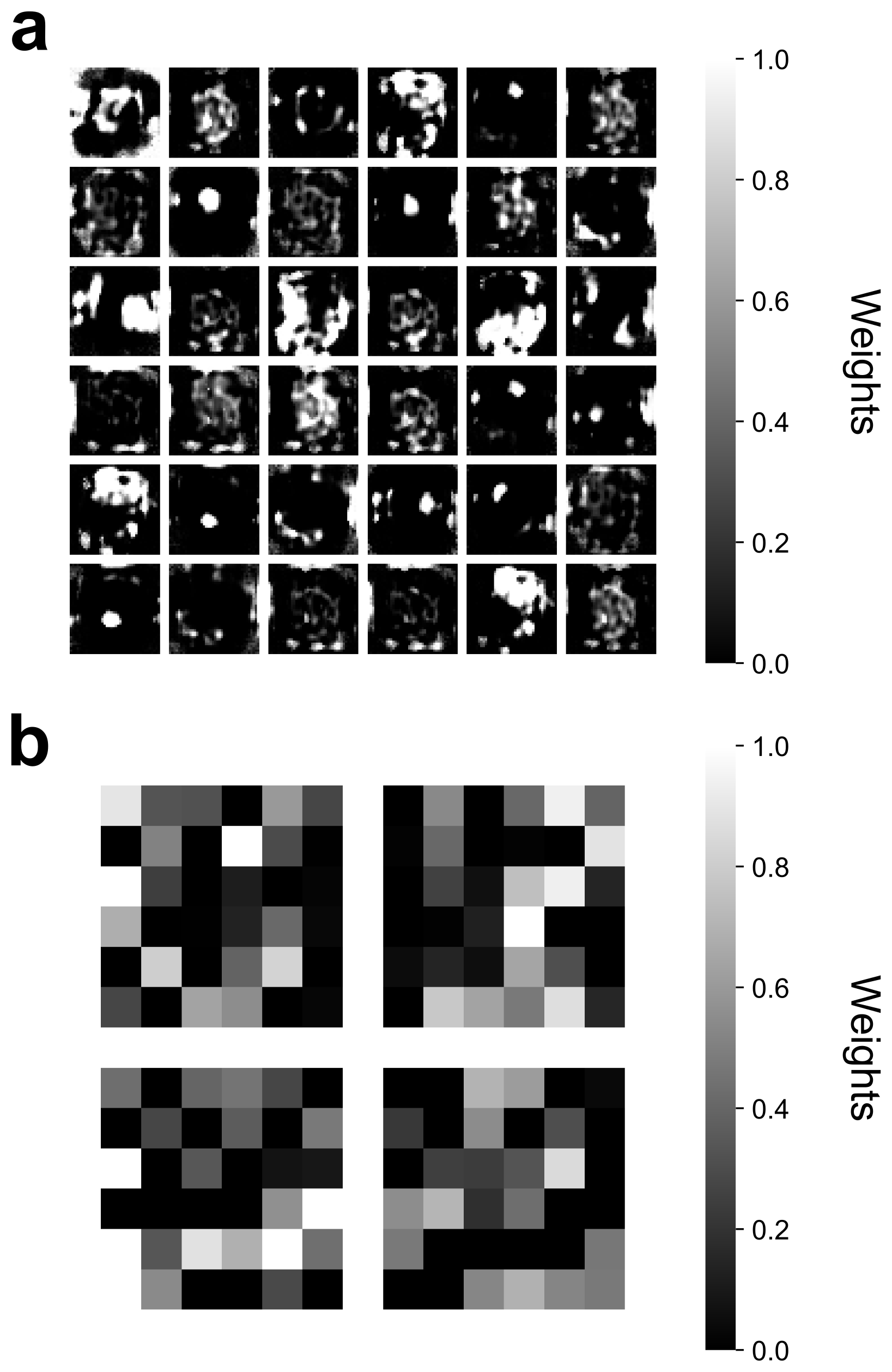}
\caption{\textbf{The trained weights of fully connected layers used in the nonlinear ONN encoder for the cell-organelle classification task.}  \textbf{a}, Non-negative weights of fully connected layer 1 (fc1), which maps the input layer consisting of $40 \times 40=1,600$ neurons to the hidden layer consisting of $6 \times 6=36$ neurons. \textbf{b}, Non-negative weights of fully connected layer 2 (fc2), which maps the hidden layer consisting of $6 \times 6=36$ neurons to the bottleneck layer consisting of $2 \times 2=4$ neurons.}
\label{cell_weights}
\end{figure}

\noindent In this task, we classified 5 classes of HeLa cell images based on which of their organelles was stained (nucleolus, cytoplasm, centrosomes, cell mask, mitochondria). The data was downloaded from a publicly available data repository (S-BSST644, available from \url{https://www.ebi.ac.uk/biostudies/}), which contains cell images taken from flow cytometry experiments performed in Ref. \cite{schraivogel2022high}. We only used the fluorescent channel of the images, which visualizes the stained organelles, for classification. The original images are 104 pixels in width with varying height, and we cropped them to a size of 100 by 100 pixels, with the cell located near the center of each image. The train, validation, and test images were taken in order from the repository according to the original sequence of images in the repository. Only images with no cell or multiple cells were skipped and replaced with the next valid images of a single cell, which agrees with the conventional practice of flow cytometry that typically discards doublet cells. The training and validation dataset contains 200 images for each class (1,000 images in total for 5 classes) that were randomly split into 160 and 40 images per class for training and validation respectively. The test dataset contains 40 additional images for each class (200 images in total for 5 classes). The modified cell-organelle dataset as described above is available for download from (\verb|/Figure_2fg/EBI_Cells.npz| in \verb|Figure_2.zip| at \url{https://doi.org/10.5281/zenodo.6888985}). As with the QuickDraw dataset, we binarized and resized the images before displaying them on the DMD, and took ground truth images for neural-network training. The neural-network structure was the same as in Supplementary Figure \ref{quickdraw_nn_structure}, and the training procedure was the same as the procedure described in \ref{onn_training}.

\section{Classification of 3D real-scene objects} \label{realscene}

\begin{figure}[h!]
\includegraphics [width=0.6\textwidth] {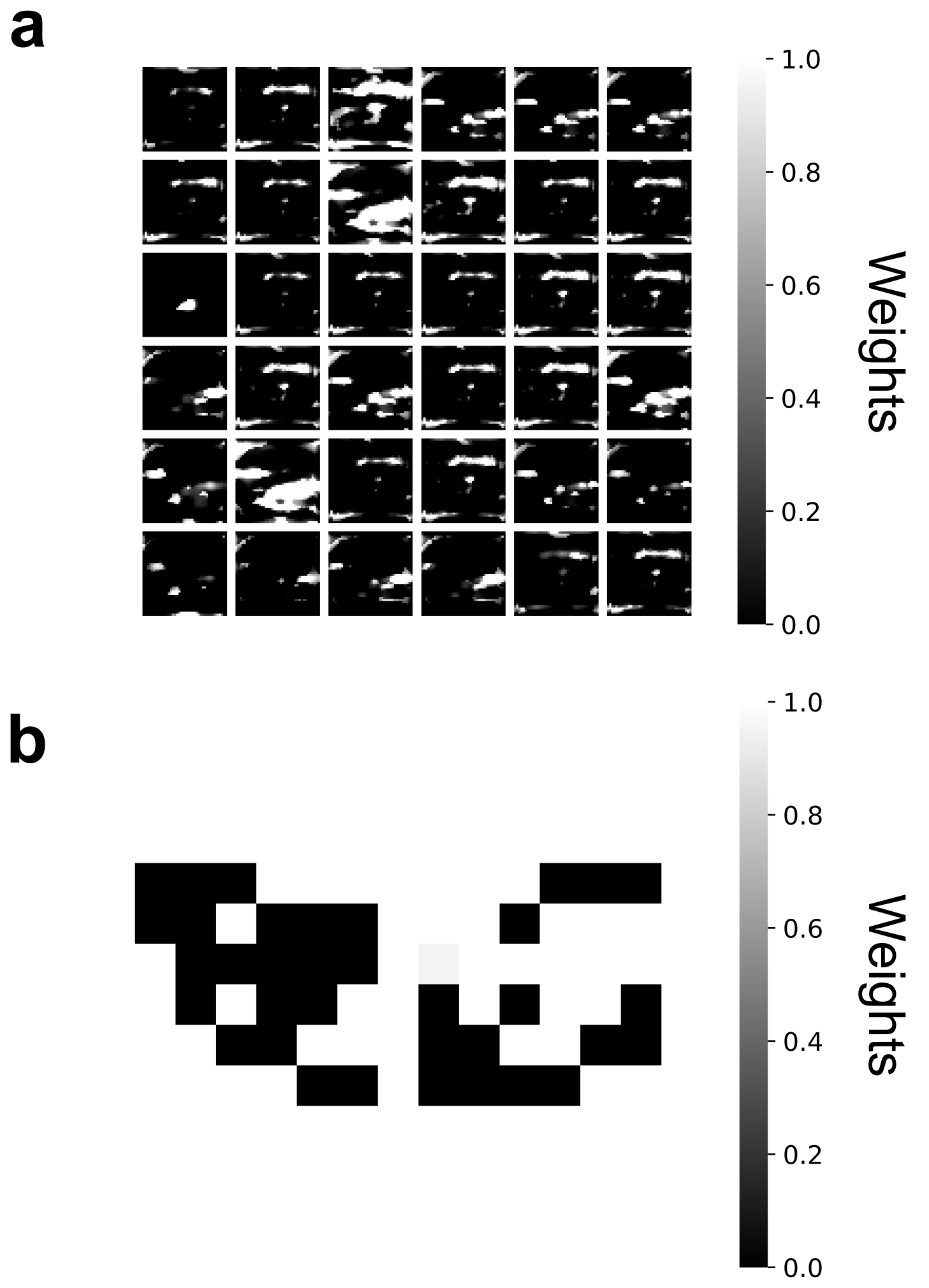}
\caption{\textbf{The trained weights of fully connected layers used in the nonlinear ONN encoder for speed-limit sign classification.}  \textbf{a}, Non-negative weights of fully connected layer 1 (fc1), which maps the input layer consisting of $40 \times 40=1,600$ neurons to the hidden layer consisting of $6 \times 6=36$ neurons. \textbf{b}, Non-negative weights of fully connected layer 2 (fc2), which maps the hidden layer consisting of $6 \times 6=36$ neurons to the bottleneck layer consisting of $1 \times 2=2$ neurons.}
\label{speedlimit_weights}
\end{figure}

\noindent In this task, we created a small 3D scene centered around a traffic sign holder (see Supplementary Figure \ref{figS_overview}). We made 8 speed-limit signs (15, 20, 25, 30, 40, 55, 70, 80) to place in that holder. We used a zoom lens to create a demagnified real image of the speed-limit sign in front of the ONN smart sensor. The zoom lens allowed us to continuously adjust the size of the image of the speed-limit sign so that the image size was about \SI{1}{\milli\meter} $\times$ \SI{1}{\milli\meter}. This image size was the same as the images displayed on the DMD in previous experiments so that the optically fanned-out copies of the speed-limit sign filled an area of $40\times40$ pixels on the LCD. Ground truth images were taken for each angle (0 to 88 degrees) for each class, thus making the entire dataset size 712. Every $4^{\text{th}}$ angle was used for validation, which resulted in 536 training images and 176 validation images. The ONN encoder had a similar structure as the one used in QuickDraw (Supplementary Figure \ref{quickdraw_nn_structure}) with the exception that the bottleneck layer had 2 instead of 4 neurons, and the digital decoder had two layers with a structure of $2 \rightarrow 40 \rightarrow 8$. The training procedure was the same as in \ref{onn_training}.    

\section{Autoencoders for image reconstruction} \label{vae}

\begin{figure}[h!]
\includegraphics [width=\textwidth] {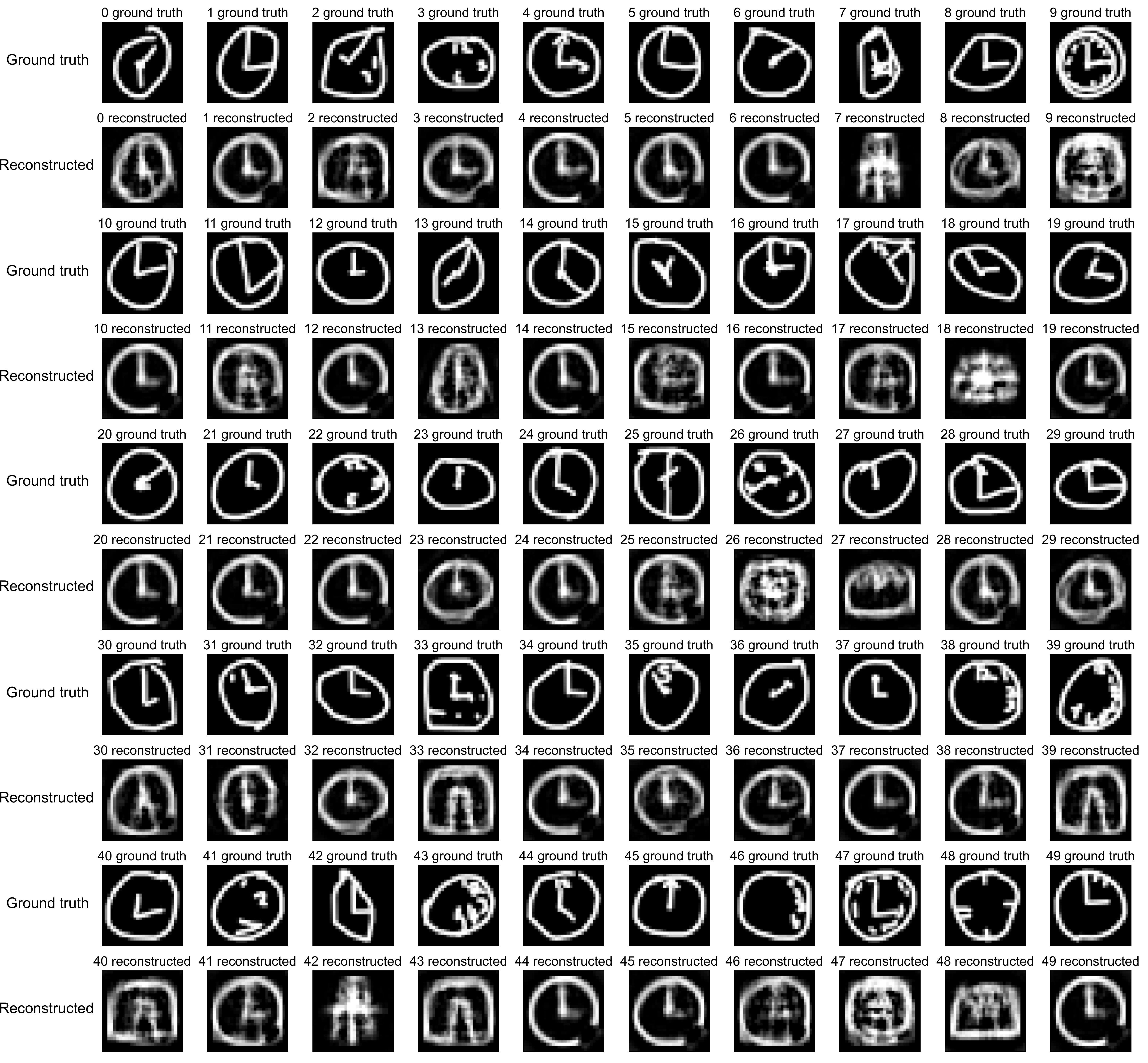}
\caption{\textbf{Reconstruction of Clocks from QuickDraw.} Reconstruction of clocks from the latent-feature space created by the ONN encoder during classification. The target reconstruction images (ground truths) are in the odd rows, and the reconstructed images are in the even rows.}
\label{clock_recon}
\end{figure}

\newpage

\begin{figure}[h!]
\includegraphics [width=\textwidth] {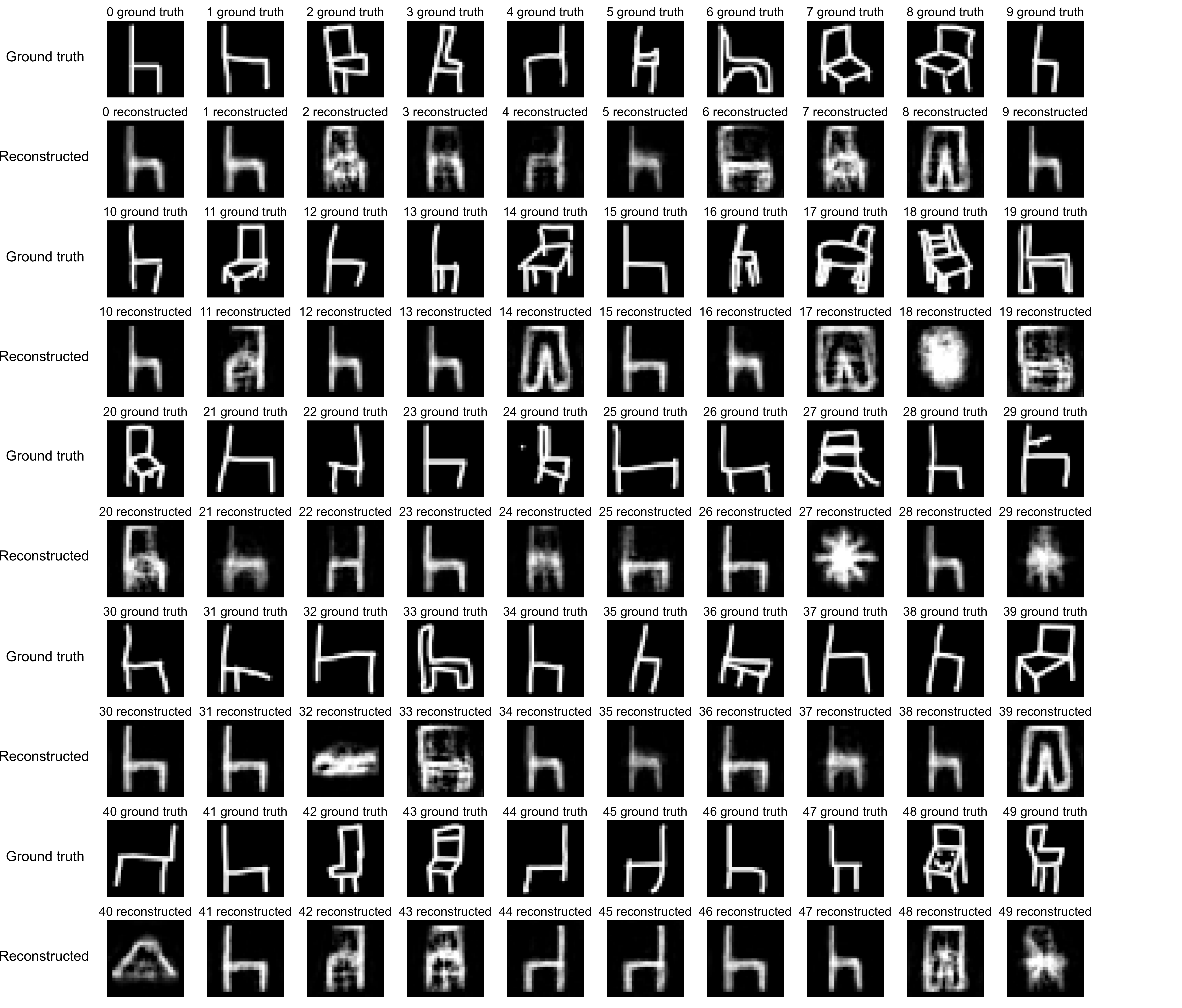}
\caption{\textbf{Reconstruction of Chairs from QuickDraw.} Reconstruction of chairs from the latent-feature space created by the ONN encoder during classification. The target reconstruction images (ground truths) are in the odd rows, and the reconstructed images are in the even rows.}
\label{chair_recon}
\end{figure}

\newpage

\begin{figure}[h!]
\includegraphics [width=\textwidth] {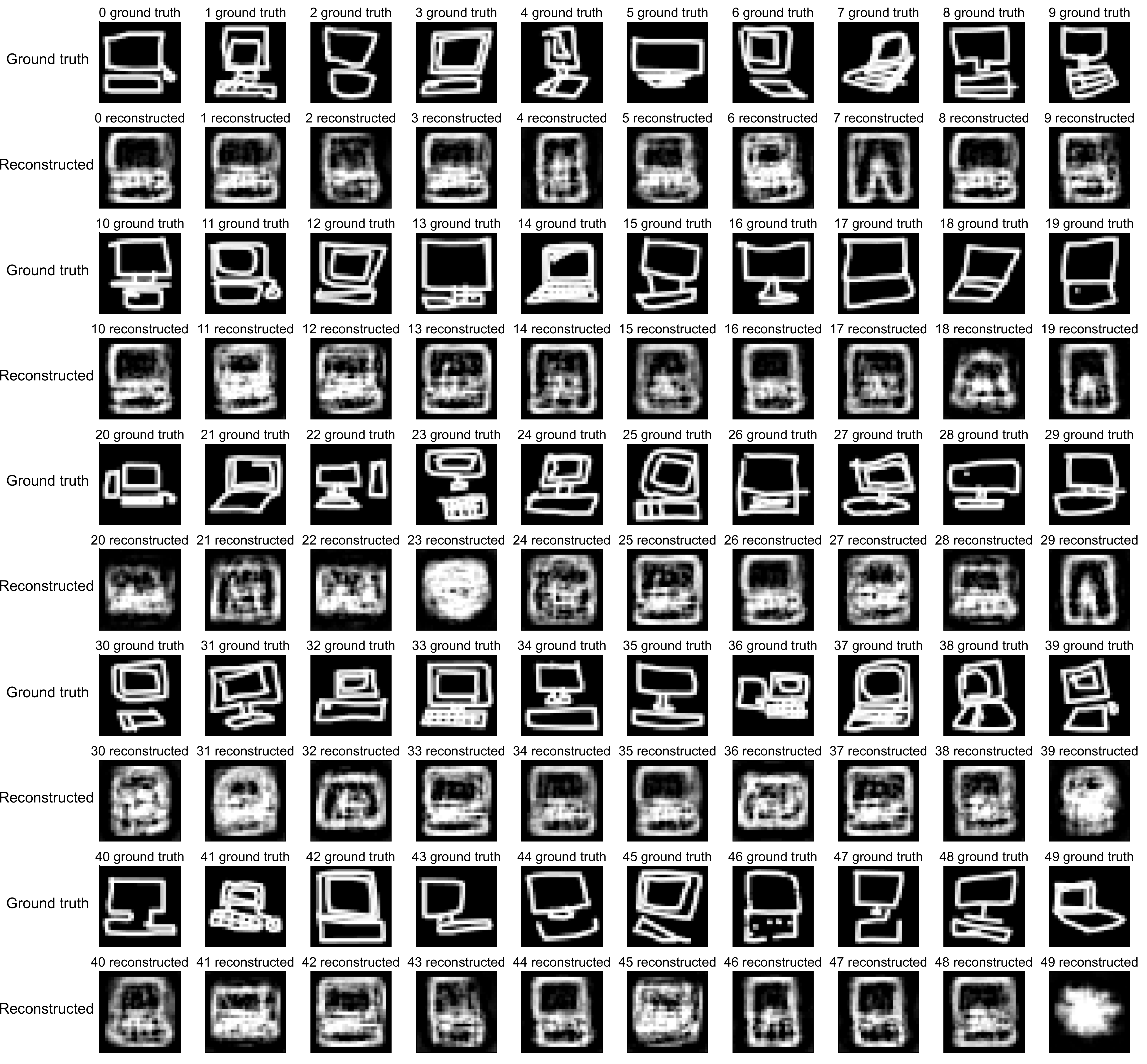}
\caption{\textbf{Reconstruction of Computers from QuickDraw.} Reconstruction of computers from the latent-feature space created by the ONN encoder during classification. The target reconstruction images (ground truths) are in the odd rows, and the reconstructed images are in the even rows.}
\label{computers_recon}
\end{figure}

\newpage

\begin{figure}[h!]
\includegraphics [width=\textwidth] {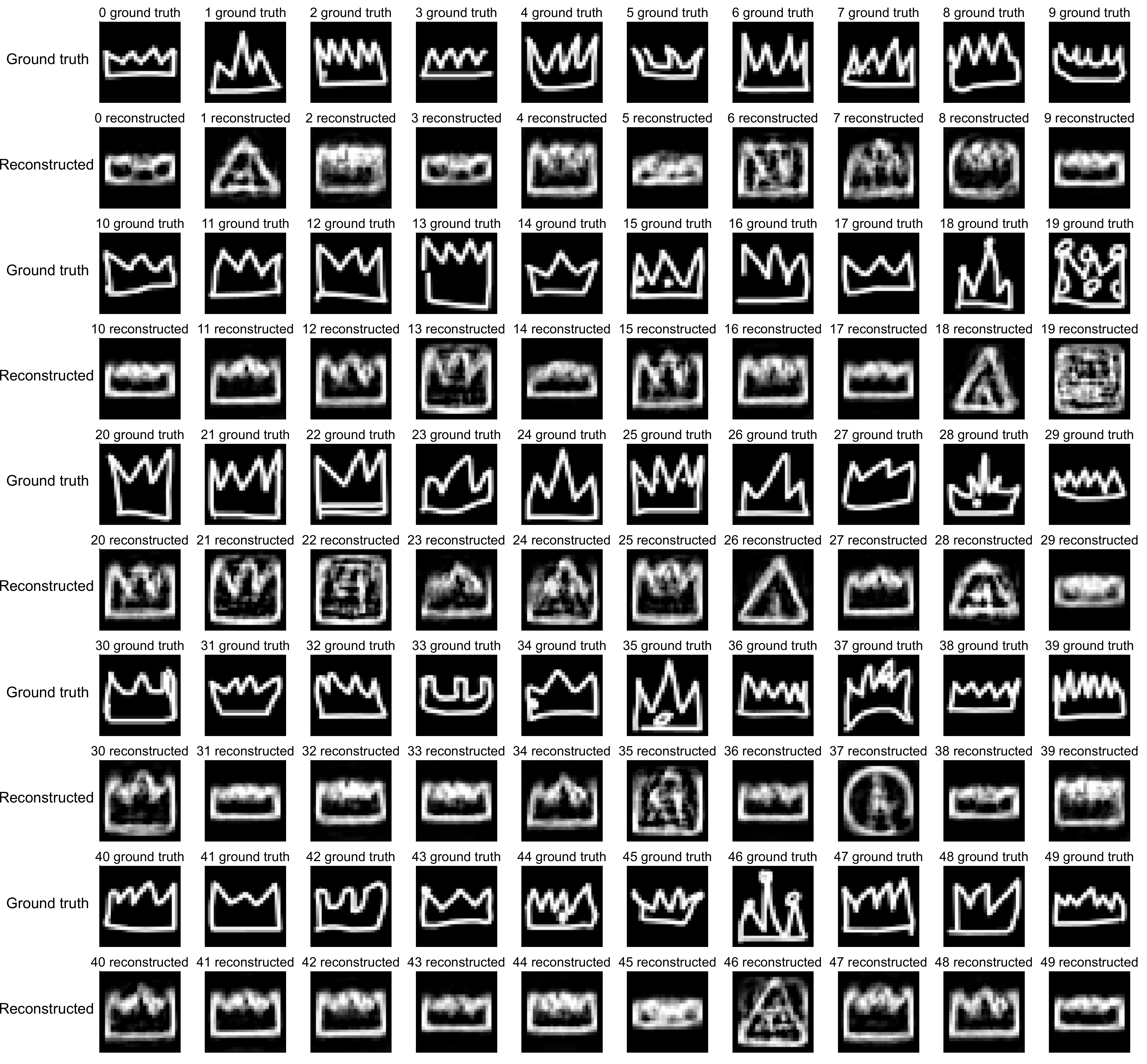}
\caption{\textbf{Reconstruction of Crowns from QuickDraw.} Reconstruction of crowns from the latent-feature space created by the ONN encoder during classification. The target reconstruction images (ground truths) are in the odd rows, and the reconstructed images are in the even rows.}
\label{crown_recon}
\end{figure}

\newpage

\begin{figure}[h!]
\includegraphics [width=\textwidth] {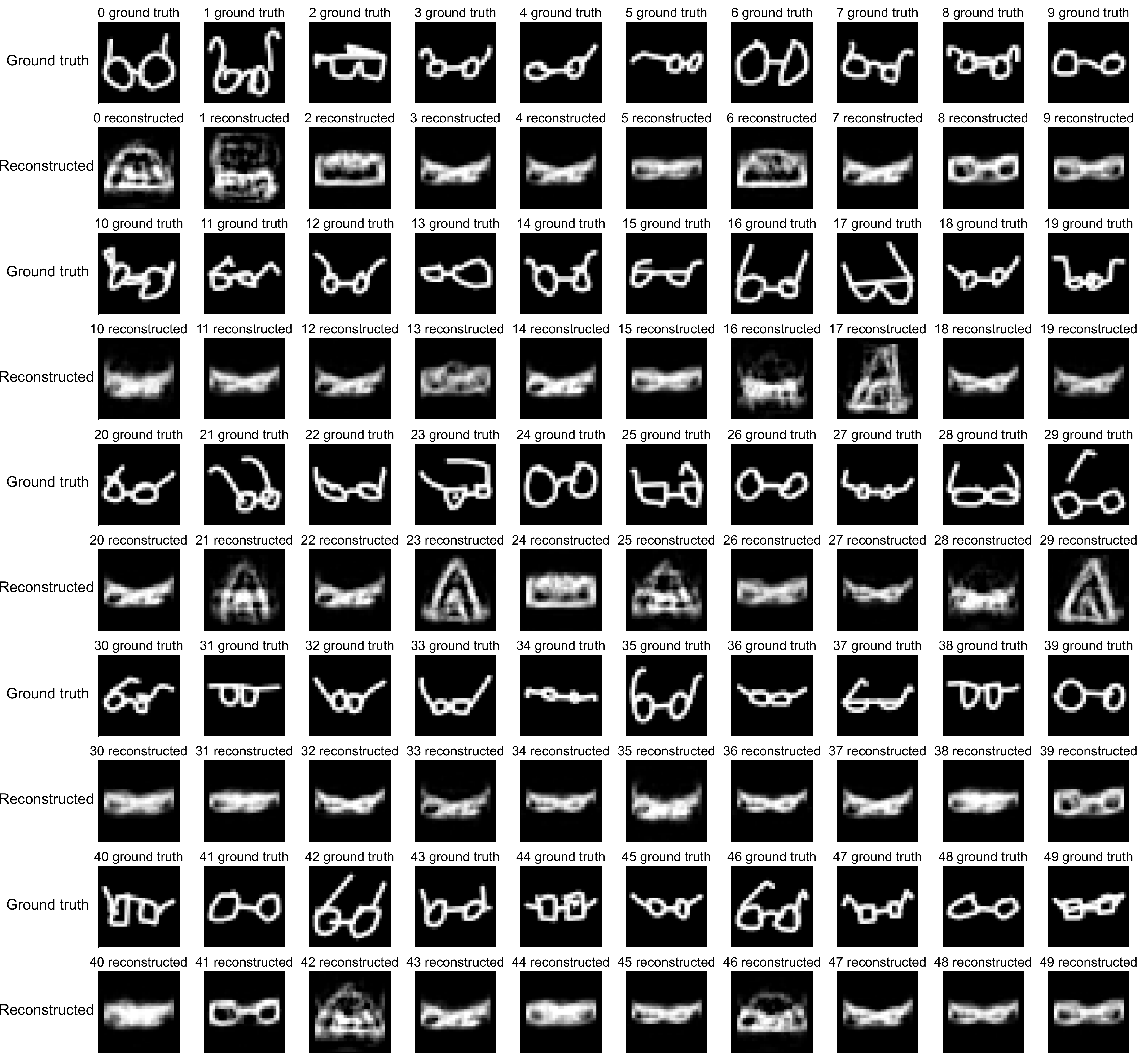}
\caption{\textbf{Reconstruction of Eyeglasses from QuickDraw.} Reconstruction of eyeglasses from the latent-feature space created by the ONN encoder during classification. The target reconstruction images (ground truths) are in the odd rows, and the reconstructed images are in the even rows.}
\label{eyeglasses_recon}
\end{figure}

\newpage

\begin{figure}[h!]
\includegraphics [width=\textwidth] {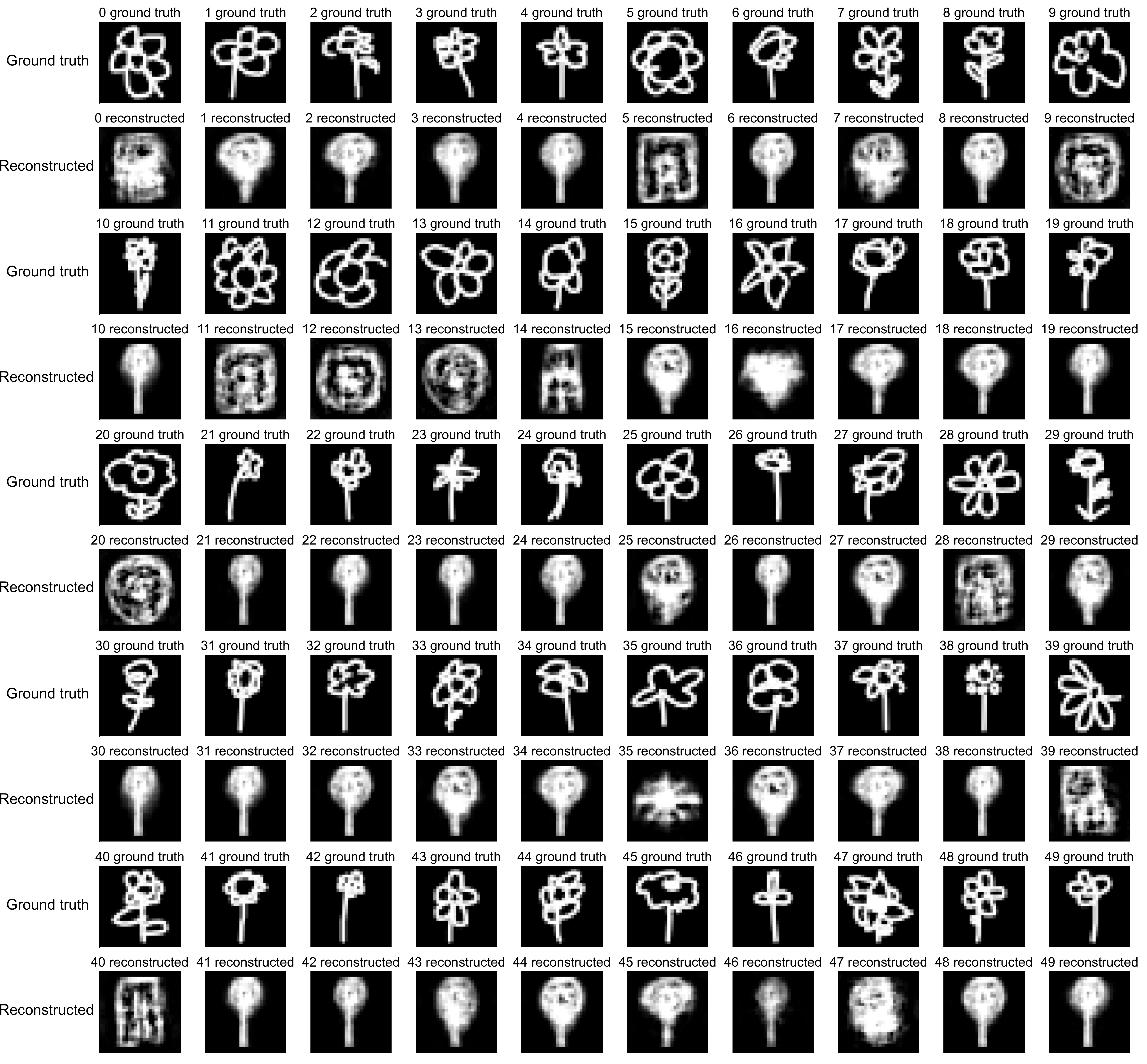}
\caption{\textbf{Reconstruction of Flowers from QuickDraw.} Reconstruction of flowers from the latent-feature space created by the ONN encoder during classification. The target reconstruction images (ground truths) are in the odd rows, and the reconstructed images are in the even rows.}
\label{flower_recon}
\end{figure}

\newpage

\begin{figure}[h!]
\includegraphics [width=\textwidth] {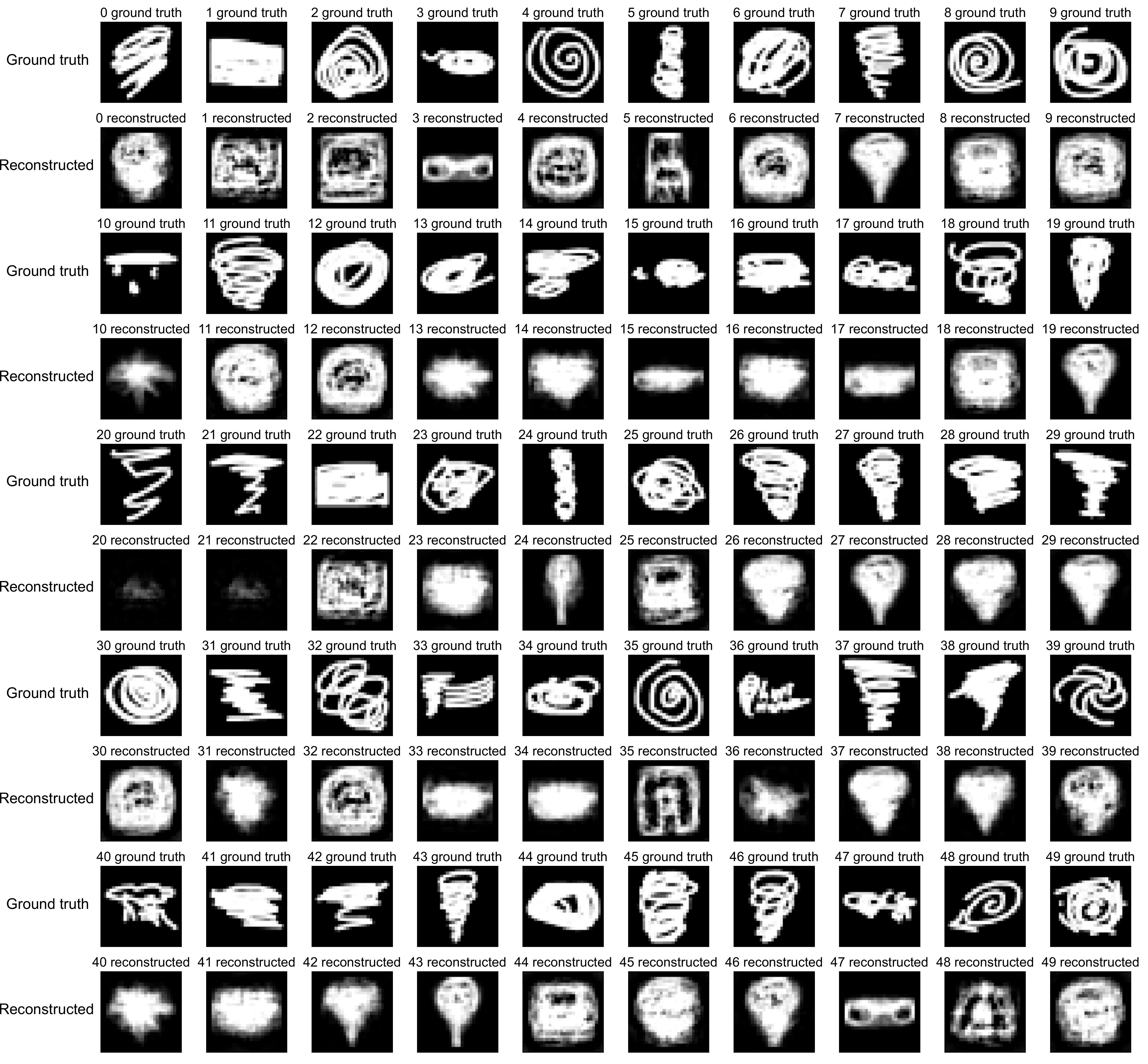}
\caption{\textbf{Reconstruction of Hurricanes from QuickDraw.} Reconstruction of hurricanes from the latent-feature space created by the ONN encoder during classification. The target reconstruction images (ground truths) are in the odd rows, and the reconstructed images are in the even rows.}
\label{hurricane_recon}
\end{figure}

\newpage

\begin{figure}[h!]
\includegraphics [width=\textwidth] {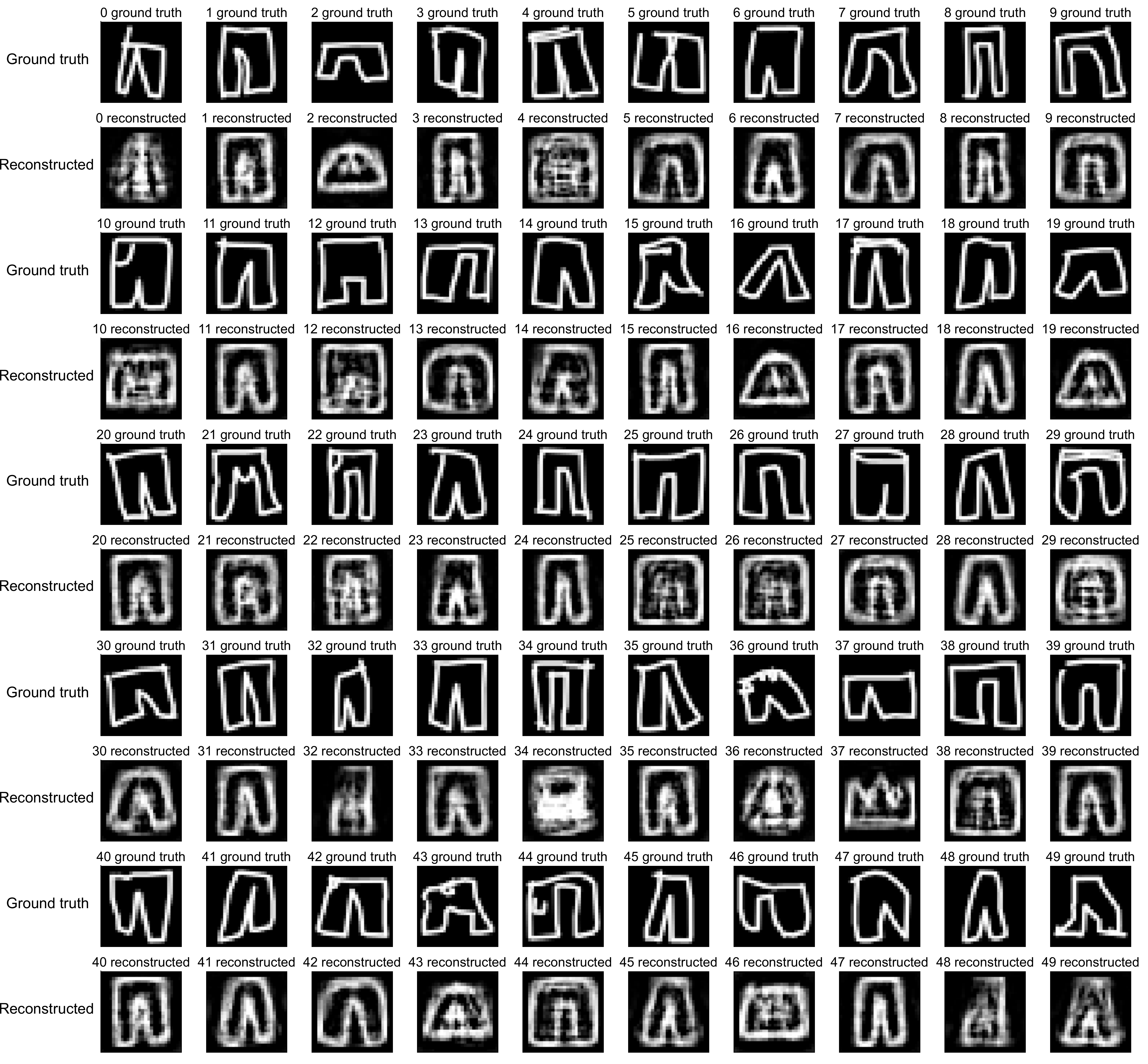}
\caption{\textbf{Reconstruction of Pants from QuickDraw.} Reconstruction of pants from the latent-feature space created by the ONN encoder during classification. The target reconstruction images (ground truths) are in the odd rows, and the reconstructed images are in the even rows.}
\label{pant_recon}
\end{figure}

\newpage

\begin{figure}[h!]
\includegraphics [width=\textwidth] {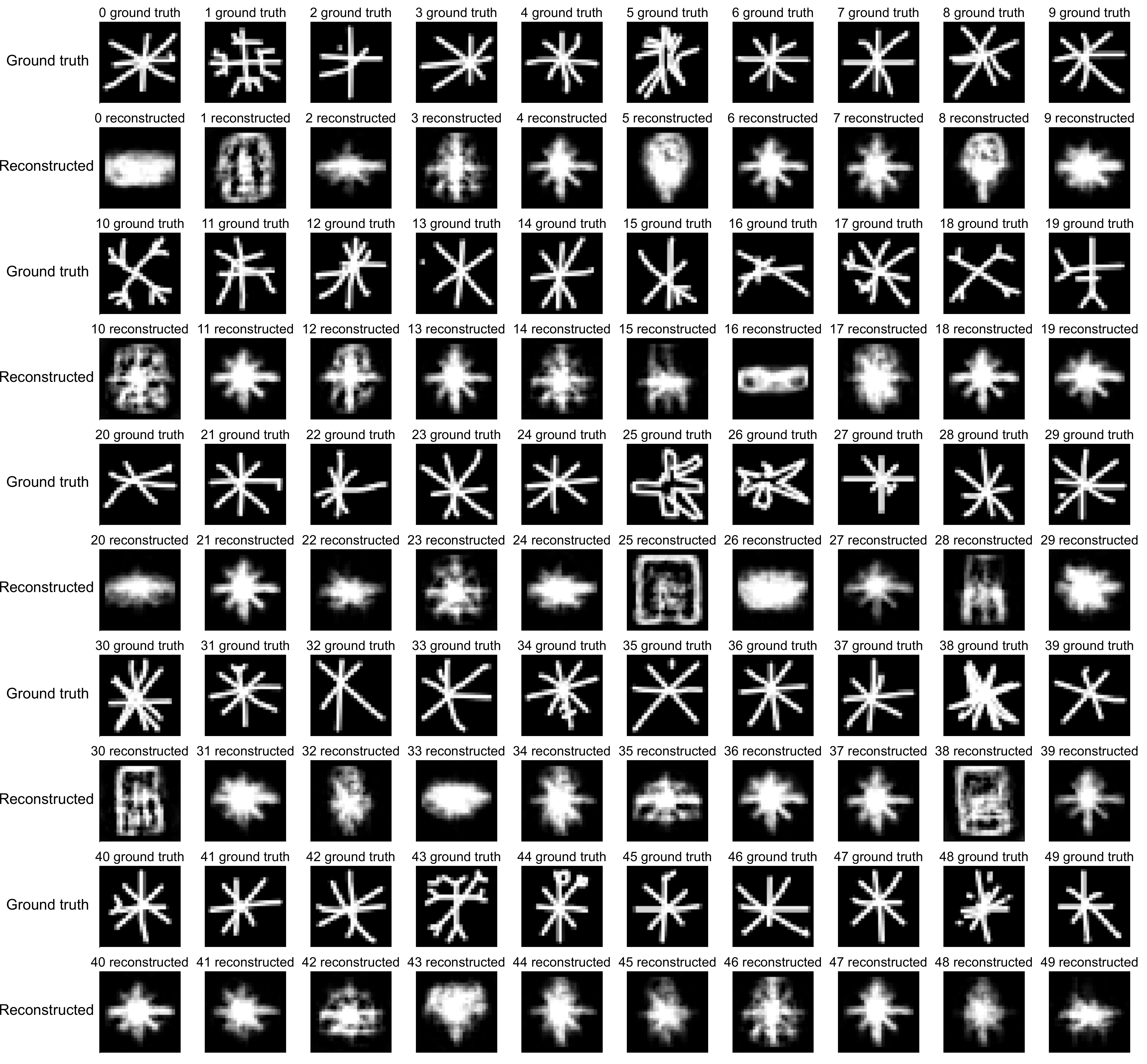}
\caption{\textbf{Reconstruction of Snowflakes from QuickDraw.} Reconstruction of snowflakes from the latent-feature space created by the ONN encoder during classification. The target reconstruction images (ground truths) are in the odd rows, and the reconstructed images are in the even rows.}
\label{snowflake_recon}
\end{figure}

\newpage

\begin{figure}[h!]
\includegraphics [width=\textwidth] {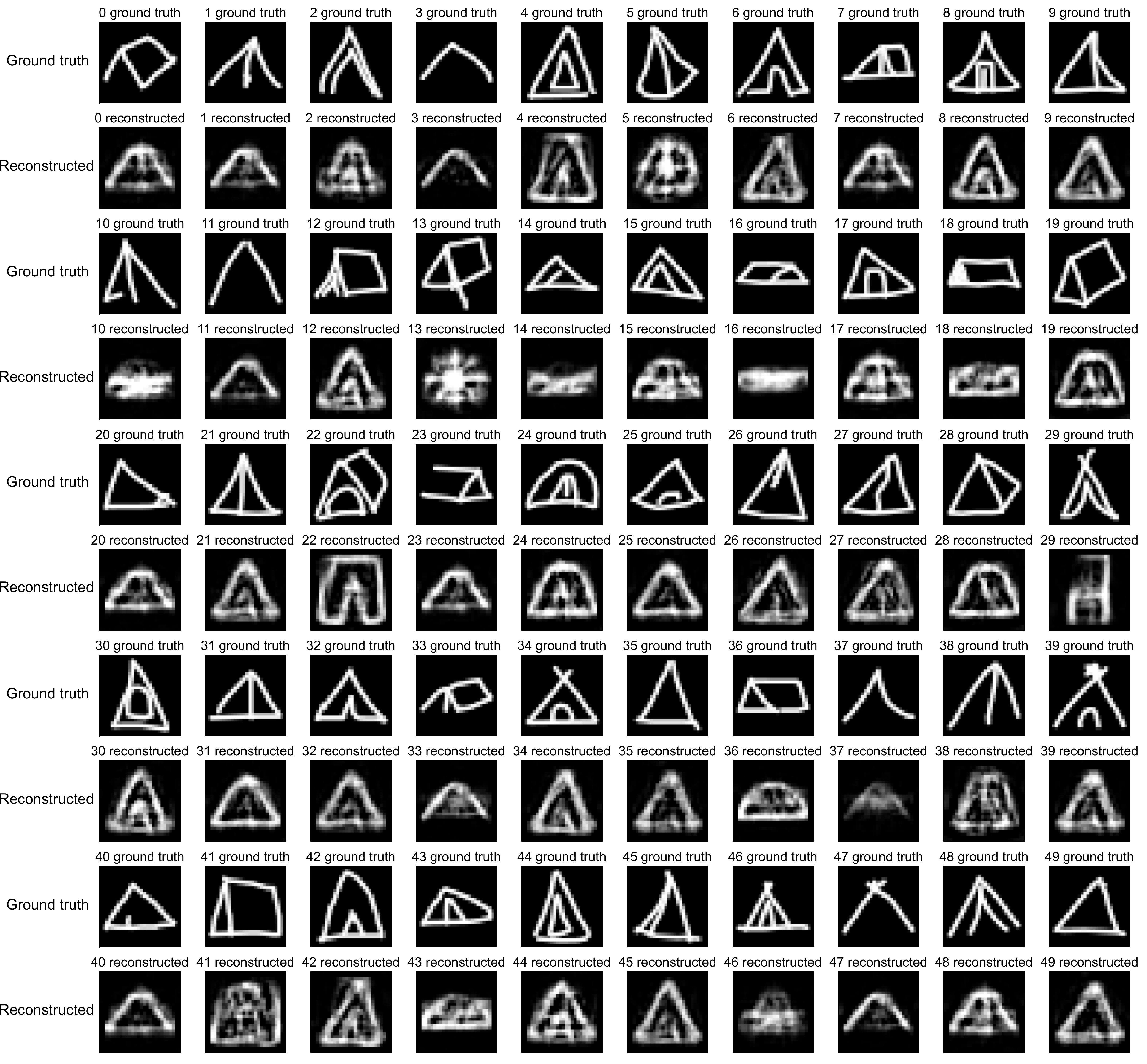}
\caption{\textbf{Reconstruction of Tents from QuickDraw.} Reconstruction of tents from the latent-feature space created by the ONN encoder during classification. The target reconstruction images (ground truths) are in the odd rows, and the reconstructed images are in the even rows.}
\label{tent_recon}
\end{figure}

\noindent To show how different digital backends can make use of the same latent-feature space for different tasks, we trained a digital decoder network to reconstruct QuickDraw Images (see Supplementary Figure \ref{clock_recon} - \ref{tent_recon}). The digital decoder network architecture used was a multilayer perceptron where the 4-dimensional output vectors created by the ONN encoder during the classification experiment (see \ref{quickdraw}) were used as the inputs. The original $28\times28$ grayscale QuickDraw images were used as the target images for reconstruction. The number and dimensions of the hidden layers were found by a random neural architecture search over a search space of one to five hidden layers with 10 to 1,000 neurons each (sampled from uniform and log-uniform distribution, respectively) and a sigmoid activation function. The best architecture found had three hidden layers and dimensions 4 $\rightarrow$ 288 $\rightarrow$ 863 $\rightarrow$ 291 $\rightarrow$ ($28\times28$). Different nonlinear activation functions and convolutional decoder architectures were also tried with no improvement to performance. Batch normalization (BatchNorm) layers were employed before each of the nonlinear activation functions, i.e., each layer was essentially linear matrix vector multiplication $\rightarrow$ BatchNorm $\rightarrow$ nonlinear activation. This improved performance significantly. The $28\times28=784$ outputs from the network were compared to the original QuickDraw images with the structural similarity index (SSIM) using the implementation from Ref. \cite{ssim}. The negative SSIM was backpropagated and minimized using an Adam optimizer \cite{kingma2014adam}.

From the reconstructions, we concluded that the latent-feature space preserved some salient features of the QuickDraw images in addition to the class information: For instance, chairs are usually facing in the correct direction (left and right facing) after reconstruction (see Supplementary figure \ref{chair_recon}), the presence or absence of a tent's floor is usually preserved (see Supplementary figure \ref{tent_recon}) and so is the outline of different types of hurricanes (conical and spherical perspectives, see Supplementary figure \ref{hurricane_recon}). Often, it is clear why the encoder-decoder network was misled into a faulty reconstruction. For example, the sole rounded crown \#37, (see Supplementary figure \ref{crown_recon}) was reconstructed into something akin to a round clock, the sprawling legs of chair \#27 (see Supplementary figure \ref{chair_recon}) were reconstructed into the arms of a star. It is possible that a more powerful digital decoder network can produce more faithful reconstructions. However, we found that larger and more complicated architectures, if at all, only yield marginal improvements to the reconstructed image quality. We therefore believe that the reconstruction is largely limited by the low-dimensional latent space with a compression ratio of $28 \times 28 /4 = 196$.

\section{Anomaly detection using unsupervised learning} \label{cluster}

\begin{figure}[h!]
\includegraphics [width=0.9\textwidth] {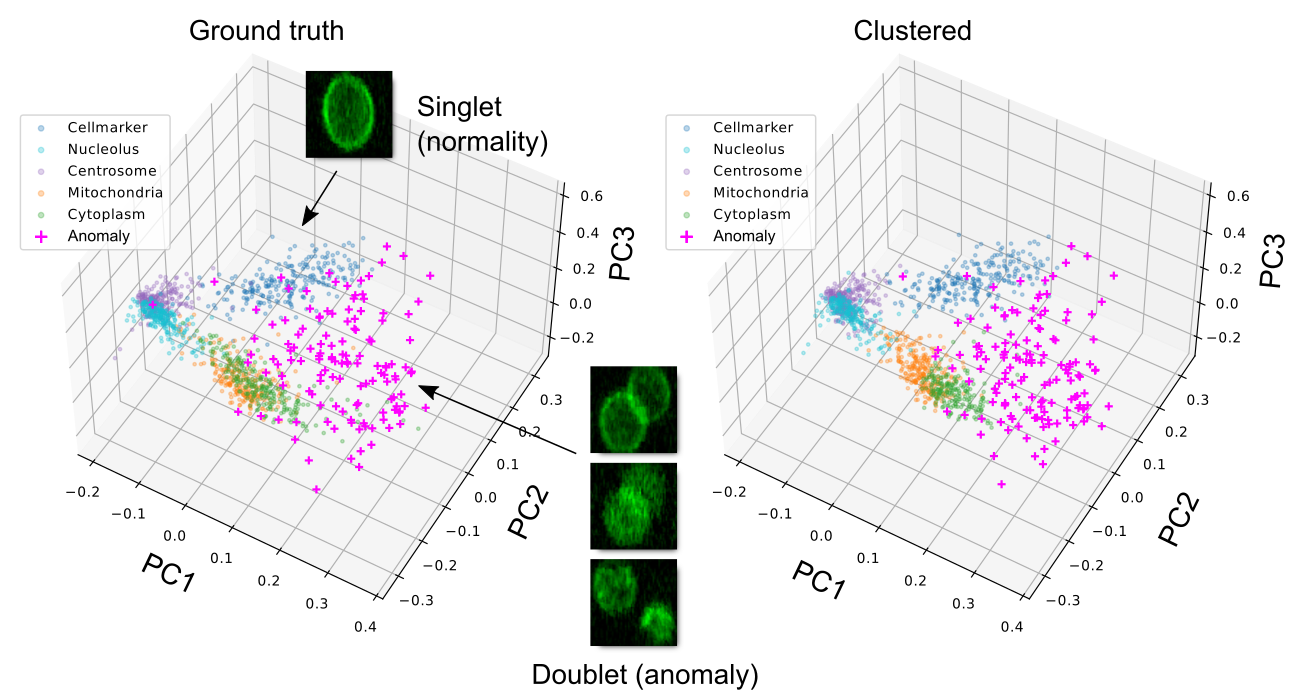}
\caption{\textbf{Visualization of the latent space of the 5-class cell-organelle dataset plus anomaly compressed by the ONN encoder and the results of spectral clustering.} The ONN encoder compresses the images from 5 normal cell classes and 1 additional anomaly class into 4-dimensional feature vectors. The first 3 principal components (PCs) are plotted. The ground truth of the 6 classes are shown in the left panel, and the clusters found by spectral clustering algorithm are shown in the right panel.}
\label{latent_space_clustering}
\end{figure}

\noindent In order to test if our ONN encoder trained for classifying regular single-cell images also has the ability to detect anomalies it had never observed before, we added to the original dataset 418 additional images, each containing at least two cells (the anomaly), from the same data repository described in \ref{cluster}. We displayed the anomaly images on the DMD in front of the ONN encoder, which was loaded with exactly the same weights previously trained to classify single-cell images only (Supplementary Figure \ref{cell_weights}), and then collected 4-dimensional feature vectors at the output of the ONN encoder. A principal component analysis (PCA) was then performed on the 4-dimensional feature vectors, and the first 3 principal components were normalized to a similar scale and plotted in Supplementary Figure \ref{latent_space_clustering}.  

The data points corresponding to the anomalous images can be identified as a cluster separate from the 5 original clusters in the latent space (Supplementary Figure \ref{latent_space_clustering} left panel). As a result, the 5 classes of single-cell images that were used for classification and the additional anomalous images could be found using unsupervised learning methods. We performed spectral clustering on the feature vectors, using a nearest neighbor distance metric to construct an affinity matrix. We found 6 clusters, each corresponding to the 5 single-cell classes plus the anomaly class. The result is shown in the plot in the right panel of Supplementary Figure \ref{latent_space_clustering}. We note that the spectral clustering algorithm only detects clusters of points without assigning class labels, but the correspondence between clusters and cell classes could be found in multiple ways (e.g., by examining original images in each cluster). After assigning the most probable class label to each cluster for the maximum overall likelihood (Supplementary Figure \ref{latent_space_clustering}), we computed the confusion matrix of classifying the original 5 classes plus the new anomaly class by comparing to the ground truth labels (Fig. 3e). The true positive rate was calculated as the percentage of anomalous images classified as anomalous images, and the false positive rate was calculated as the percentage of normal images in the total number of images classified as anomaly. 

\section{Nonlinear parameter regression} \label{npr}
\noindent As another example of how different digital backends can make use of the same latent-feature space for different tasks, we performed nonlinear parameter regression with the real-scene setup. Using the latent-feature space created by the classifier for the speed-limit signs (see \ref{realscene}), we used a 2 $\rightarrow$ 50 $\rightarrow$ 100 $\rightarrow$ 1 digital neural network backend to infer the viewing angle that the speed-limit sign was seen from. For this task (i.e., just training of the aforementioned digital backend), we used every even angle as the training set and every odd angle as the validation set. The loss function used was an L1 loss, i.e., $|\theta_{\text{predicted}} - \theta_{\text{true}}|$. The nonlinear regression can be seen in Figure 3g in the main text. We note here that the angle predictions performance is reduced if the network is required to predict viewing angle for all, rather than just one, speed-limit class at a time (see Supplementary Figure \ref{nonlinear_regression_all})

\begin{figure}[h!]
\includegraphics [width=\textwidth] {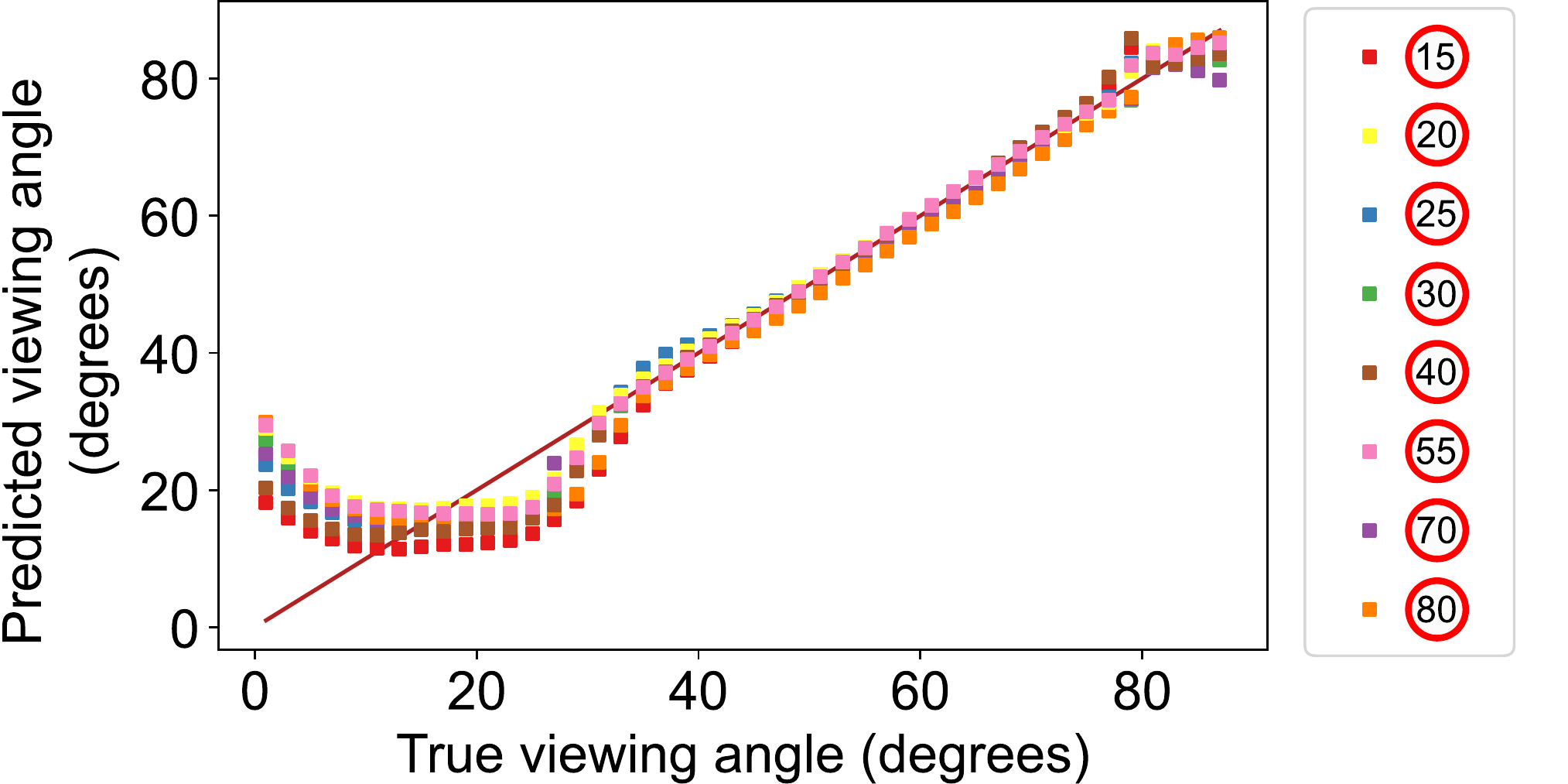}
\caption{\textbf{Viewing angle prediction for the speed-limit signs.} A digital neural architecture of 2 $\rightarrow$ 50 $\rightarrow$ 100 $\rightarrow$ 1 was used to predict the viewing angles of the speed-limit signs from the latent-feature space created by the multilayer ONN encoder for classification.}
\label{nonlinear_regression_all}
\end{figure}

\part{Simulation of Deeper Optical-neural-network Encoders for More Complex Tasks}

\section{Deeper optical neural networks for the 10-class cell-organelle classification task} \label{full_cell}

\begin{figure}[h!]
\includegraphics [width=0.8\textwidth] {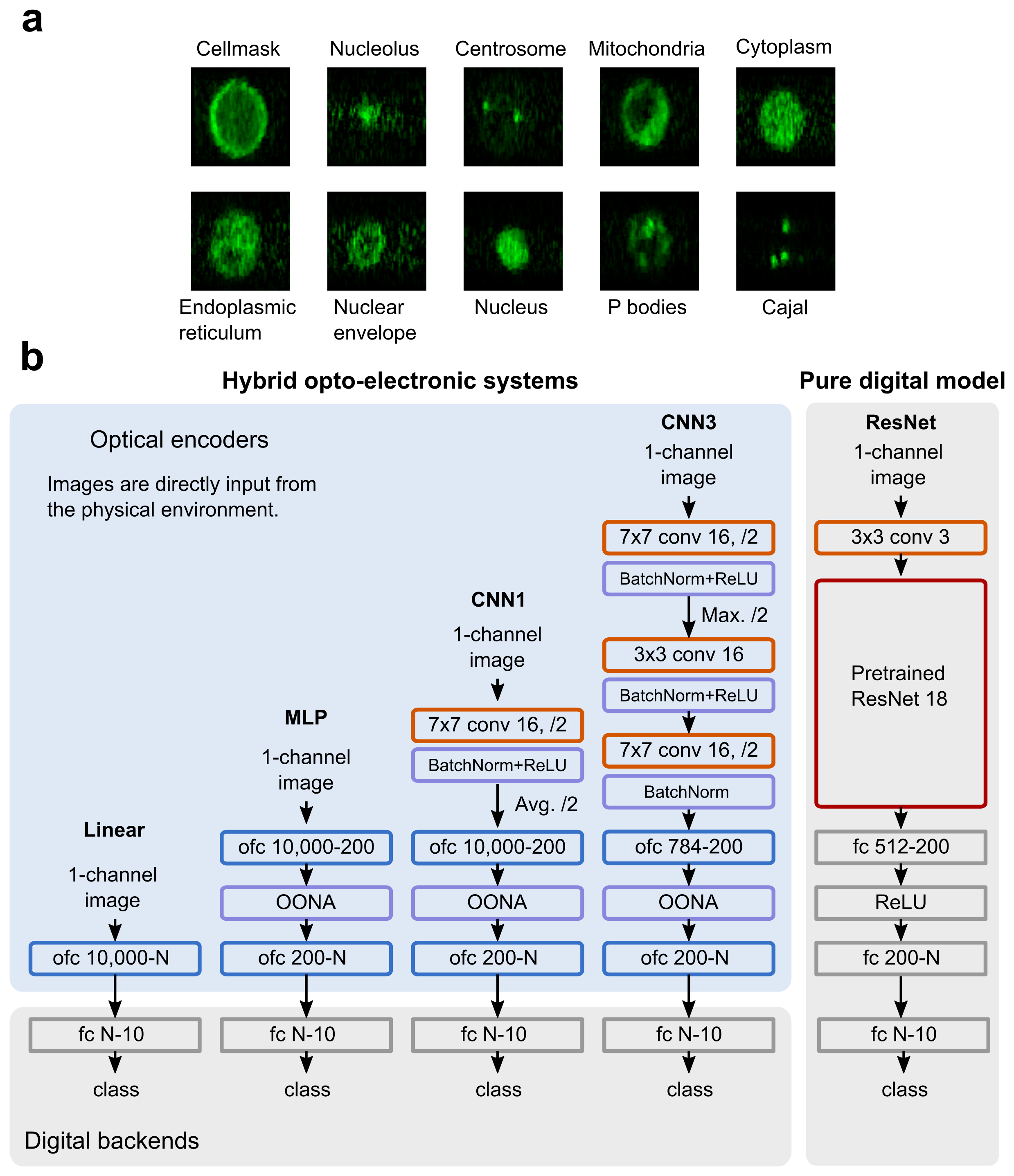}
\caption{\textbf{Simulation of the extended cell-organelle classification task.} \textbf{a}, Example images of each of the 10 cell-organelle classes. \textbf{b}, The architecture of ONN encoders simulated for the extended cell organelle classification. ofc: optical fully connected layers; OONA: optical-to-optical nonlinear activation; conv: convolutional layers; fc: (digital) fully connected layer; ReLU: Rectified Linear Unit; BatchNorm: batch normalization layer. }
\label{full_cell_NN_structure}
\end{figure}

\noindent In this section, we use realistic simulation of ONN encoders to show that they can achieve competitive performance on a more complex cell sorting task with near-term plausible architectures. We simulated several ONN-encoder models to classify images from 10 different classes of cell organelles (cellmask, nucleolus, centrosome, mitochondria, cytoplasm, endoplasmic reticulum, nuclear envelope, nucleus, P bodies, and Cajal). The cell images were downloaded from the same data repository as in \ref{cell} (S-BSST644, available from \url{https://www.ebi.ac.uk/biostudies/}). All the original images from the data repository were included for classification, including images with no cell or multiple cells in the field-of-view. Only two classes were excluded from the classification task because they either contain too few images for training (Golgi) or no fluorescent signal (control). All the images were cropped to a size of 100 by 100 pixels, centered at the center of mass of the image. There are 93,050 images for the total of 10 classes, among which 83,750 images were used for training, and 9,300 images were set aside as the test dataset. During the training session, the 83,750 images were split into train and validation sets, each containing 74,450 and 9,300 images respectively. 

We trained several models to study how additional layers affect the classification accuracy for varying image compression ratios (Supplementary Figure \ref{full_cell_NN_structure}). The image compression ratio is determined by the bottleneck dimension $N$, which determines how many photodetector pixels are needed in principle to read out the outputs of the optical encoder and convert them to digital electronic signals. It is preferred that the bottleneck dimension is as small as possible since the data throughput rate increases as $1/N$ given the total electronic data transmission bandwidth is constant. Meanwhile, it's also desired that the digital backends are as small as possible so that the total digital processing latency is minimized and the post-processing may potentially be performed in the sensor or by edge-computing devices. For all the models, their digital backends only constitute a fully connected layer of size $N$ to 10, where each output neuron corresponds to 1 of the 10 classes (Supplementary Figure \ref{full_cell_NN_structure}a). The digital backend uses real-number weights and biases. All the optical layers (i.e., in the shaded block of `Optical encoders' in Supplementary Figure \ref{full_cell_NN_structure}b) have non-negative weights. The rationale is that the input light to the system will usually be incoherent (e.g., fluorescence) and the nonlinear layer (e.g., image intensifier) also usually emits incoherent light, although coherent versions may be possible, for instance, with VCSEL arrays \cite{heuser2020developing, chen2022deep}. Therefore, the simulated models with non-negative optical weights present a lower bound on the performance achievable by ONN encoders. If one or more layers of the ONN encoder is real-valued by coherent light implementation, the performance of the model can be further improved \cite{chang2018hybrid}. In addition, $\sim 2\%$ relative noise was added to the input of each optical layer in the same manner as described in \ref{onn_training} to realistically simulate physical noise. 

The first model we simulated (Linear) is a wide fully connected layer with $100 \times 100 = 10,000$ dimensional to $N$ with non-negative weights and no bias. The linear optical encoder serves as the baseline for the accuracy achievable by any ONN without nonlinearity ($\sim$70\% on this dataset). For comparison, the linear optical encoder implemented in experiment had a fully connected layer of size $40 \times 40=1,600$ to 4.

The second model we simulated is a multilayer perceptron (MLP), which consists of two fully connected layers and one element-wise optical-to-optical nonlinear activation (OONA) layer. The OONA assumes the saturating nonlinear function measured from the image intensifier used in this experiment (Supplementary Figure \ref{figs_nonlinear_curves}). The input layer of the neural network is 10,000 dimension, and the hidden layer has 200 dimension. Besides both the input and hidden dimension
being much larger, this network is similar to the 2-layer fully connected ONN we realized experimentally.

The third model we simulated (CNN1) is a 3-layer convolutional neural network. Compared to MLP, it has one additional convolutional layer in front of the MLP. The convolutional layer used in this simulation has a non-negative-valued kernel of size $7 \times 7$: it takes 1 input channel (i.e., green fluorescence) and outputs 16 channels. Multi-channel optical convolutional layers of this kind have been realized before with 4f systems \cite{chang2018hybrid}. The nonlinear function used after the convolutional layer is a shifted ReLU activation (i.e., trained Batch Norm followed by ReLU), which could be realized with either a slight modification of the image intensifier electronics, or by the threshold-linear behavior of optically controlled VCSEL \cite{heuser2020developing} or LED arrays. We have primarily assumed pooling operations are AvgPool (we note here that this did not reduce performance in comparison to max pooling), which are straightforwardly implemented with optical summation. CNN1 leads to significant improvement of test accuracy in comparison to MLP ($>90\%$ vs $84\%$), and can be realistically implemented once a scalable technical solution to ReLU function is found.

The fourth simulated model (CNN3) is a 5-layer convolutional neural network. Compared to CNN1, CNN3 has two more non-negative convolutional layers which enabled it to reach higher accuracy, especially at a high compression ratio. We use MaxPool operation once in CNN3 after the first convolutional layer. Admittedly, this is more challenging but could plausibly be realized effectively by using a broad-area semiconductor laser or placing a master limit on the energy available to a VCSEL or LED array, such that the first unit to rise above threshold would suppress activity in others.

The fifth simulated model (ResNet) is purely digital, and is used to estimate classification results obtained on this dataset by a state-of-the-art digital machine learning model. The model uses a pretrained ResNet 18 backend \cite{he2016deep} provided by Pytorch v1.11.0 \cite{paszke2019pytorch}. Four additional layers were added to the ResNet backend to adapt input images or convert to output classification results. The entire model was retrained on the full cell-organelle dataset to finetune the weights. 

While these ONN designs are ultimately speculative; In general, we anticipate that practically realizing more powerful ONN encoders will require jointly designing compact, low-cost ONN hardware components and developing optics-friendly DNN architectures, rather than simply directly adapting existing digital DNN architectures.

The training of all the models was performed in Pytorch v1.11.0 with the AdamW optimizer. The training code is available at: \url{https://github.com/mcmahon-lab/Image-sensing-with-multilayer-nonlinear-optical-neural-networks} The 10-class cell-organelle dataset is downloadable from Zenodo: (\verb|EBI_Cells_grey_w_anomaly3.npz| in \verb|Figure_4.zip| at \url{https://doi.org/10.5281/zenodo.6888985}). 

\bibliographystyle{npjqi.bst}
\bibliography{references}